\documentclass[12pt]{article}

\usepackage{a4wide}
\usepackage{amsfonts}
\usepackage{color,colortbl}
\usepackage{epic,eepic}
\usepackage{graphicx}
\usepackage{multicol,multirow}
\usepackage[margin=1cm]{caption}
\usepackage{verbatim}
\usepackage{pgfplots}
\pgfplotsset{compat=1.5}
\usepackage{filecontents}
\usepackage{graphicx}
\usepackage{caption}
\usepackage[margin=1cm]{caption}
\usepackage{tikz} 
\usepackage{bbold}
\usepackage{graphicx}
\usepackage{amssymb}
\usepackage{amsfonts}
\usepackage{subcaption}
\usepackage[inline]{enumitem}	
\usepackage{tensor}
\usepackage{braket}
\usepackage{lmodern}
\usepackage{setspace}
\usepackage[margin=2.5cm]{geometry}
\usepackage{amsthm}
\usepackage{mathtools}
\usepackage{color}
\usepackage{sansmath}
\usepackage{accents}
\usepackage{authblk}
\usepackage[numbers]{natbib}

\usepackage{ifxetex}
\ifxetex
   \usepackage{polyglossia}
   \setdefaultlanguage{english}
   \usepackage{fontspec}
\else
   \usepackage[utf8]{inputenc}
\fi
\usepackage{amsmath}
\usepackage{graphicx,color,overpic,mathtools}
\usepackage{bm}
\usepackage{amsmath}
\usepackage{upgreek}
\usepackage{amssymb}
\usepackage{setspace}
\usepackage{hyperref}
\usepackage{xcolor}
\PassOptionsToPackage{normalem}{ulem}
\usepackage{ulem}

\usepackage{tensor}

\usepackage{overpic}
\usepackage{hyperref}
\usepackage{pgf}
\usepackage{tikz}
\usepackage{bbm}
\usepackage{bbold}
\usepackage{dsfont}

\usepackage{braket}

\newcommand{\ii}{\mathrm{i}}

\renewcommand{\d}{\mathrm{d}}
\graphicspath{ {./images/} }

\title{Eliminating the `impossible':  Recent progress on local measurement theory for quantum field theory}

\author[1]{Maria Papageorgiou}
\author[2]{Doreen Fraser}
\affil[1]{Department of Applied Mathematics and Institute for Quantum Computing, University of Waterloo, Canada and
Division of Theoretical and Mathematical Physics, University of Patras, Greece mepapageorgiou@uwaterloo.ca}
\affil[2]{Department of Philosophy, University of Waterloo, Canada dlfraser@uwaterloo.ca}

\begin{document}

\maketitle

\noindent Arguments by Sorkin \cite{sorkin1993impossible} and Borsten, Jubb, and Kells \cite{Borsten:2019cpc} establish that a natural extension of quantum measurement theory from non-relativistic quantum mechanics to relativistic quantum theory leads to the unacceptable consequence that expectation values in one region depend on which unitary operation is performed in a spacelike separated region. Sorkin \cite{sorkin1993impossible} labels such scenarios `impossible measurements'. We explicitly present these arguments as a no-go result with the logical form of a reductio argument and investigate the consequences for measurement in quantum field theory (QFT). Sorkin-type impossible measurement scenarios clearly illustrate the moral that Microcausality is not by itself sufficient to rule out superluminal signalling in relativistic quantum theories that use L{\"u}ders' rule. We review three different approaches to formulating an account of measurement for QFT and analyze their responses to the `impossible measurements' problem. Two of the approaches are:  a measurement theory based on detector models proposed in Polo-G{\'o}mez, Garay, and Mart{\'i}n-Mart{\'i}nez \cite{Polo_Gomez_2022} and a measurement framework for algebraic QFT proposed in Fewster and Verch \cite{fewster2020quantum}. Of particular interest for foundations of QFT is that they share common features that may hold general morals about how to represent measurement in QFT. These morals are about the role that dynamics plays in eliminating `impossible measurements', the abandonment of the operational interpretation of local algebras $\mathcal{A}(O)$ as representing possible operations carried out in region $O$, and the interpretation of state update rules. Finally, we examine the form that the `impossible measurements' problem takes in histories-based approaches and we discuss the remaining challenges.

\bigskip

\noindent \textbf{Statements and Declarations}: DF and MP gratefully acknowledge support from a Social Sciences and Humanities Research Council of Canada Insight Grant number 435-2019-0548. MP acknowledges support of the ID$\#$ 62312 grant from
the John Templeton Foundation, as part of the \href{https://www.templeton.org/grant/thequantum-
information-structure-ofspacetime-qiss-second-phase}{‘The Quantum Information
Structure of Spacetime’ Project (QISS)}.

\medskip

\noindent \textbf{Acknowledgements}:  Special thanks to Jos{\'e} de Ram{\'o}n Rivera for continuous feedback on this work, and for co-organizing with us a pandemic reading group on the FV framework (and to the participants). Many thanks to Maximilian Ruep for useful correspondence. We would also like to thank Rafael Sorkin, Eduardo Mart{\'i}n-Mart{\'i}nez, Achim Kempf, Charis Anastopoulos, Luis Garay, Chris Fewster, Rainer Verch, Ian Jubb, Adamantia Zampeli, Jos{\'e} Polo-G{\'o}mez, Dan Grimmer, Jason Pye, Robert Oeckl, Francisco Calder{\'o}n, Laura Ruetsche, John Earman, Gordon Belot, Dave Baker, Emily Adlam, Jeremy Butterfield, Tein van der Lugt, Simon Saunders, and Nick Huggett. Thanks also to audiences at the QISS 2022 conference at Western, RQI-N 2022, Foundations of QFT conference at Western, and the Oxford Philosophy of Physics seminar. DF and MP gratefully acknowledge support from a Social Sciences and Humanities Research Council of Canada Insight Grant. MP acknowledges support of the ID$\#$ 62312 grant from
the John Templeton Foundation, as part of the \href{https://www.templeton.org/grant/thequantum-
information-structure-ofspacetime-qiss-second-phase}{‘The Quantum Information
Structure of Spacetime’ Project (QISS)}.

\newpage

\section{Introduction}\label{introduction}

Non-relativistic quantum mechanics (NRQM) has an accepted measurement theory for laboratory measurements (see, e.g., Busch et al. \cite{busch_book_2016}, a standard reference on Quantum Measurement Theory). There is no consensus about how to interpret NRQM (including this measurement theory); this is the infamous Measurement Problem. However, the recipe for extracting probabilistic predictions for measurement outcomes from NRQM is uncontroversial. In contrast, it is far from straightforward to formulate an analogous measurement theory for local laboratory measurements in relativistic quantum field theory (QFT). As Sorkin's 1993 paper ``Impossible measurements on quantum fields'' \cite{sorkin1993impossible} illustrates, the natural generalization of the non-relativistic measurement scheme to relativistic quantum theory fails because it entails superluminal signalling. Sorkin uses a minimal theoretical framework for relativistic quantum theory to construct his examples of impossible measurements, but articulating an adequate measurement theory has also been a longstanding problem in more comprehensive axiomatic formulations of QFT such as algebraic QFT. Philosophers Earman and Valente declare that the lack of a measurement theory for algebraic QFT is ``a major scandal in the foundations of quantum physics'' \cite[p.17]{earman2014relativistic}. In a recent paper that aims to make amends for this, Fewster and Verch \cite{fewster2020quantum} remark that there has been a ``gap" between the fields of quantum measurement theory and algebraic QFT that ``has--surprisingly--lain open for a long time."

We take Sorkin's examples of `impossible measurements' and the recent extensions by Borsten, Jubb, and Kells in \cite{Borsten:2019cpc} as a starting point for investigating possible formulations of a measurement theory for QFT. We approach the problem of formulating a measurement theory for QFT from the disciplinary perspectives of philosophy and relativistic quantum information (RQI). Our ultimate goal is to explore the landscape of recent proposals for a measurement theory for QFT. There has been intense recent interest in this topic, so we will not be able to offer a comprehensive survey. We will instead focus our attention on a few well-developed proposals that are different in spirit, and will clarify both their differences and their similarities. The two main approaches that we will consider are the detector models approach prominent within Relativistic Quantum Information that motivated the proposal for a detector-based measurement theory by Polo-G{\'o}mez, Garay, and Mart{\'i}n-Mart{\'i}nez in \cite{Polo_Gomez_2022} (see also Chapter 3 of \cite{SmithAlexander2017}) and the framework for measurement in algebraic QFT (AQFT) presented by Fewster and Verch in \cite{fewster2020quantum}. Both of these proposals successfully address the `impossible measurements' problem. We will also consider history-based formalisms, which is Sorkin's preferred approach. As we shall discuss, how to eliminate `impossible measurements' in this framework is an open problem. Our aim is not to advocate for any one of these proposals. On the contrary, one of our conclusions is that the proposals are designed to address different problems and are currently each suitable for different purposes. Measurement theory for QFT is still an area of active research, so there is no one settled formulation that satisfies all desiderata. 

Sorkin \cite{sorkin1993impossible} and Borsten et al. \cite{Borsten:2019cpc} furnish an excellent starting point for understanding the current situation because addressing this `impossible measurement' problem is widely regarded as the first order of business for establishing a measurement theory for QFT. As we shall explain, these impossible measurement scenarios rely on assumptions that can be framed as an informal no-go result. This \textit{reductio ad absurdum} argument is useful because it identifies an apparently reasonable set of premises that lead to an unacceptable conclusion. The premises include the basic elements of ideal measurement theory for quantum mechanics, including L{\"u}ders' rule for state update for non-selective measurements. This measurement theory is adapted to Minkowski spacetime by making the natural assumption that causal order defines a partial temporal order. The Microcausality principle that operators associated with spacelike separated regions commute is also imposed. When the system is not being measured, the Heisenberg picture representation for the dynamics is used. There are examples of `impossible measurement' scenarios that comply with all of these requirements, and yet the expectation values for a measurement confined to one bounded region depend on which non-selective measurement is carried out in a spacelike separated bounded region. This conclusion is clearly unacceptable because it violates the prohibition on superluminal signalling or information transfer that is typically understood to be a hallmark of relativistic theories.

Different approaches to formulating an account of measurement for QFT can be classified according to how they respond to this \textit{reductio} argument by revising, rejecting, or adding premises. The detector models approach has the pragmatic goal of representing detectors that are actually used to theoretically and experimentally investigate the measurement of relativistic fields, typically in quantum optics and quantum information (e.g., Unruh-DeWitt dectectors). The response to the `impossible measurements' reductio argument is to use NRQM, and not QFT, to model the detectors. This allows ideal measurement theory for NRQM to be applied to the detector model without (for all practical purposes) leading to superluminal signalling, provided that care is taken to satisfy relativistic constraints when coupling the detector and field. Essentially, the addition of these assumptions for a concrete detector model to Sorkin's premises is what excludes FAPP the possibility of `impossible measurements' in the detector model's regime of applicability. The Fewster-Verch (FV) framework for measurement in algebraic QFT presented in \cite{fewster2020quantum} takes a different approach that begins with general physical principles for QFT. Fewster and Verch adopt axioms for AQFT that go beyond the minimal set of physical principles assumed by Sorkin. These additional physical principles entail that ideal measurement theory cannot be extended from NRQM to QFT in the straightforward manner posited by Sorkin. The FV framework also rejects some of Sorkin's premises about ideal measurement theory. In particular, the assumption that L{\"u}ders' rule is applied to determine the post-measurement state of the system is rejected. The FV framework introduces a new measurement theory for AQFT with new state update rules that are informed by the physical principles of AQFT. Histories-inspired approaches also reject the assumption that L{\"u}ders' rule is applicable, but do not aim to introduce state update rules that describe the measurement process `step-by-step'; instead, probabilities are directly assigned to entire histories.

It is worth emphasizing, as Sorkin also emphasizes, that the `impossible measurements' problem is a separate issue from the Measurement Problem. In NRQM, the Measurement Problem originally arose after the standard theory of NRQM (including dynamics) was formulated and a rudimentary measurement theory was introduced. In brief, one version of the Measurement Problem in NRQM is that the unitary quantum dynamics (e.g., as given by the Schr{\"o}dinger equation) is inconsistent with the prescription for state update after measurement (e.g., as given by L{\"u}ders' rule). In general, the Schr{\"o}dinger equation determines that the composite of the system and measuring device ends up in an entangled state that is not an eigenstate of the measured quantity, while the L{\"u}ders' rule for selective measurement assigns an eigenstate of the measured quantity. (Furthermore, it is the state update rule that seems to be correct about the post-measurement state.) The Measurement Problem in QFT should be set up in an exactly analogous way. This means that before the Measurement Problem in QFT can be posed, the physical theory of QFT (including dynamics) and a measurement theory for QFT must be fixed. The `impossible measurements' problem pertains to how the physical theory of QFT and a measurement theory for QFT are formulated. In this sense, the `impossible measurements' problem for QFT arises prior to the Measurement Problem for QFT and the ensuing interpretational issues. This means that addressing the `impossible measurements' problem does not require a solution to the Measurement Problem for QFT. However, the solution to the `impossible measurement' problem that is adopted may well affect the form that the Measurement Problem takes in QFT. In general, both the physical theory of QFT and the measurement theory for QFT differ from NRQM; therefore, the Measurement Problem may take different forms in QFT and NRQM.

It is also useful to consider the historical context of the `impossible measurement' problem. Of course, measurement \textit{is} possible in QFT; it is commonplace to use QFTs to make theoretical predictions for measurements conducted on relativistic quantum systems. In QED, for example, standard predictions take the form of scattering amplitudes, which involve asymptotic states. This is in contrast to NRQM, in which predicted quantities typically take the form of properties of an instantaneous state at a finite time or a stationary state. Blum \cite[p.46]{blum_state_2017} characterizes this historical shift from a focus on states in NRQM to scattering theory in QED as a paradigm shift because it constitutes a significant change in the paradigmatic problem of what is to be calculated from the theory. Blum offers an illuminating account of how the quantum state ``withers away" in two lines of development of relativistic quantum theory in the 1930's and 1940's, one that originates with Heisenberg's S-matrix theory and the other with Wheeler-Feynman electrodynamics. As Blum explains, this paradigm shift was prompted both by the desideratum of obtaining an explicitly relativistic formulation of quantum theory and by the need for a calculationally tractable theory. (See \cite{Fraser2023} for further discussion of the historical background.)

Asymptotic scattering theory works well for predictions for many experimental scenarios, especially in particle physics, but does not cover all cases of interest. Relativistic quantum information is a field in which finite time processes that occur in a local laboratory environment are theoretically and experimentally investigated. Consequently, a measurement theory for local measurements that applies to relativistic quantum theory at finite times is needed. Sorkin \cite{sorkin1993impossible} concerns precisely this problem. For this reason, \cite{sorkin1993impossible} has been influential in the relativistic quantum information community more broadly \cite{dowker2011useless, PhysRevD.65.065022, PhysRevA.64.052309, Benincasa_2014,deramon2021relativistic}. Recently, the issue of how to formulate a measurement theory for QFT has attracted renewed attention, in part due to Borsten et al.'s \cite{Borsten:2019cpc} sharpening of Sorkin's results and Fewster and Verch's proposed measurement theory for algebraic QFT \cite{fewster2020quantum}. 

We begin by explicitly formulating the reductio argument that underlies Sorkin-type impossible measurement scenarios. After rehearsing Sorkin's \cite{sorkin1993impossible} original examples of impossible measurement scenarios and Borsten, Jubb, and Kells' \cite{Borsten:2019cpc} recent examples, we use the reductio argument to analyze the root causes of the `impossible measurements' problem and to classify different approaches to formulating an account of measurement for QFT. Sec. \ref{FVsection} and \ref{detectormodelssec} offer overviews of the FV framework and the detector models approach, respectively, focusing on analysis of how each approach addresses the reductio argument and rules out impossible measurement scenarios. In Sec. \ref{comparisonsection}, we compare the FV and detector-based measurement theories. Our comparison does not focus exclusively on the important differences; we also identify substantial similarities between the two measurement theories. The similarities are of particular interest because, given the differences in goals and strategies employed by the two approaches, similarities suggest morals about general features that might be shared by any measurement theory for QFT. In Sec. \ref{consistenthistories} we consider how the histories-inspired approaches favored by Sorkin address impossible measurement scenarios. Sec. \ref{conclusion} summarizes our conclusions. We hope that this paper lays the groundwork for productive dialogue among the many communities of physicists and philosophers who are working on theoretical, practical, and interpretative issues surrounding the treatment of local measurements in QFT.

\section{Sorkin \cite{sorkin1993impossible} and Borsten, Jubb, and Kells \cite{Borsten:2019cpc} `impossible measurements'}

Sorkin's original paper \cite{sorkin1993impossible} presents examples of `impossible measurement' scenarios that exhibit dependence of the expectation values of a measurement in one region on the identity of the measurement operation performed in a spacelike separated region. The purpose of these examples is to show that ideal measurement theory cannot be na\"{i}vely extended  from NRQM to relativistic quantum theory. Borsten, Jubb, and Kells \cite{Borsten:2019cpc} explicate Sorkin's assumptions and introduce additional examples of `impossible measurements'. In \ref{reductio} we begin by reviewing the set of assumptions that gives rise to these `impossible measurement' scenarios and explicitly cast the argument in the logical form of a \textit{reductio ad absurdum} argument. We then take up examples of `impossible measurement' scenarios in Sec. \ref{sorkinsec} and \ref{borstensec}. This will be followed by analysis of conclusions that are supported by the reductio argument and a survey of strategies for responding to the reductio argument in Sec. \ref{Secadhoc}. 

\subsection{The `impossible measurements' reductio argument}\label{reductio} 

The `impossible measurement' scenarios presented by Sorkin \cite{sorkin1993impossible} and Borsten, Jubb, and Kells \cite{Borsten:2019cpc} are a type of no-go result. No-go results such as Bell's theorem have played an important role in foundations of quantum theory because they identify a set of assumptions that cannot all be true. (Conditional on the conclusion being false, which in the case of Bell's theorem is established by closing the loopholes in the experimental tests of the Bell inequalities.) Similarly, the `impossible measurement' scenarios are valuable because they play the role of identifying a set of assumptions that cannot all be true. (Conditional on the conclusion being false, which in this case is established by experimental tests that rule out superluminal signalling.) In relativistic quantum information (RQI), `impossible measurement' results have played the heuristic roles of motivating the formulation of models of local measurement that are suited to QFT and serving as a criterion of adequacy for proposed local measurement models for QFT (i.e., they must not permit detectable signalling in Sorkin-type measurement scenarios). We will also use the `impossible measurements' results to classify different approaches to formulating a measurement theory for QFT (Sec. \ref{Secadhoc}). For these heuristic purposes, it is useful to extract a \textit{reductio ad absurdum} argument from the examples of `impossible measurements'. A reductio argument is an argument in which an apparently acceptable set of premises leads by apparently acceptable reasoning to an apparently unacceptable conclusion. 

Both Bell's theorem and the argument based on `impossible measurement' examples that is set out below  take the form of reductio arguments. An important difference between these two arguments is that Bell's theorem is a no-go theorem provable using mathematics from mathematically-stated premises (i.e., a deductive argument) while the `impossible measurements' reductio argument is a more informal no-go result (i.e., the conclusion is established by producing examples of scenarios in which superluminal signalling is possible). This is a significant difference, and one that makes Bell's theorem much more powerful, but for our purposes what is important is that the informal `impossible measurement' reductio argument serves the heuristic functions of motivating and guiding the formulation of a measurement theory for QFT. 

Relatedly, Sorkin assumes only a minimal, informal framework for relativistic quantum theory. Assume that (when no measurements occur) there is a Heisenberg picture representation of some quantum field $\Phi$ (e.g., a free scalar quantum field) and an observable $A_k$ is associated with a region of Minkowski spacetime $O_k$ by restriction of the field $\Phi$ to $O_k$. Microcausality is the only principle from QFT that is assumed. This starting point is intended to invoke only generally agreed upon features of relativistic quantum theory. Following the presentation of `impossible measurement' examples in Borsten, Jubb, and Kells \cite{Borsten:2019cpc}, here is the reductio argument:

\begin{quote}

\textbf{P1 Local degrees of freedom} An observable $A_k$ is associated with a region of Minkowski spacetime $O_k$ by restriction of the field $\Phi$ to $O_k$. \cite{sorkin1993impossible}

\smallskip

\textbf{P2 Dynamics} When measurements are not being performed, use the Heisenberg picture representation (i.e., time-dependence is carried by the observables).

\smallskip

\textbf{P3 Ideal measurement theory for relativistic quantum theory}

\textbf{(a) Detection assumptions}:  

(i) eigenstate-eigenvalue link:  ``the measurement outcomes are the eigenvalues of the self-adjoint operator corresponding to the observable" \cite{Borsten:2019cpc}

(ii) Born rule:  In a state $\rho$, the probability of an outcome $n$ that corresponds to a projector $E_n$ is given by $\text{Prob}(n)=\text{tr}(\rho E_{n})$.

\smallskip

\textbf{(b) Preparation assumption}:  The state $\rho (t ^\prime)$ at time $t ^\prime$ after a non-selective measurement is determined by applying L{\"u}ders' rule (for non-selective measurement) to the state at time $t$ prior to the measurement.

\smallskip

\textit{L{\"u}ders' rule for non-selective measurement for arbitrary self-adjoint observables}\footnote{This is a generalisation of L{\"u}ders' rule for non-selective measurement for discrete observables: For a compact self-adjoint observable A, $A=\sum_n \lambda_n E_n$, where $\lambda_n$ are distinct eigenvalues and $E_n$ are associated projectors onto associated eigenspaces that resolve the identity. A selective measurement is conditioned on obtaining the outcome $\lambda_n$. A non-selective measurement is not conditioned on obtaining any particular outcome. L{\"u}ders' rule for non-selective measurement for discrete observables: $\rho(t) \rightarrow \rho (t^\prime) = \sum_n E_n \rho(t) E_n$}

By the spectral theorem, $A=\int^{\infty} _{- \infty} \lambda dE(\lambda)$ where $E(\cdot)$ maps Borel subsets $B \subseteq \mathbb{R}$ to projectors on $\mathcal{H}$. For a set of mutually disjoint Borel sets $\mathcal{B}=\{B_n \}_{n \in \mathcal{I}}$ that covers $\mathbb{R}$ (with $\mathcal{I}$ some countable indexing set), each $B_n$ represents a possible bin for a measurement outcome. The corresponding projectors\footnote{not necessarily rank-1} $E_n  := E(B_n)$ resolve the identity $\sum_{n \in \mathcal{I}} E_n = \mathds{1}_\mathcal{H}$. L{\"u}ders' rule for non-selective measurement for arbitrary self-adjoint observables:

\begin{equation}
\rho(t) \rightarrow \rho (t^\prime) :=  \mathcal{E}_{A, \mathcal{B}} \left( \rho(t) \right) = \sum_n E_n \rho(t) E_n \label{luders}
\end{equation}
\textbf{(c) Relativistic temporal ordering}:   Define the temporal ordering relation $O_j \prec O_k$ iff some point of $O_j$ causally precedes some point of $O_k$ in the spacetime. Take the transitive closure of $\prec$. Regions must be chosen such that this extended $\prec$ is a partial order (i.e., cannot have both $O_j \prec O_k$ and $O_k \prec O_j$).\footnote{As Sorkin \cite[p.3]{sorkin1993impossible} notes, $\prec$ may be extended to some non-unique linear order. P4 Microcausality ensures that different choices of linear order do not affect the expectation values for any sequence of projective measurements associated with the set of regions.}

\textbf{P4 Microcausality}: If $O_j$ and $O_k$ are spacelike separated, then $[ A_j , A_k ] = 0$ for all $A_j \in \mathcal{A}(O_j)$, $A_k \in \mathcal{A}(O_k)$.

\textbf{C Conclusion} There are bounded, spacelike separated regions $O_1$ and $O_3$ for which the expectation values of a measurement confined to $O_3$ depends on which unitary operation is performed in $O_1$.

\end{quote} 

That the existence claim in the conclusion of the argument is compatible with the premises is established by the examples set out in the following two subsections. Premise P3 sets out the assumptions of ideal measurement theory for relativistic quantum theory. Parts (a) the detection assumption and (b) the preparation assumption are carried directly over from NRQM. The relativistic ingredient of P3 is (c), which specifies a temporal ordering relation for regions in Minkowski spacetime. P4 Microcausality is an uncontroversial assumption within QFT.

The conclusion is that there are bounded, spacelike separated regions $O_1$ and $O_3$ for which the expectation values of a measurement confined to $O_3$ depends on which unitary operation is performed in $O_1$. In both of the responses to Sorkin-type impossible measurement scenarios discussed in Sec. \ref{FVsection} and \ref{detectormodelssec} below the operation performed in region $O_1$ is implemented by a non-selective measurement. A further argument can be made that the conclusion of the reductio argument is unacceptable because it allows for superluminal signalling. As a consequence of the detection assumptions P3(a), the probabilities for measurement outcomes in $O_3$ are dependent on which measurements are carried out in spacelike separated region $O_1$. If we assume that parties can make multiple measurements on identically prepared systems to build up statistics following Borsten et al. \cite{Borsten:2019cpc}, then in principle an observer in $O_{3}$ could determine whether a measurement was carried out in spacelike separated region $O_{1}$.This violates the prohibition on superluminal signalling or information transfer that is typically understood to be a hallmark of relativistic theories. The reason for labeling the regions $O_1$ and $O_3$ will become apparent in the next section:  the examples of superluminal signalling involve measurements in another intermediate region $O_2$. We will refer to this reductio argument as the `impossible measurements' reductio argument, but it should be appreciated that while Sorkin was (as far as we are aware) the first to raise this problem, Borsten, Jubb, and Kells \cite{Borsten:2019cpc} make important contributions that refine the argument.

\subsection{Sorkin's examples of impossible measurements} \label{sorkinsec}
 
 Sorkin offers two versions of his no-go result, a QFT version and a QM version with qubits on Minkowski spacetime. Since we will argue that the QM version is not compelling, we will begin by reviewing the QFT version. Consider $O_1$ and $O_3$, two bounded spacelike separated regions of Minkowski spacetime, and a unitary element of the local algebra $\mathcal{A}(O_1)$ that is characterized by a parameter $\lambda$, i.e., $U_{\lambda}\in \mathcal{A}(O_1)$. This can be thought of as a local unitary `intervention' that will transform the state of the field $\ket{\psi_0}\rightarrow U_{\lambda}\ket{\psi_0}:=\ket{\psi_1}$. Independent of the interpretation of this `local kick', prohibition of superluminal signalling entails that expectation values of observables outside the causal future of $O_1$ should not depend on the value of $\lambda$. In this case of two spacetime regions, this is guaranteed by Microcausality, which imposes $[U_{\lambda},C]=0 \,\,\, \forall \lambda$ and for all $C \in \mathcal{A}(O_{3})$. As a result of Microcausality
\begin{align}
   \langle \psi_1 |C| \psi_1\rangle &= \langle \psi_0| U_{\lambda}^*\, C \,U_{\lambda}| \psi_0\rangle \nonumber \\
   &= \langle \psi_0 | C  U_{\lambda}^* U_{\lambda} | \psi_0 \rangle \nonumber \\
   &= \langle \psi_0 |C| \psi_0\rangle \label{nolambda}.
\end{align}
 This expectation value is independent of $\lambda$, and so the value of $\lambda$ cannot be used to signal to spacelike separated regions.
 
\begin{figure}
\centering
\includegraphics[width=0.8\textwidth]{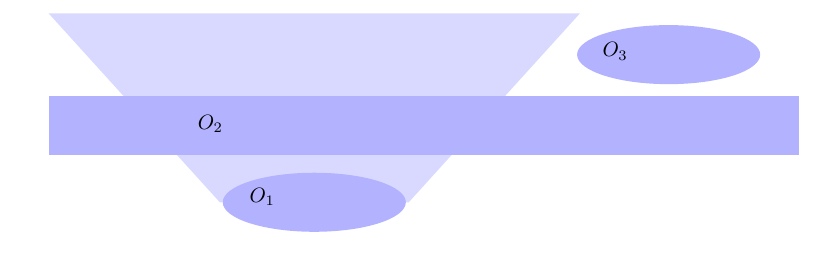}
\hspace{-2cm}
\caption{Region $O_2$ is a thickened hypersurface between $O_1$ and $O_3$.}
\label{fig1}
\end{figure}

\begin{figure}
\centering
\includegraphics[width=0.8\textwidth]{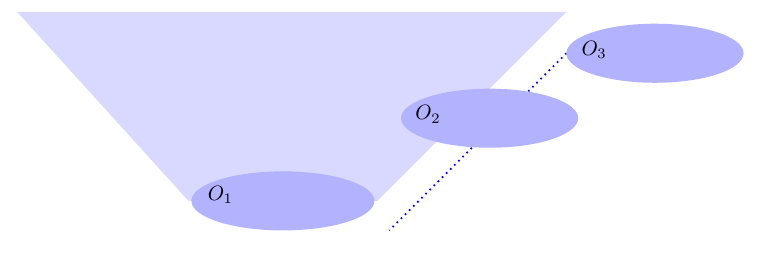}
\hspace{-2cm}
\caption{Region $O_2$ partially invading the future lightcone of $O_1$ and the past lightcone of $O_3$.}
\label{fig2}
\end{figure}

The situation changes dramatically if one considers a third region $O_2$ `between' $O_1$ and $O_3$ that is partially in the causal future of $O_1$ and partially in the causal past of $O_3$. Roughly speaking, this third region can `link' the first two in counter intuitive ways. The region $O_2$ is chosen by Sorkin to be a thickened hypersurface that lies in the chronological future of $O_1$ and in the chronological past of $O_3$ (see Figure \ref{fig2}). Associated with $O_2$ is a non-selective measurement of the projector $P_2=\ket{\psi_2}\bra{\psi_2}$. Applying the non-selective L{\"u}ders' rule to the state $\ket{\psi_1}= U_{\lambda}\ket{\psi_0}$, it is easy to see that the expectation values of $C$ is
 \begin{equation}
     \langle C \rangle= \langle U_{\lambda}^* P_2 C P_2U_{\lambda}\rangle_0+ \langle U_{\lambda}^* (\mathbb{1}-P_2) C (\mathbbm{1}-P_2)U_{\lambda}\rangle_0, \label{lambda}
 \end{equation}
 where we denote with $\langle...\rangle_0$ the expectation value over the state $\ket{\psi_0}$. This expression is equal to $\text{prob}(P_2=1)\text{Exp}(C,P_2=1)+ \text{prob}(P_2=0)\text{Exp}(C,P_2=0)$ and will generally depend on $\lambda$. Sorkin is choosing a particular state $\ket{\psi_2}$ to be a superposition of the vacuum and a one-particle state to demonstrate the $\lambda-$dependence, but the details of the derivation are not important. One simply has to notice that, in general, the $\lambda-$dependence on the r.h.s. of \eqref{lambda} will not drop out (as it did in \eqref{nolambda}) since $U_{\lambda}$ is guaranteed to commute with $C$ but \textit{not} with $P_2\in \mathcal{A}_2(O_2)$, because $O_1$ and $O_2$ are not spacelike separated. This $\lambda$-dependence instantiates the conclusion of the no-go result, since it allows for superluminal signalling between the spacelike separated regions $O_1$ and $O_3$ (because, in principle, a signal can be encoded in the value of $\lambda$).

To fully appreciate the no-go result, it is important to analyse the role of region $O_2$ in terms of the premises that are laid down in the previous section. Microcausality (P4) provides the ground for thinking of regions $O_1$ and $O_3$ as `separate' or statistically independent in a \emph{bipartite} scenario. By invoking a third `intervention region' $O_2$ we open up the possibility of signalling between $O_1$ and $O_3$ (Conclusion). This is because the non-selective measurement that is associated with region $O_2$ updates or `prepares' the state over which the expectation value of $C$ is evaluated, in accordance with the standard rules set out in P3. More explicitly, the preparation assumption (b) (L\"{u}ders' rule) is used for the measurement over $O_2$, while the detection assumptions (a) go into the evaluation of the expectation value of $C$. A temporal ordering relation $t<t^{\prime}$ is needed to apply L{\"u}ders' rule. Premise (c) defines a relativistic temporal ordering relation, which reflects the causal structure of Minkowski spacetime, i.e., $O_i \prec O_j$ if $O_j$ is partially in the causal future of $O_i$. Before the transitive closure is taken, this ordering relation does not apply for regions $O_1$ and $O_3$, that is, they are not `comparable' and it does not hold that $O_1 \prec O_3$. Based on this ordering relation we can only claim that $O_1 \prec O_2$ and $O_2 \prec O_3$. Once we take the transitive closure to obtain a partial order, $O_1 \prec O_2$ and $O_2 \prec O_3$ implies $O_1 \prec O_3$ and we can apply the measurement rules accordingly. Then, the influence, or signalling, between the regions $O_1$ and $O_3$ is `mediated' by region $O_2$, and it was made possible through taking the transitive closure (as Sorkin points out in \cite{sorkin1993impossible}). Perhaps the involvement of a third region would partially demystify the conclusion, but from a local perspective of the observers that one could associate with $O_1$ and $O_3$, the \emph{non-selective} measurements over $O_2$ should be irrelevant. Thus, one of the problems posed by this example is the consistent description of multi-partite measurements (involving more than two parties) in relativistic spacetimes.\footnote{For an arbitrary number of measurements, one would have to extend the partial order to a total order (which always exists, but is not unique).}

Sorkin also offers a `baby' QM version of the no-go result. In this case, there are two qubits that one can think of as embedded over regions $O_1$ and $O_3$ in Minkowski spacetime. The two qubits are initially in an entangled state, and the first one can potentially be flipped by a local unitary operation (analogue to the local unitary over region $O_1$) before a global projector is applied to the total system (analogue to the non-selective measurement over $O_2$). Evidently the expectation values of observables of the second qubit (analogue to $O_3$) will generally depend on whether the first qubit was flipped or not before the global operation. This is not surprising because the global projection presupposes some notion of global access to the total system.  Sorkin suggests that this example is ``[i]n a sense ... all we need, since one would expect to be able to embed it in any quantum field theory which is sufficiently general to be realistic" (p.7).

While it is true that NRQM should somehow be related to QFT, it does not follow that the QM example is sufficient for Sorkin's purposes. Precisely how NRQM relates to QFT is a non-trivial and somewhat controversial matter, as the discussion below of detector models and the FV framework for algebraic QFT will highlight. It is not obvious which features of Sorkin's QM example should be expected to carry over to QFT. The value of Sorkin's quantum field theoretic example is that it clearly demonstrates which set of assumptions adapted from NRQM cannot be transferred to QFT. Furthermore, there are disanalogies between the two examples that seem relevant. In the case of the two qubits there is no third `disjoint' party. The `third' system is simply the total system. Of course, operations over the total system are by definition global. In the QFT example, there is a non-trivial third party $O_2$, seemingly `disjoint' from $O_1$ and $O_3$. Nevertheless, that third party is responsible for an operation which, loosely speaking, would also `connect' $O_1$ and $O_3$. Relatedly, Weinstein \cite{weinstein_superluminal_2006} points out that Sorkin's QM example implicitly assumes that ideal measurements are instantaneous. Another disanalogy is that the initial state of the QM system must be entangled over the qubits, while there are no restrictions on the initial state in the QFT example.\footnote{This is more obvious in an example given by Borsten et al. \cite{Borsten:2019cpc} that is presented in the next section, which uses a factorized state. See Sec. \ref{Secadhoc} for discussion.} These disanalogies seem to undermine the usefulness of the QM example.

\subsection{Borsten, Jubb, and Kells' examples of impossible measurements}\label{borstensec} 

 At first glance, Sorkin's QFT example that illustrates his no-go result seems to be very particular to the choice of $O_2$ to be a thickened hypersurface (Figure \ref{fig1}). It is definitely bothersome, but not really surprising, that such \emph{global} operations, like the one over region $O_2$, can cause signalling between the two spacelike separated parties. The global projector represents an operation that presupposes some notion of global access to the total system. Sorkin recognizes this shortcoming of his example, but insists that there is still a genuine problem for QFT:  ``[i]n a way it is no surprise that a measurement such as of [$A_2$], which occupies an entire hypersurface, should entail a physical non-locality; but surprising or not, the implications seem far from trivial...What then remains of the apparatus of states and observables, on which the interpretation of quantum mechanics is traditionally based?"
 
 Unfortunately, the problem raised by Sorkin cannot be easily dismissed by simply excluding global operations. Borsten, Jubb, and Kells \cite{Borsten:2019cpc} supply examples that establish that the problem persists for general \emph{bounded} regions $O_2$ that partially invade the future lightcone of $O_1$ and the past lightcone of $O_3$, i.e., $J^+(O_1)\bigcap O_2 \neq \emptyset$ and $J^-(O_3)\bigcap O_2 \neq \emptyset$\footnote{The causal future/past $J^{+/-}(x)$ of a spacetime point $x$ is the set of all points reached from $x$ by smooth future-directed causal curves. For a spacetime region $O$ we write $J^{^\pm}(O)= \bigcup_{x\in O}J^{^\pm}(x)$ \cite{fewster2020quantum}.} (Figure \ref{fig2}). As we shall discuss in the next section, Borsten et al. posit a general condition on allowed local operators that guarantees no-signalling for non-selective measurement. 
 
Some examples of seemingly innocent locally implementable operations that lead to `impossible measurements' are given in \cite{Borsten:2019cpc} (and also \cite{PhysRevA.64.052309}). For finite dimensional Hilbert spaces, it is particularly interesting for quantum information purposes to analyse the causal behaviour of operations that correspond to measuring observables of the type $\hat{A}\otimes \mathbbm{1}+ \mathbbm{1} \otimes \hat{B}$ versus $\hat{A}\otimes \hat{B}$ on the tensor product of two local subsystems $\mathcal{H}_1 \otimes \mathcal{H}_2$. In \cite{Borsten:2019cpc} it was shown that the latter can be problematic, despite the expectation that such a `factorised' operation should be locally implementable in the Hilbert space sense (by means of LOCC, local operations and classical communication). They provide the following concrete example of a bipartite system that starts out in the factorised state $\ket{\psi}= \ket{0}\otimes\frac{1}{\sqrt{2}}(\ket{0}+\ket{1})$. First, a local unitary `kick' $\hat{U}=e^{i\gamma \hat{\sigma}_x\otimes \mathbb{1}}$ is applied, and then a non-selective measurement of the observable $\ket{1}\bra{1}\otimes \hat{\sigma}_z$ is performed. The outcome is that expectation values of observables over $\mathcal{H}_2$ will generally depend on $\gamma$, e.g., $\langle \mathbb{1} \otimes \hat{\sigma}_x\rangle= \text{cos}^2(\gamma)$.\footnote{By applying the causality condition \eqref{Borstencondition} that Borsten et al. derive (see next section) for the case $A_1= \hat{U}$, $A_2=\ket{1}\bra{1}\otimes \hat{\sigma}_z$ and $A_3=\mathbbm{1}\otimes \hat{\sigma}_z$ one can appreciate the role that the degeneracy of $A_2$ plays in failing to satisfy \eqref{Borstencondition}. That is, the non-selective measurement of a degenerate observable can enable superluminal signalling, even if this observable is factorised.} A similar QFT example of this problematic case is the measurement of a product of fields such as $\hat{\phi}(f_1)\hat{\phi}(f_2)$ where $f_1,f_2$ are supported in disjoint spacetime regions (see \cite{PhysRevD.105.025003}).\footnote{Details:  where non-selective measurements are implemented by unitary `kicks' or operations involving 1-parameter families of Kraus operators. Jubb also shows that expectation values involving products of fields can be recovered using only `possible measurements' of smeared fields and the identity.}
 
\subsection{Discussion}\label{Secadhoc} 

\subsubsection{Analysis of the `impossible measurements' reductio argument}

Sorkin-type impossible measurement scenarios clearly illustrate the moral that P4 Microcausality is not by itself sufficient to rule out superluminal signalling in relativistic quantum theories. The question of how to respond to this reductio argument---i.e., how to revise the set of assumptions so that superluminal signalling is excluded for all possible measurements---will be taken up in the next subsection. First we will analyze the argument itself.

Reductio arguments perform the useful function of pinpointing a set of assumptions that lead to a problematic conclusion. However, the implied negative information about issues that are \textit{not} relevant to resolving the problem stated in the conclusion can be valuable. The `impossible measurements' reductio argument is particularly valuable in this respect because it is distinct from some of the other recognized foundational issues that face QFT.\footnote{It is worth pointing out that Sorkin's quantum `impossible measurement' scenarios may also have analogues in classical relativistic theories \cite{MuchVerch}.}

An obvious first line of response to the `impossible measurements' reductio might seem to be to make L{\"u}ders' rule manifestly Lorentz covariant. However, this response does not address the impossible measurements problem. L{\"u}ders' rule P3(b) for a state update from $t$ to $t ^\prime$ is applied to a sequence of measurements in relativistic spacetime by using the temporal ordering relation in P3(c). P3(c) imposes a temporal order for a given set of regions (such as those identified in Fig. \ref{fig2}), but the transitive closure operation is not Lorentz covariant in the sense that it relies on a preferred foliation of spacetime. The order induced by transitive closure can be extended to a non-unique linear order, which is equivalent to the choice of a preferred hypersurface of simultaneity. However, as Sorkin notes, different choices of linear order do not affect the expectation values for any sequence of projective measurements associated with the regions as a result of P4 Microcausality. Since the conclusion of the reductio argument concerns the expectation values for $O_3$, which are assigned in a Lorentz covariant way, making L{\"u}ders' rule manifestly Lorentz covariant seems unlikely to solve the problem. (Of course, adopting a manifestly Lorentz covariant alternative to L{\"u}ders' rule may well be part of the solution, as the FV measurement framework demonstrates.)

Another indication that the fact that L{\"u}ders' rule is not manifestly Lorentz covariance is not the root cause of the impossible measurements problem is that making L{\"u}ders' rule Lorentz covariant is not sufficient to solve the problem. Hellwig and Kraus \cite{PhysRevD.1.566} proposed a manifestly Lorentz covariant version of L{\"u}ders' rule back in 1970.\footnote{Hellwig and Kraus argue that their proposal is physically equivalent to a more elaborate proposal for a Lorentz covariant version of L{\"u}ders' rule made by Schlieder in \cite{schlieder}.} Hellwig and Kraus stipulate that L{\"u}ders' rule only updates the state of the field in the causal future and causal complement of the region in which the measurement is performed; the state of the field in the causal past of the measurement region remains unchanged. Clearly, making L{\"u}ders' rule manifestly Lorentz covariant in this way does not address the problem of Sorkin-type impossible measurements because this state update rule still applies in region $O_2$, which is contained in the causal future and causal complement of $O_1$. Sorkin himself suggests (but does not endorse) an alternatiave Lorentz-covariant modification, which is to restrict L{\"u}ders' state update to the causal future of a measurement region. However, this does not by itself address the impossible measurement problem, which involves measurement regions such as $O_3$ in Fig. \ref{fig2} that are not strictly contained in either the causal future or the causal complement of $O_2$. Sorkin considers the restriction of allowed measurement regions to those that are strictly causally ordered (e.g., $O_3$ in Fig. \ref{fig2} must be entirely in the causal future or entirely in the causal complement of $O_2$), but this restriction seems to lack an independent physical motivation, as we discuss in Sec. \ref{Secreductiostrategies}.\footnote{See \cite{Borsten:2019cpc} for further discussion of shortcomings of other proposals along these lines in different spacetime contexts.} The upshot is that impossible measurement scenarios cannot be blamed on the fact that L{\"u}ders' rule is not manifestly Lorentz covariant. Of course, a measurement theory for QFT would ideally contain manifestly Lorentz covariant state update rules, but making L{\"u}ders' rule manifestly Lorentz covariant is not by itself sufficient to address the `impossible measurements' reductio argument.

As Sorkin also points out, the failure of collapse interpretations to be manifestly Lorentz covariant is also not the source of the impossible measurements problem:

\begin{quote}

It is often objected that the idea of state-vector reduction cannot be Lorentz-invariant, since ``collapse” will occur along different hypersurfaces in different rest-frames. However we have just seen that well-defined probability rules can be given without associating the successive collapses to any particular hypersurface. Thus the objection is unfounded to the extent that one regards the projection postulate as nothing more than a rule for computing probabilities. (p.4)

\end{quote}

\noindent In general, proposed solutions to the Measurement Problem that offer a physical interpretation of quantum theory will not address the impossible measurements problem if they leave all of the premises of the `impossible measurements' reductio argument intact insofar as their implications for the assignment of probabilities upon measurement are concerned. The `impossible measurements' reductio argument relies on the uncontroversial formal recipe for extracting probabilities for measurement outcomes from NRQM; the problem is that the attempt to extend this formal recipe to the relativistic context leads to superluminal signalling. There is consensus about the irrelevance of the Measurement Problem in the responses to Sorkin-type impossible measurement scenarios that are discussed below (see p.3 of \cite{fewster2020quantum} and Sec. II of \cite{grimmer_measurements_2021}).\footnote{In the detector models approach, Grimmer, Torres, and Mart{\'i}n-Mart{\'i}nez \cite{grimmer_measurements_2021} agree with this assessment of the irrelevance of the Measurement Problem in NRQM to Sorkin-type impossible measurement scenarios, but in addition suggest that there is a `relativistic cut' between NRQM and QFT that is analogous to the Heisenberg cut. See the discussion in Sec. \ref{FurtherCompSec} below.} The Measurement Problem is thus not directly relevant to the resolution of the Sorkin-type no-go result.

Interpretations of quantum theory proposed as solutions to the Measurement Problem are not relevant to addressing the `impossible measurements' reductio argument. However, the implication could run in the other direction:  resolutions of the impossible measurements reductio could have implications for the possible interpretations of relativistic quantum theory. ``The" Measurement Problem is actually a collection of problems (see \cite{maudlin_three_1995,muller_six_2023}), but a historically important and intuitive variant is the following:  the unitary quantum dynamics (e.g., as given by the Schr{\"o}dinger equation) is inconsistent with the prescription for state update after measurement (e.g., as given by L{\"u}ders' rule). Rejecting some of the premises of the `impossible measurements' reductio argument and/or adding new assumptions to block impossible measurement scenarios may involve changes to both halves of this Measurement Problem for NRQM. That is, the representation of relativistic dynamics in QFT and the state update rules in the accompanying measurement theory for QFT could both be different from NRQM in ways that affect the form taken by the Measurement Problem in QFT. For example, the FV measurement framework for AQFT involves both the addition of an assumption about relativistic dynamics and a revised measurement theory for QFT (see Sec. \ref{FVsection}).

Another feature of the Sorkin reductio argument is that it makes no assumptions about the state of the relativistic quantum system that is measured. This is another important piece of negative information about factors that are not relevant to ruling out impossible measurement scenarios. In particular, while Sorkin's QFT example chooses the vacuum state as the initial state, it is not necessary that the system be in the vacuum state or any other state that satisfies the assumptions of the Reeh-Schlieder theorem. In fact, it is interesting to note that the `impossible measurements' reductio argument assumes Microcausality, but the Reeh-Schlieder theorem does not (see \cite{clifton_entanglement_2001} for discussion). As a result, Sorkin's impossible measurements are not related to state-dependent phenomena, such as the entanglement of a state across spacelike separated regions. Furthermore, impossible measurement scenarios are not caused by the special properties of the Type III von Neumann algebras that are ubiquitous in QFT; the `impossible measurements' reductio argument applies in principle to von Neumann algebras of any type.\footnote{Borsten et al. \cite{Borsten:2019cpc} draw attention to the fact that the projectors in the P3(b) version of L{\"u}ders' rule are not necessarily of rank one. For discussion of limitations on the extent to which it is possible to apply L{\"u}ders' rule in the context of Type III algebras, see \cite{busch_luders_2009} and \cite{ruetsche_earman_2011}.} Of course, given that Type III von Neumann algebras are often physically relevant in QFT, the particular interpretive issues that they raise will need to be addressed, as will the implications of the Reeh-Schlieder theorem.\footnote{For example, the FV framework and detector models approach have been deployed to analyze the theoretical limitations on harvesting entanglement from Reeh-Schlieder states in \cite{ruep_weakly_2021} and \cite{grimmer_measurements_2021}, respectively. The Reeh-Schlieder property is also mentioned below in the context of discussions of selective measurements and properties of Type III algebras, which are not directly relevant to impossible measurement scenarios (which concern only non-selective measurements).} However, it is important to recognize that Sorkin-type impossible measurement scenarios raise a separate set of foundational issues.\footnote{The interaction picture for representing the dynamics of QFT is used in one of Sorkin's examples and some of the literature responding to Sorkin-type impossible measurement scenarios. The interaction picture is also problematic due to Haag's theorem (See \cite{EarmanandFraser2006} for discussion.) These issues will be set aside in this paper because they are not directly related to the impossible measurements problem insofar as simply adopting an acceptable alternative to the interaction picture is insufficient to block impossible measurements. Moreover, the `impossible measurements' reductio argument relies on a different set of premises from proofs of Haag's theorem.}

\subsubsection{Strategies for responding to the `impossible measurements' reductio argument}\label{Secreductiostrategies}

We will proceed to evaluate the lessons of the `impossible measurements' reductio argument by working under the assumption that the conclusion is genuinely unacceptable (i.e., that the expectation values of a measurement performed in one region cannot depend on which unitary operation is performed in a spacelike separated region). This means that the avenues of response to the reductio argument can be distinguished by their rejection of different sets of premises and/or their addition of different sets of premises to the argument. Assuming that the conclusion of the `impossible measurements' reductio argument is deemed unacceptable, responding to the reductio requires blocking the derivation of the conclusion by rejecting one or more premises or adding one or more premises. 

The most straightforward response is an ad hoc\footnote{This is not a pejorative use of the term ad hoc. Both Sorkin \cite{sorkin1993impossible} and Borsten et al. \cite{Borsten:2019cpc} use this term to describe their proposals.} one:  target P1 and P3, which taken together entail that the measurable observables include all $A_k$ that can be obtained by restricting the field $\Phi$ to any region $O$. An ad hoc resolution of the reductio can be obtained by simply excluding any observable that can lead to superluminal signalling. Sorkin proposes (but does not endorse) restricting the regions to which observables may be assigned. For example, imposing the restriction that measurable observables may only be defined on regions that are strictly causally ordered (i.e., for regions $O_j$ and $O_k$, \textit{all} $x \in O_j$ causally precede \textit{all} $y \in O_k$ or vice versa) (p.9). As Sorkin notes, it is difficult to imagine how the possibility of performing a measurement operation could depend on spacetime in this way (see also \cite{Benincasa_2014}). There are presumably not `spacetime police' to ensure that laboratory measurements are only carried out when they are strictly causally ordered. 

Borsten et al. \cite{Borsten:2019cpc} propose a different ad hoc resolution of the reductio that imposes a restriction directly on the observables rather than the associated regions (see also \cite{PhysRevA.64.052309,PhysRevD.34.1805}). They argue that the following condition rules out superluminal signalling by non-selective measurements in Sorkin-type scenarios:
\begin{gather}\label{Borstencondition}
    \text{An operator }A_{2} \in \mathcal{A}(O_{2})\text{ with resolution }\mathcal{B}\text{ will not enable signalling iff } \\ \label{causality condition} [\mathcal{E}_{A_{2},\mathcal{B}}(A_{3}),A_{1}]=0 \text{, as an operator equation, for all }A_{1,3} \in \mathcal{A}(O_{1,3})\text{.} \notag
\end{gather}
\noindent Again, the logic is that this condition is imposed for the purpose of excluding superluminal signalling. The condition can be enforced by `banning' observables $A_2$ that do not satisfy it, or else bringing in some notion of coarse-graining that entails a measurement resolution that is large enough for the criterion to be met.\footnote{In \cite{PhysRevD.105.025003} the measurement resolution is introduced by considering Gaussian measurements. It is pointed out that, in particular examples, the allowed accuracy of a Gaussian measurement is determined by all future experiments. Some mechanism would have to constrain future experiments accordingly.} Both options are ad hoc, as long as they are demanded only to avoid superluminal signalling, and would have to be further motivated on physical grounds. 

In this paper, we focus our attention on more comprehensive, physically motivated proposals for modeling measurement in QFT that address the `impossible measurements' reductio. However, we do not wish to diminish the value of this ad hoc approach, which is further developed by Jubb in \cite{PhysRevD.105.025003}. This approach may be regarded as complementary to both the detector models approach and the FV framework. Borsten et al. emphasize that their no-signalling criterion is model-independent (``a theoretical limit on what is possible, independent of how it is attempted" (p.3)), in contrast to the detector models approach. They also argue that their condition is applicable to any state update rule, not only L{\"u}ders' rule. As we shall discuss in Sec. \ref{detectormodelssec}, condition \ref{Borstencondition} holds FAPP for important examples of detector models. Borsten et al. note that their condition (\ref{Borstencondition}) is general enough to be applied to the Type III von Neumann algebras that are physically relevant in QFT. Condition (\ref{Borstencondition}) can also be compared with the results of applying the FV framework for AQFT. (See \cite{PhysRevD.105.025003,fewster_measurement_2023} for further discussion.)  
 
In Sec. \ref{FVsection}--\ref{comparisonsection}, we examine in depth two proposals for representing measurement in QFT:  the FV framework for measurement in AQFT and the detector-based measurement theory for QFT. We have chosen to focus on these two proposals because each makes substantial, physically motivated revisions to the premises of the reductio argument. In contrast to the ad hoc resolutions, these revisions do not rule out impossible measurement scenarios automatically; non-trivial arguments are required to show that, in each of these measurement theories, non-selective measurements of Sorkin-type cannot be used to signal. 

In Sec. \ref{FVsection}, we will consider the Fewster-Verch (FV) measurement framework for AQFT. Fewster and Verch adopt a `top down' approach that aims to treat measurements in general and quantum field systems in general. Both the quantum field system and the measurement probe are modeled using AQFT. The initial motivation for this approach was to provide a framework in which the localization properties of observables of Unruh-DeWitt detectors could be studied \cite[p.5]{fewster_measurement_2023}. Subsequently the FV framework was used to addresses the `impossible measurements' problem \cite{PhysRevD.103.025017}. The strategy involves rejecting many of the premises of the `impossible measurements' reductio argument as well as adding as premises axioms from AQFT. In particular, a new measurement theory for AQFT is formulated to replace much of P3. In the axiomatic context of AQFT, it is recognized that Microcausality by itself is insufficient to rule out superluminal signalling for reasons unrelated to impossible measurements (see \cite{redei_how_2010,earman2014relativistic} and the discussion in Sec. \ref{axiomsection} below); additional axioms or assumptions are needed. An important goal of this approach is the principled one of determining which physical principles are needed to consistently represent relativistic quantum systems and the measurements performed on them.  

In contrast, the detector models approach that is examined in Sec. \ref{detectormodelssec} adopts a `bottom up' strategy of constructing models for different types of detectors (e.g., an Unruh-DeWitt detector), each of which represents the interaction between the detector and a quantum field system. A main plank of this strategy is to model the detector using NRQM and (as far as possible) the measurement theory set out in P3 (including L{\"u}ders' rule for ideal measurements). However, the detector models approach rejects Sorkin's assumption that the measurement theory set out in P3 can be applied directly to the quantum field system. The primary goal of this approach is the pragmatic one of obtaining practically applicable models of realistic detectors, typically used in quantum optics and quantum information. We offer an in-depth comparison of the detector models approach and the FV framework in Sec. \ref{comparisonsection}.

Clearly, the ad hoc, FV, and detector models approaches are not the only possible approaches to addressing the `impossible measurements' reductio; the reductio functions as a useful heuristic for suggesting alternative approaches to formulating a measurement theory for QFT. Different approaches might involve rejecting or revising other premises and/or adding premises that reflect missing relativistic and/or quantum principles. The open-ended nature of the project means that our review will not be comprehensive, but we will discuss Sorkin's own preferred response to the `impossible measurements' problem, which is to shift to a histories-based formulation of QFT. In Sec. \ref{consistenthistories}, we review recent progress on a histories-inspired formulation of QFT and the remaining challenges to resolving the `impossible measurements' problem in this framework. Another example of an approach to formulating a measurement theory for QFT is the positive formalism proposed in Oeckl \cite{oeckl_local_2019} which adopts the strategy of abstracting an operational framework based on
probes and composition from non-relativistic quantum mechanics and then developing a concrete implementation for QFT. The resolution of the `impossible measurements' problem in this framework is the subject of current research.

\section{Principled approach:  Fewster-Verch framework for measurement in AQFT}\label{FVsection} 

The Fewster-Verch (FV) framework \cite{fewster2020quantum,fewster_measurement_2023} adopts a `top down' strategy for formulating a measurement theory for QFT:  first general principles for QFT in the algebraic framework are posited and then a compatible measurement theory is devised. Sec. \ref{axiomsection} draws on philosophical analysis of AQFT by Earman and Valente \cite{earman2014relativistic} and Ruetsche \cite{ruetsche2011interpreting} to trace the implications of the physical principles that Fewster and Verch adopt as axioms. Our aim is to foster an appreciation for the role that these principles play in ruling out Sorkin-type impossible measurement scenarios and also in informing the measurement theory for AQFT introduced by Fewster and Verch. Sec. \ref{FVmeasurementsec} introduces the measurement theory component of the FV framework. Sec. \ref{bostelmann} reviews the proof by Bostelmann et al. \cite{PhysRevD.103.025017} that Sorkin-type impossible measurements are excluded by the FV framework, highlighting the role of the Time-Slice Property axiom. Sec. \ref{stateupdatesection} is devoted to discussion of the physical interpretation of the FV measurement theory, emphasizing points of contrast with both Quantum Measurement Theory for NRQM and the detector-based measurement theory for QFT.

\subsection{The principles of AQFT and their implications}\label{axiomsection}

Algebraic QFT (AQFT) associates algebras of observables $\mathcal{A}$ with open-bounded regions $O$ of a spacetime $M$. An algebraic state $\omega$ is a positive,\footnote{In a *-algebra, an element is positive if is a finite convex combination of elements of the form $A^{*}A$} normalized, linear functional from $\mathcal{A}(O)$ or $\mathcal{A}(M)$ to $\mathbb{C}$. For any $A \in \mathcal{A}$, $\omega (A)$ represents the expected value of a measurement of $A$ in state $\omega$. A set of axioms for AQFT is chosen that represents both quantum and relativistic principles. For an introduction to AQFT, see \cite{fewster2019algebraic} or \cite{ruetsche2011interpreting}. Our discussion will focus on the relevant details of Fewster and Verch's version of AQFT. Fewster and Verch formulate their axioms for AQFT on globally hyperbolic spacetime.\footnote{A spacetime is globally hyperbolic if and only if it has no closed causal curves and the causal hull of any compact set is compact. The causal hull of a subset $S$ of spacetime $M$ is the intersection of its causal past and future, $J^{+}(S) \bigcap J^{-}(S)$  \cite{fewster2020quantum}.} As a result, features that are particular to the special context of Minkowksi spacetime are not included in the axioms. Fewster and Verch prefer a set of axioms that is inspired by the locally covariant approach to AQFT proposed by Brunetti, Fredenhagen and Verch in \cite{brunetti_generally_2001}. Their main motivation for considering globally hyperbolic spacetimes is, of course, to open up the possibility of incorporating gravity. Fewster and Verch's axioms are devised to apply to a collection of globally hyperbolic spacetimes; however, only the special case of a single globally hyperbolic spacetime is needed for representing `impossible measurement' scenarios. The FV axiomatization liberalizes traditional algebraic QFT in one more respect:  the algebras $\mathcal{A}$ are taken to include not only self-adjoint operators, but also effects. This generalization is made to bring the measurement theory for AQFT in line with Quantum Measurement Theory (e.g., \cite{busch_book_2016}), in which projection-valued measurements are a special case \cite[p.7]{fewster2020quantum}. Fewster and Verch similarly include Effect Valued Measures (EVMs), which are known as Positive Operator Valued Measures (POVMs) in Quantum Measurement Theory. An EVM is a map $E$ from effects of the probe to the effects of the system: $E:\chi \rightarrow \mathcal{A}$, where $\chi$ is a $\sigma$-algebra and $\mathcal{A}$ is a *-algebra, such that $E$ has properties of a measure and takes values $A \in \mathcal{A}$ such that $A$ and $\mathbb{1}-A$ are both positive \cite[p.7]{fewster2020quantum}.

Following \citet{PhysRevD.103.025017}, here are the axioms for a single globally hyperbolic spacetime (with information about the general case of a collection of spacetimes in footnotes):

\begin{enumerate}

    \item \textbf{(Global algebras)} The theory specifies a unital *-algebra $\mathcal{A}$ associated with a globally hyperbolic spacetime $M$.\footnote{In general, the theory can admit a collection of globally hyperbolic spacetimes $\mathbf{M}$ and specifies a unital *-algebra for each $M$ in $\mathbf{M}$. This opens up the possibility for comparison of a theory on different spacetimes even when one spacetime is not embeddable in the other \cite{fewster_locally_2015}.}
    
    \item \textbf{(Compatibility)} A region is an open, causally convex subset $N$ of $M$. $\mathcal{A}(N)$ is a unital sub-*-algebra of $\mathcal{A}(M) \coloneqq \mathcal{A}$.\footnote{A subset $N$ is causally convex if it is equal to its causal hull $J^{+}(N) \bigcap J^{-}(N)$. For a collection of globally hyperbolic spacetimes, this becomes a compatibility requirement that $\mathcal{A}$ can be defined on $N$, taken to be a globally hyperbolic spacetime in its own right with metric and time orientation inherited from $M$. }

    \item \textbf{(Isotony)} For regions $N_{1} \subseteq N_{2}$: $\mathcal{A}(N_{1}) \subseteq \mathcal{A}(N_{2})$.
    
    \item \textbf{(Time-Slice Property)} If $N$ contains a Cauchy surface for $M$, then $\mathcal{A}(N)=\mathcal{A}(M)$. That is, there is a local embedding isomorphism $\alpha_{M;N}$: $\mathcal{A}(N) \rightarrow \mathcal{A}(M)$.
    
    \item \textbf{(Microcausality)} If regions $N_{1}$ and $N_{2}$ are causally disjoint, then the elements of $\mathcal{A}(N_{1})$ commute with the elements of $\mathcal{A}(N_{2})$.
    
    \item \textbf{(Haag Property)} Let $K$ be a compact subset of $M$. If an element $A \in \mathcal{A}(M)$ commutes with every element of $\mathcal{A}(M)$ for every region $N$ in the causal complement $K^{\perp}$ of $K$, then $A \in \mathcal{A}(L)$ whenever $L$ is a connected, open, causally convex subset containing $K$.\footnote{This is a weakened form of Haag duality. See \cite{fewster2020quantum} for discussion.}

\end{enumerate}

In this axiomatization, the global algebra $\mathcal{A}(M)$ is posited and (Compatibility) is used to define local algebras:  associated with every casually convex subset $N \subseteq M$ is an algebra $\mathcal{A}(N)$ that (Compatibility) ensures is compatible with the global algebra $\mathcal{A}(M)$. (Compatibility) and (Time-Slice Property) entail a local version of (Time-Slice Property):

\noindent \textbf{(Local Time-Slice Property)} If $N_{1} \subset N_{2}$ and $N_{1}$ contains a Cauchy surface for $N_{2}$, then $\mathcal{A}(N_{1}) = \mathcal{A}(N_{2})$.

\noindent Axioms such as these time-slice properties have a long history in AQFT. Axioms in the same family include Primitive Causality, Local Primitive Causality, and Second Causality \cite{dimock_algebras_1980}. (See \cite{earman2014relativistic,ruetsche2011interpreting,Calderon2022} for discussion.)

The only obvious overlap with the assumptions for the `impossible measurements' reductio argument set out in Sec. 2 is (Microcausality). (Local Time-Slice Property) and (Time-Slice Property) are independent of (Microcausality) (i.e., there are models in which (Microcausality) is satisfied but not either of the other two axioms, and vice versa) \cite{haag_postulates_1962}. All of the axioms posited by Fewster and Verch play a role in the FV framework, but the key axioms that Bostelmann et al. \cite{PhysRevD.103.025017} invoke to demonstrate that the FV framework does not allow Sorkin-type impossible measurements to occur are (Isotony) and (Local Time-Slice Property).  (Isotony) is the natural requirement that the inclusion relations among algebras reflect the spacetime relations among spacetime regions. (Local Time-Slice Property) is the one that does the nontrivial work in shielding the FV framework from Sorkin-type measurement scenarios. 

To appreciate the significance of (Time-Slice Property) for ruling out superluminal signalling, we will draw on the comprehensive analysis of relativistic causality assumptions in AQFT presented in Earman and Valente \cite{earman2014relativistic}. Earman and Valente's main conclusion is that in Minkowski spacetime the ``most direct expression of relativistic causality" in AQFT is (Local Time-Slice Property).\footnote{\label{footnote}Earman and Valente adopt the (Local Primitive Causality) axiom in the (Time-Slice Property) family (and (Isotony), which strengthens it). Aside from the restriction to Minkowski spacetime and the use of concrete von Neumann algebras, the main difference between (Local Primitive Causality) axiom and Fewster and Verch's (Local Time-Slice Property) is that the former applies to any open-bounded regions while the latter is restricted to open, causally convex regions. The definition of a region as a causally convex region in the FV (Compatibility) axiom is a consequence of the more general spacetime context of collections of globally hyperbolic spacetimes. For example, Fewster and Verch \cite[pp.24--25]{fewster2020quantum} argue by appeal to an example that the minimal localization region for a system observable induced by measurement is the entire causal hull of the region in which the system and probe interact because in any smaller localization region whether induced observables commute would be sensitive to changes in geometry. See \cite{fewster_locally_2015} for a more general argument along similar lines. The arguments offered by Earman and Valente apply to causally convex regions, so this difference in the definition of regions will be set aside here, but it may have interpretative consequences. (Thanks to Laura Ruetsche for pointing this out.)} Consideration of how the FV framework uses (Local Time-Slice Property) to block Sorkin-type impossible measurements strengthens Earman and Valente's argument for the relative importance of (Local Time-Slice Property), but one does not have to accept their conclusion that (Local Time-Slice Property) is the most direct expression of relativistic causality in order to follow their analysis. Earman and Valente distinguish two aspects of relativistic causality that are relevant to our discussion:  no superluminal signalling (i.e., by performing local operations on quantum fields) and no superluminal propagation of quantum fields. We will focus on their positive analysis of the relationship between (Local Time-Slice Property) and relativistic causality.   

Earman and Valente argue that a dynamical axiom is needed in order to enforce relativistic causality. (Microcausality) is a kinematical axiom that imposes an independence or separability requirement (p.3). In contrast, (Time-Slice Property) concerns dynamics. As Fewster and Verch explain, ``[t]he timeslice assumption is one of the lynch-pins of the structure and encodes the idea that the theory \textit{has} a dynamical law, although \textit{what} it is is left unspecified" \cite[p.9]{fewster_algebraic_2015}. That is, (Time-Slice Property) states that there exist local embedding isomorphisms $\alpha_{M;N}$ that reflect the dynamics. Axioms in this family are sometimes labelled ``Existence of Dynamics" to make their role transparent \cite[p.14]{fewster2019algebraic}. Positing an axiom that imposes a dynamical constraint can exclude spacelike dependencies between expectation values in one region and unitary operations performed in a spacelike separated region by enforcing the requirement that fields cannot propagate faster than the speed of light. Intuitively, if the fields cannot propagate faster than the speed of light, then the effects of local operations on the fields should not be able to propagate faster than the speed of light either. 

Earman and Valente \cite[p.19]{earman2014relativistic} argue that this intuition about needing a dynamical axiom like (Time-Slice Property) to exclude superluminal signalling is supported by considering classical field theories. In classical relativistic field theories, the prohibition on superluminal field propagation is typically enforced by the field equations. More specifically, the field equations are a system of symmetric, quasi-linear, hyperbolic partial differential equations that are associated with a set of causal cones that typically\footnote{In atypical cases the causal cones of the hyperbolic partial differential equations could differ from the null cones of the spacetime, which would in principle permit superluminal signalling \cite{geroch2010faster,earman2014relativistic}.} do not permit superluminal propagation of the field \cite{geroch2010faster}. Determinism keeps the fields propagating within the causal cones. Consider the initial value problem for a system of field equations. The specification of `initial' data on a closed subset $S$ of points in Cauchy surface $\Sigma$ picks out a unique solution of the field equations in the future and past domains of dependence of $S$, $D(S)$.\footnote{Assuming that the solution of the field equations does not `blow up' at future or past times (i.e., global existence and uniqueness conditions are satisfied) \cite{earman2014relativistic}. The domain of dependence $D(S)$ of $S$ is the set of points $p$ such that every inextendible causal curve through $p$ meets $S$ \cite{earman2014relativistic}.} Note that determinism is a fact about what the initial state and dynamical laws entail about future states, not an epistemic matter of what we can know or predict. As Earman and Valente explain, the prohibition on superluminal propagation ``follows from a mild form of verificationism":  ``the local nature of determinism means that there is no way to detect the effects of any alleged superluminal propagation since once the relevant initial data on $S$ are fixed, data at points relatively spacelike to $S$ and to $D(S)$ can be varied in any manner one likes (consistent with the constraints (if any) on initial data) without making any difference at all for the solution in $D(S)$" \cite[p.19]{earman2014relativistic}.

For AQFT on Minkowski spacetime, \textit{Local Quantum Determinism} is the the analogue of the initial value problem for classical fields:

\begin{quote}
Local Quantum Determinism: For any physical states $\omega$ and $\omega ^{\prime}$, and any $O \subset M$, if $\omega \mid_{\mathcal{A}(O)} = \omega ^{\prime} \mid_{\mathcal{A}(O)}$ then $\omega \mid_{\mathcal{A}(D(O))} = \omega ^{\prime} \mid_{\mathcal{A}(D(O))}$, where $\omega \mid_{\mathcal{A}(O)}$ represents the restriction of the state $\omega$ to the local algebra $\mathcal{A}(O)$.
\end{quote}

 \noindent (Local Time-Slice Property) ensures that Local Quantum Determinism holds when $O$ is a causally convex subset $N$ of $M$ by identifying $\mathcal{A}(N)$ and $\mathcal{A}(D(N))$. Modulo the restriction to causally convex regions,\footnote{See footnote \ref{footnote}.} (Local Time-Slice Property) is the axiom that Earman and Valente regard as the most direct expression of relativistic causality. Note that the prohibition on superluminal field propagation does not follow automatically from the quantization of a classical theory with hyperbolic field equations. The interpretation of the algebras of observables as having localization regions that include their own domain of dependence, as required by (Local Time-Slice Property), is also necessary \cite[p.24]{earman2014relativistic}.\footnote{Earman and Valente \cite[pp.23--24]{earman2014relativistic} cite Segal's quantization scheme for the Klein-Gordon field as an example with hyperbolic classical field equations in which Local Primitive Causality is violated and superluminal field propagation appears to be possible. The slight complication is that Segal does not explicitly interpret a local algebra $\mathcal{A}(O)$ as representing observables measurable in $O$, so strictly speaking superluminal propagation is not a verifiable prediction of Segal's scheme.}  Again, determinism in the sense captured by Local Quantum Determinism is a fact about the theory, not an epistemic matter. However, epistemic considerations can provide motivation for the adoption of (Local Time-Slice Property). Bostelmann et al. \cite[p.3]{PhysRevD.103.025017} declare ``Morally:  If one knows the initial conditions of a quantum field on a Cauchy surface, then one knows the quantum field everywhere."

 Before moving on to discuss the details of the FV measurement theory in the next subsection, we will pause to consider why AQFT needs its own measurement theory and how the axioms inform the representation of measurement in AQFT. Earman and Valente \cite[p.14]{earman2014relativistic} offer a straightforward argument that L{\"u}ders' rule cannot be na{\"i}vely extended to AQFT. Apply the GNS construction using state $\omega$ to obtain a Hilbert space representation on which the algebraic state $\omega$ is an expectation-valued map. L{\"u}ders' rule for a non-selective measurement of $A \in \mathcal{A}(O)$ can be formally applied to an algebraic state $\omega$ as follows:

\begin{equation}
\omega ^{\prime} ( \cdot) = \sum _{j} \omega ( P^{A} _{j} \cdot P^{A} _{j} ),
\end{equation} 

\noindent where $P^{A} _{j}$ are projections. However, $\omega ^{\prime}$ cannot be interpreted as the state of the system after measurement:  in general, $\omega ^{\prime}$ differs from $\omega$ in the causal past of region $O$ (as well as the causal future). Therefore, $\omega \rightarrow \omega ^{\prime}$ cannot represent a physical change of state---that is, a transition from a state $\omega$ before the measurement of A to a state $\omega ^{\prime}$ after the measurement of A. In other words, the projection postulate is inapplicable; the measurement preparation assumption P3(b) of ideal measurement theory is not a legitimate assumption in the context of AQFT. Earman and Valente recognize that there is a need for the type of measurement theory for AQFT that Fewster and Verch develop: ``assuming AQFT is an empirically adequate theory, there must be within the AQFT description of the combined object-measurement apparatus system...a description of measurement processes and their outcomes'' \cite[p.17]{earman2014relativistic}. 

Given that algebraic QFT has traditionally been given an operational interpretation in terms of local measurement operations, one might wonder why it needs a separate measurement theory. In their seminal presentation of axioms for algebraic QFT, Haag and Kastler offer the following as the core of their operational interpretation: ``[a]n operation in the space-time region $B$ corresponds to an element from [local algebra] $\mathcal{U}(B)$" \cite[p.851]{haag_kastler_1964}. Fewster and Verch point out two respects in which this operational interpretation falls short of their goals for their own measurement theory. First, Haag and Kastler do not ``set out how exactly one measures an observable or performs an operation within a region of spacetime" \cite[p.8]{fewster2020quantum}.  Second, they ``were reluctant to interpret elements of the local algebras as observables (which they considered to arise as limits of local algebra elements)" (p.8). That is, AQFT is not connected to predictions via the interpretation of local algebras in terms of laboratory procedures; instead, the connection to predictions is still made by relating the algebras of operators to collision cross sections using asymptotic limits via Haag-Ruelle scattering theory. (See \cite{Fraser2023} for further discussion of the history of operational interpretations of AQFT.)
 
 The axioms of AQFT inform and constrain the interpretation of the formalism. (Local Time-Slice Property) and (Isotony) carry immediate consequences for the localization of algebras of observables. Fewster and Verch emphasize that, as a consequence of these axioms, an observable can be localized in different regions (p.7). Moreover, recognition of the larger localization regions staked out by the domains of dependence of regions is not optional. Due to determinism, the enlargement of the localization region from $N$ to $D(N)$ needs to be taken into account. Ruetsche \cite[pp.115, 110--111]{ruetsche2011interpreting} endorses this ``resolute reading" of (Local Time-Slice Property)\footnote{Ruetsche's discussion is based on the Primitive Causality axiom for globally hyperbolic spacetime in the Time-Slice Property family \cite[p.107]{ruetsche2011interpreting}. Primitive Causality entails Time-Slice Property, and Primitive Causality plus Isotony entail Local Time-Slice Property.} as mandating the larger localization regions. She argues that, by (Local Time-Slice Property), an algebraic state $\omega$ on $\mathcal{A}(O)$ is also automatically a state on $\mathcal{A}(D(O))$. Appealing to states does not reduce the size of the localization region because in AQFT states inherit their localization properties from the algebras. Another reason that the larger localization regions $D(N)$ need to be taken into account to block Sorkin-type impossible measurement scenarios, as we shall see in Sec. \ref{bostelmann}. 
 
 The association of an algebra $\mathcal{A}(O)$ with more than  one localization region $\mathcal{A}(D(O))$ is in tension with the traditional operational interpretation of local algebras $\mathcal{A}$ as representing operations that can be performed in a laboratory in region $O$. If the localization region associated with $\mathcal{A}(O)$ can be expanded to include the domain of dependence $D(O)$, then the algebra of observables will typically be associated with spatiotemporal regions outside of the lab and a duration longer than the duration of the measurement. As we shall discuss in Sec. \ref{FVmeasurementsec}, the FV measurement framework does not rely on local algebras of observables to represent local laboratory operations.

(Local Time-Slice Property) also precludes a natural strategy for representing dynamical evolution in a globally hyperbolic spacetime. In NRQM, there is a unique foliation of the spacetime into spacelike hypersurfaces that are paramaterized by absolute time. The dynamics is represented by a unitary operator $U(t)$ that induces time evolution by acting on either the operators (Heisenberg picture) or the states (Schr{\"o}dinger picture). One might try to represent the dynamics in AQFT by proceeding by analogy with NRQM. A globally hyperbolic spacetime can be foliated into a family of spacelike Cauchy surfaces. Pick a preferred foliation $\{ \Sigma_{t} \}$ and then try to interpret the associated set of algebras ${\mathcal{A}(\Sigma_{t})}$ as representing time evolution in the Heisenberg picture. However, it is not possible to do this. As Ruetsche \cite[pp.110--111]{ruetsche2011interpreting} argues, the root of the problem is that there is no `set' of algebras $\{ \mathcal{A}(\Sigma_{t}) \}$ that can be associated with a foliation of hypersurfaces $\{\Sigma_{t} \}$ due to (Local Time-Slice Property) (and (Isotony)): $\mathcal{A}(\Sigma_{t})=\mathcal{A}(\Sigma_{t ^{\prime}})=\mathcal{A}(M)$ because $D(\Sigma _{t}) = D(\Sigma _{t ^{\prime}}) = M$. The Schr{\"o}dinger picture is not possible either. Furthermore, Ruetsche \cite[pp.110--111]{ruetsche2011interpreting} argues, it is doubly problematic to interpret the algebraic states as evolving in time. Algebraic states are time-independent; they are associated with spacetime regions indirectly, as functionals of algebras directly associated with spacetime regions. The first problem is that physically significant states such as (in Minkowski spacetime) the vacuum state $\omega_{0}$ are global states in the sense of being states for all of space and time. The global vacuum state $\omega_{0}(M)$ is not a state that can figure in a time evolution for a system---e.g., a transition from $\omega_{0}$ to $\omega_{0} ^{\prime}$---because $\omega_{0}$ already generates the expectation values for the system on all of spacetime. The second problem is that---for any algebraic state $\omega$---a time evolution cannot be introduced by restrictions of $\omega$ to a foliation of Cauchy surfaces $\{ \Sigma_{t} \}$. Once again, (Local Time-Slice Property) (and (Isotony)) undermine this strategy:  $\omega \mid_{\mathcal{A}(\Sigma_{t})}$ and  $\omega \mid_{\mathcal{A}(\Sigma_{t ^{\prime}})}$ cannot be interpreted as states at times $t$ and $t ^{\prime}$, respectively, because $\mathcal{A}(\Sigma_{t}) = \mathcal{A}(\Sigma_{t ^{\prime}}) = \mathcal{A}(M)$. 

Of course, AQFT \textit{does} represent relativistic dynamics; the dynamics is encoded in the assignment of algebras to spacetime regions in accordance with the axioms, including (Time-Slice Property). Once the dynamics is represented in this manner, one can choose a foliation of (thickened) Cauchy surfaces and then infer the algebra of observables associated with each of these hypersurfaces. The point is that the dynamics is not stipulated by associating an algebra of observables with one of these (thickened) hypersurfaces and then applying a time evolution operator to determine the algebras of observables associated with the other hypersurfaces in the foliation. Instead, (Time-Slice Property) supplies local embedding isomorphisms that represent the dynamical relations among algebras defined on different spacetime regions. The FV measurement theory presented in the next section will provide an example of how dynamics can be represented in this manner. (See Adlam \cite{adlam2022laws} for a general discussion of representations of dynamical laws that do not rely on time evolution.) 

In sum, the positive conclusion of this subsection is that, as Earman and Valente argue, there are principled reasons to expect that (when supplemented with a compatible measurement theory) the axioms for AQFT adopted by Fewster and Verch will exclude superluminal signalling. The negative conclusions are that the FV axioms place constraints on both the representation of localization and the representation of relativistic dynamics. First, an algebra $\mathcal{A}(O)$ is in general associated with more than one localization region. Furthermore, the traditional operational interpretation of an algebra of observables in AQFT needs to be given up:  $\mathcal{A}(O)$ cannot be straightforwardly interpreted as representing a set of operations that can (in principle) be performed in a lab in region $O$.\footnote{Fewster and Verch \cite[p.11]{fewster_measurement_2023} suggest a similar interpretative moral in their recent article:  ``the message to be drawn is that algebra elements localisable in a region
$O$ need not represent local operations that can be undertaken there. On the other hand, hermitian algebra elements have a good interpretation as observables."} Second, the dynamics of the system in relativistic spacetime cannot be represented by using the usual method of choosing a preferred foliation of Cauchy surfaces with respect to which a time evolution operator is defined. The measurement theory proposed by Fewster and Verch uses alternative representations of local operations and the relativistic dynamics that are informed by the FV axioms. It also delivers on the expectation that superluminal signalling is excluded, as Bostelmann et al.'s proof that `impossible measurements' are actually impossible demonstrates.

\subsection{The FV measurement theory for AQFT}\label{FVmeasurementsec}

Fewster and Verch \cite{fewster2020quantum} adopt a three-pronged strategy for devising their measurement theory for AQFT. First, they introduce abstract AQFT models to represent the system and the measurement probe(s). This means that both the system and the probe(s) are assigned algebras of observables that satisfy the axioms set out in the previous subsection. A probe is coupled to a system in some region and measurement of the probe outside of this region, when it is not coupled to the system, is used to infer properties of the target system. Second, an abstract version of scattering theory for finite regions that is adapted to AQFT is applied. Finally, the same formal approach as Quantum Measurement Theory (QMT), exemplified by Chapter 10 of \cite{busch_book_2016}), is applied to this dynamical representation of the measurement process in AQFT in order to derive state update rules. While the same reasoning is applied at each step in the derivation, key differences are that abstract algebras are used instead of Hilbert spaces (and Type I von Neumann algebras) and that spacetime dependence is made explicit.\footnote{Thanks to Chris Fewster for emphasizing this point.} As a result, the FV measurement theory differs both formally and in physical interpretation from QMT. 


\begin{figure}
\centering
\includegraphics[width=0.6\textwidth]{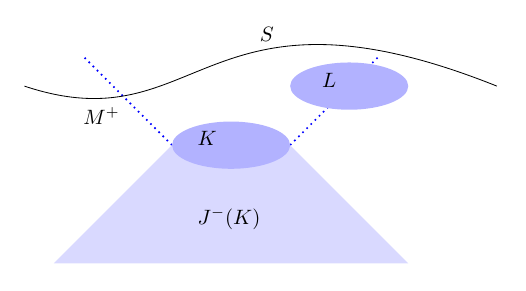}
\hspace{-2cm}
\caption{The FV definition of scattering morphisms (Based on Fig. 1 in \cite{fewster2020quantum}). Note that $J^{-}(K)$, the causal past of the system-probe interaction region $K$, includes $K$. The dotted lines represent the lightlike boundaries of $J^{+}$. $M^{+}$ is the complement of $J^{-}(K)$. $S$ is an example of a Cauchy surface contained in $M^{+}$.}
\label{FVfig1}
\end{figure}

To represent the dynamical process of measurement, Fewster and Verch formulate a sophisticated abstract version of scattering theory within AQFT. Consider a system and a single probe that are coupled in compact region $K \subseteq M$ and uncoupled outside of $K$. (See Fig. \ref{FVfig1}.) Let the algebra $\mathcal{S}$ represent the uncoupled system, the algebra $\mathcal{P}$ the uncoupled probe, and the algebra $\mathcal{C}$ the coupled system and probe. Fewster and Verch then adopt a scattering theoretic picture to represent the measurement interaction. The `in' region is $M^{-}$, the complement of $J^{+}(K)$ (i.e., the causal future of $K$, including $K$). The `out' region is $M^{+}(K)$, the complement of $J^{-}(K)$ (i.e., the causal past of $K$, including $K$). The coupled algebra $\mathcal{C}$ is identified with the uncoupled system-probe algebra $\mathcal{U} = \mathcal{P} \otimes \mathcal{S}$\footnote{Assumption to avoid technical detail:  assume the algebras have discrete topology and use the algebraic tensor product (p.8).} in $M^{-}$ and $M^{+}$, but not in the causal hull of coupling region $K$. That is, for each region $L$ in $M^- \bigcup M^+$ there is an isomorphism $\chi$: $\mathcal{U}(L) \rightarrow \mathcal{C}(L)$ that commutes with both the local embeddings $\alpha_{M;N} \otimes \beta_{M;N}$ for $\mathcal{U}$ and $\gamma_{M;N}$ for $\mathcal{C}$ that are guaranteed by (Time-Slice Property).\footnote{The existence of such isomorphisms can be checked for a specified interaction by constructing a model. The assumption that such isomorphisms exist is viable in general in perturbative AQFT \cite[p.9]{fewster2020quantum}.} Algebraic states are defined on each of the algebras. For example, the analogue of an `in' state in which there is no system-probe interaction is a state $\omega \otimes \sigma$ over $\mathcal{U}$. To track the difference between states for different algebras, we will use the symbol $\varpi$ to represent a state over $\mathcal{C}$, which is an actual state in which the system and probe interact.

As in scattering theory, the central object is $\Theta$, the scattering morphism, which relates representations of the system and probe at `early' and `late' times. $\Theta$ is an algebraic isomorphism defined as a combination of isomorphisms $\chi$: $\mathcal{U}(L) \rightarrow \mathcal{C}(L)$ and the local embedding isomorphisms that are guaranteed by (Time-Slice Property), $\alpha_{M;N} \otimes \beta_{M;N}$ (for uncoupled algebra $\mathcal{U}$) and $\gamma_{M;N}$ (for coupled algebra $\mathcal{C}$). More precisely, the scattering morphism $\Theta$ is defined as an automorphism of $\mathcal{U}(M)$ by introducing advanced and retarded maps $\tau^\pm$:

\begin{center}
    $\Theta = (\tau^{-})^{-1} \tau^{+}: \mathcal{U}(M) \rightarrow \mathcal{U}(M)$
\end{center}

\noindent $(\tau^{-})^{-1} \tau^{+}$ is a composition of six isomorphisms that map the algebras as follows:

\begin{enumerate}\label{scatteringlist}
    \item $\mathcal{U}(M) \rightarrow \mathcal{U}(M^{+})$
    \item $\mathcal{U}(M^{+}) \rightarrow \mathcal{C}(M^{+})$
    \item $\mathcal{C}(M^{+}) \rightarrow \mathcal{C}(M)$
    \item $\mathcal{C}(M) \rightarrow \mathcal{C}(M^{-})$
    \item $\mathcal{C}(M^{-}) \rightarrow \mathcal{U}(M^{-})$
    \item $\mathcal{U}(M^{-}) \rightarrow \mathcal{U}(M)$
\end{enumerate}

\noindent Intuitively, $\Theta$ takes the uncoupled observable $A^\prime \in \mathcal{U(M^{+})}$ associated with coupled observable $C^{\prime} \in \mathcal{C}(M^{+})$ at `late' times, spacetime translates $C^\prime$ to its earlier counterpart, the coupled observable $C \in \mathcal{C}(M^{-})$, and then maps this observable to the corresponding uncoupled observable $A \in \mathcal{U}(M^{-})$. 
  
Though inspired by conventional scattering theory, there are some important differences. First, `early' and `late' do not refer to asymptotic times; moreover, they do not refer to times picked out by Cauchy surfaces at all. `Early' and `late' refer to the entire spacetime regions $M^{-}$ and $M^{+}$, respectively, that are outside of the causal past and the causal future, respectively, of coupling region $K$. This departure from conventional scattering theory enables the FV framework to represent finite time measurement processes. A second difference is that in the FV framework the `in' and `out' systems are not required to be free systems \cite[p.4]{fewster2020quantum}. In contrast, in conventional scattering theory the `in' and `out' systems are taken to be free; typically, `in' and `out' systems are represented using Fock space representations for free systems. A third difference is that Fewster and Verch's scattering theory is abstract:  the scattering map $\Theta$ is an isomorphism at the algebraic level; concrete Hilbert space representations (and the GNS construction) are not used. As we have already emphasized, the dynamics is treated abstractly by positing (Time-Slice Property) to impose a dynamical constraint---the existence of algebraic isomorphisms implementing local embeddings---that any concrete dynamical model (e.g., for a specified Lagrangian) must obey. Fewster and Verch note that this is advantageous because it avoids one of the main challenges of constructive QFT:  constructing an explicit dynamics for a specified interacting theory. Though, of course, the ultimate goal is to apply the FV framework to realistic measurement scenarios for interacting systems. With this end in view, the FV framework is designed to be compatible with perturbative AQFT \cite[p.9]{fewster2020quantum}.

The next step is to obtain state update rules that are adapted to the algebraic formulation of AQFT and the scattering morphism $\Theta$. The set up and derivations closely parallel the presentation of QMT in Chapter 10 of Busch, Lahti, Pellonp\"{a}\"{a}, and Ylinen \cite{busch_book_2016}. Fewster and Verch first present a measurement scheme for a system observable induced by a measured probe observable and then define an instrument that represents state updates after measurements. As already noted, the algebras of observables include not only self-adjoint operators, but what FV label EVMs, which are known as POVMs in QMT. In both the FV framework and QMT, projection-valued measurements are a special class of measurements. In QMT, L{\"u}ders' rule applies to the special case of repeatable, ideal, nondegenerate measurements. 

Consider the measurement scheme for the FV framework. As in QMT, a measurement scheme specifies the correspondence between probe observables and the corresponding system observables that are measured. A measurement of probe observable $B$ (when the probe is prepared in state $\sigma$) can be interpreted as a measurement of induced system observable\footnote{Recall that ``observable" is understood in the permissive sense of an EVM---$A$ need not be a self-adjoint operator.} $A = \varepsilon_{\sigma}(B)$, where $\varepsilon_{\sigma}(B)$: $\mathcal{P}(M) \rightarrow \mathcal{S}(M)$ and $\sigma$ is the state in which the probe is prepared. By construction \cite[p.7]{fewster_measurement_2023}, $\omega(A)$---which represents the expectation value of a measurement of uncoupled system observable $\varepsilon_{\sigma}(B) = A$---is the same as $\underaccent{\tilde}{\varpi} _{\sigma} (\Tilde{B})$---the expectation value of the corresponding coupled system-probe observable $\Tilde{B}$ in the corresponding state $\underaccent{\tilde}{\varpi} _{\sigma}$ over $\mathcal{C}$. This is achieved by imposing the following condtion (Eq. (3.9) in \cite{fewster2020quantum}):
\begin{gather}
    \underaccent{\tilde}{\varpi} _{\sigma} (\Tilde{B}) = \omega (A) \text{ for all states $\omega$ of $\mathcal{S}(M)$} \\
    \text{where $\Tilde{B} \in \mathcal{C}(M)$ and $\Tilde{B}$ is the observable at `late' times that corresponds to $B \in \mathcal{P}(M)$}\notag\\
    \text{and $\underaccent{\tilde}{\varpi}_{\sigma}$ is the state at `late' times that corresponds to $\omega \otimes \sigma$ at `early' times}\notag
\end{gather}

\noindent (That is, $\underaccent{\tilde}{\varpi}_{\sigma}=(\tau^{-})^{-1*}(\omega \otimes \sigma)$ and $\Tilde{B}=(\tau^{-})^{-1}B)$.) Applying this condition results in a definition of the measurement scheme for the FV measurement framework in terms of $\Theta$:  
\begin{equation}
   \omega (\varepsilon_{\sigma}(B)) = (\omega \otimes \sigma)(\Theta (\mathfrak{1} \otimes B)) =  \underaccent{\tilde}{\varpi} _{\sigma} (\Tilde{B}) 
\end{equation}

As in QMT, CP-instruments are introduced to describe the effect on the system of a measurement satisfying this measurement scheme. More specifically, Fewster and Verch adopt the same criterion of adequacy for the updated state as QMT (cf. \cite[pp.230-1]{busch_book_2016}):  ``[w]e would like to obtain a new system state that is conditioned on the observation of this effect, which means that the new state correctly predicts the conditional probability for the joint observation of B together with any system effect, given that B is observed.'' \cite[p.13]{fewster2020quantum}.

Since `impossible measurement' scenarios involve only non-selective measurements, we will only introduce the FV state update rule for non-selective measurements in this section. Recall also (from Sec. \ref{Secadhoc}) that the `impossible measurements' problem concerns how to formulate a measurement theory for QFT, not how to interpret the formalism. We will therefore discuss how the FV framework resolves the `impossible measurements' problem in the next subsection and then offer a physical interpretation of the FV measurement framework in the following subsection, where we will also consider the selective state update rule. The non-selective state update rule is an instance of the selective rule applied to the `always true' effect. The FV state update rule for non-selective measurements (i.e., when there is no filtering conditional on which probe effect is observed) is
\begin{equation}\label{nonselLR} 
    \omega ^{\prime} _{ns} (A)= (\Theta ^{*} (\omega \otimes \sigma)) (A \otimes 1)
\end{equation}

\noindent where $^{*}$ denotes the adjoint map.\footnote{That is, in general ($\zeta ^{*} \omega)(X) = \omega(\zeta (X))$.} For comparison, the analogous instrument for QMT gives the non-selective state update rule $\rho_{f}(\Omega)=tr_{\kappa}[U(\rho \otimes \sigma) U^{*}]$, where the pointer variable $Z$ has a value in $\Omega$ (i.e., a non-selective measurement is performed), $\kappa$ is the Hilbert space for the probe, and $\rho \otimes \sigma$ is the initial system-probe state \cite[p.231]{busch_book_2016}. The expression in Eq. \eqref{nonselLR} is the algebraic version of tracing out the probe. Moreover, the scattering morphism $\Theta$ represents the dynamics, unitary time evolution operator $U$.

We can now appreciate how the FV measurement represents local operations and dynamics in a manner that complies with the negative morals noted in Sec. \ref{axiomsection}. (Time-Slice Property) and (Isotony) imply that an algebra of observables $\mathcal{A}(N)$ can be localized in any region in the domain of dependence of $N$. The FV framework squares this with the fact that experiments happen in local labs by explicitly introducing a representation for the probe and a region $K$ in which the system and probe interact. A coupled system-probe algebra $\mathcal{C}$ is assigned to the causal hull of this region $K$, but this algebra can also be localized in any region $D(K)$. The localization of operations performed in the lab is instead reflected in the assumption that the algebras  $\mathcal{C}$ and $\mathcal{U}=\mathcal{P} \otimes \mathcal{S}$ can only be identified outside of $K$, where and when the system and probe are uncoupled. Second, the dynamics of the system, probe, and coupled system-probe are \textit{not} represented by choosing a foliation of Cauchy surfaces with respect to which to define a unitary operator to represent the time evolution. The FV framework demonstrates that the local embedding isomorphisms underwritten by (Time-Slice Property) suffice to represent not only the dynamics of the system, but also the dynamics of the system-probe measurement interaction. For the latter, the scattering morphism $\Theta$ is the key.

\subsection{How the FV framework blocks Sorkin-type impossible measurement scenarios}\label{bostelmann}

\begin{figure}
\centering
\includegraphics[width=0.6\textwidth]{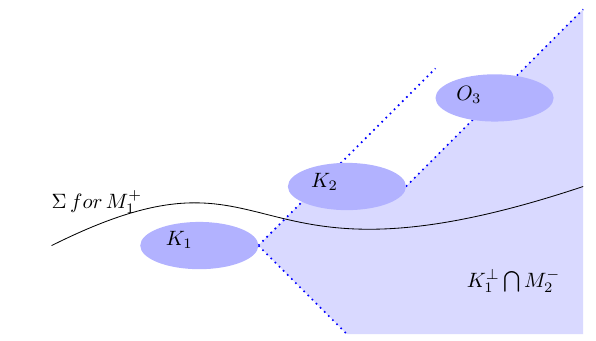}
\hspace{-2cm}
\caption{Figure for Bostelmann et al.'s proof (Based on Fig. 2 in \cite{PhysRevD.103.025017})}
\label{FigBostelmann}
\end{figure}

Bostelmann, Fewster, and Ruep \cite{PhysRevD.103.025017} prove that the application of the FV framework to Borsten et al.'s \cite{Borsten:2019cpc} `impossible measurement' scenario does not allow superluminal signalling. That is, the expectation value of a non-selective measurement in $O_3$ is independent of which unitary `kick' is implemented in $K_1$ (see Fig. \ref{FigBostelmann}). $K_{1}$ is the region in which the interaction occurs between the system $\mathcal{S}$ and the first probe $\mathcal{\mathcal{P_1}}$ and $K_{2}$ is the compact region in which the interaction occurs between the system $\mathcal{S}$ and the second probe $\mathcal{P_2}$. Outside of the causal hulls of $K_{1}$ and $K_2$ the uncoupled algebra $\mathcal{U}=\mathcal{S} \otimes \mathcal{P_1} \otimes \mathcal{P_2}$ is isomorphic to coupled algebra $\mathcal{C}$. $O_{3}$ is the region in which the superluminal signal is allegedly received. Two scattering morphisms are introduced: $\hat{\Theta}_{1}$  for the first measurement in $K_{1}$ and $\hat{\Theta}_{2}$ for the second measurement in $K_{2}$. $\hat{\Theta}_{1}$ is defined by extending the scattering morphism $\Theta_{1}: \mathcal{S} \otimes \mathcal{P_1} \rightarrow \mathcal{S} \otimes \mathcal{P_1}$ to $\mathcal{U}=\mathcal{S} \otimes \mathcal{P_1} \otimes \mathcal{P_2}$, and similarly for $\hat{\Theta}_{2}$.

Bostelmann et al. \cite{PhysRevD.103.025017} prove that the expectation value for any system observable C in $\mathcal{S}(O_{3})$ is independent of which non-selective measurement is performed in $K_{1}$. Recall that in the FV framework the state update rule for non-selective measurement of system observable $A$ is 

\begin{equation}
    \omega ^{\prime} _{ns} (A)= (\Theta ^{*} (\omega \otimes \sigma)) (A \otimes 1)
    \tag{\ref{nonselLR}}
\end{equation}

\noindent Applying this update rule to the measurement scenario in Fig. \ref{FigBostelmann}, the updated state of the system (for a system observable $C \in \mathcal{S}(O_{3})$) conditional on the specified non-selective measurement operations $A$ and $B$ performed in $K_{1}$ and $K_{2}$, respectively, is 

\begin{equation}
    \omega_{AB} (C)=(\omega \otimes \sigma_1 \otimes \sigma_2)((\hat{\Theta}_{1} \circ \hat{\Theta}_{2})(C \otimes \mathbb{1} \otimes \mathbb{1})) 
\end{equation}

\noindent Consequently, the result that the expectation value for the measurement of any system observable C in $\mathcal{S}(O_{3})$ is independent of the measurement performed in $K_{1}$ can be established by proving that the following algebraic equation involving the scattering morphisms holds:

\begin{equation}\label{scatteringeq}
   \forall C \in \mathcal{S}(O_{3}): (\hat{\Theta}_{1} \circ \hat{\Theta}_{2})(C \otimes \mathbb{1} \otimes \mathbb{1}) = \hat{\Theta}_{2}(C \otimes \mathbb{1} \otimes \mathbb{1})
\end{equation}

As anticipated, (Local Time-Slice Property) is the crucial ingredient in the proof. Bostelmann et al. rely on the following two properties of the scattering morphism that are established in Proposition 1 of \cite{fewster2020quantum}:

\begin{enumerate}
    \item For every region $N \subseteq K^{\perp}$:  $\Theta$ acts trivially on $\mathcal{S} \otimes \mathcal{P}(N)$
    
    \item For every region $N \subseteq M^{+}$ and every region $N^{-} \subseteq M^{-}$ with $N \subseteq D(N^{-})$:  $\Theta(\mathcal{S} \otimes \mathcal{P})(N) \subseteq (\mathcal{S} \otimes \mathcal{P})(N^{-})$
\end{enumerate}

Essentially, both theorems are established by using (Time-Slice Property) and (Compatibility) (which, recall, entail (Local Time-Slice Property)) to show that the scattering morphism preserves the localization properties of the local algebras on which it acts (see Appendix 1 of \cite{fewster2020quantum}). The proofs involve tracing the localization properties of algebras mapped by each of the two types of component morphisms of the scattering morphism:  the isomorphisms $\chi ^\pm$ between uncoupled algebra $\mathcal{U}$ and coupled algebra $\mathcal{C}$ in regions $M^{\pm}$ and the morphisms for local embeddings $\gamma$ of $\mathcal{C}$ and $\alpha \otimes \beta$ of $\mathcal{U}$ that are guaranteed by (Time-Slice Property). 

Bostelmann et al.'s proof that Eq. (\ref{scatteringeq}) holds has two main parts. They first establish that there exists a Cauchy surface $\Sigma$ for $M_{1}^{+}$ that is in the causal future of $K_1$ and the causal pasts of both $K_2$ and $O_3$ (see Fig. \ref{FigBostelmann}). The second step is to use the resulting information about domains of dependence to apply the two scattering properties. The key is that region $O_{3}$ is in the domain of dependence of the region $K^{\perp}_{1} \bigcap M^{-}_{2}$. By the second property of scattering morphisms, applying $\hat{\Theta}_2$ to $\mathcal{U}=\mathcal{S} \otimes \mathcal{P_1} \otimes \mathcal{P_2}(O_3)$ maps the uncoupled algebra to $K^{\perp}_{1} \bigcap M^{-}_{2}$. As a result of the first property, applying $\hat{\Theta}_{1} \circ \hat{\Theta}_{2}$ then gives the same result as applying $\hat{\Theta}_{2}$ because the localization region $K^{\perp}_{1} \bigcap M^{-}_{2}$ is in the causal complement of $K_{1}$. Therefore, any interactions between probe 1 and the system in $K_{1}$ do not affect the expectation values of system observables localized in $O_{3}$.

As we have emphasized, postulation of (Local Time-Slice Property) is the main reason that the FV framework does not allow Sorkin-type impossible measurements. In terms of the reductio argument in Sec. \ref{reductio}, Fewster and Verch's resolution involves adding this principle to the listed set of premises. Of course, the FV framework also adds other axioms for AQFT and a formulation of measurement theory that is suited to relativistic QFT. With the introduction of new state update rules, L{\"u}ders' rule for non-selective measurement is \textit{not} applied to obtain the updated state for the system in $O_3$. Instead, the state update rule for non-selective measurement in the FV framework (Eq. (\ref{nonselLR})) is applied. An obvious difference between the rules is that L{\"u}ders' rule is applied in a concrete Hilbert space representation and the FV state update rule is formulated in abstract algebraic terms. More specifically, the salient difference between the two rules is that L{\"u}ders' rule invokes (a sum of) projectors while the FV state update rule for non-selective measurement invokes $\Theta ^*$, which is (the adjoint of) a composite of algebraic isomorphisms. Recall that the algebraic isomorphisms composing $\Theta$ are of two types:  dynamical isomorphisms underwritten by (Time-Slice Property) and isomorphisms between algebras $\mathcal{U}$ and $\mathcal{C}$ in $M^+$ and $M^-$. The main point is that the FV state update rule for non-selective measurement depends on the algebraic dynamics; it does not depend on the action of operators in the algebra. The physical interpretation of this state update rule will be taken up in the next section. Fewster and Verch \cite[p.5]{fewster_measurement_2023} also draw the following moral from Sorkin-type impossible measurements:  ``the fact that a unitary operator is localisable in some region $O$ does not imply that it induces an operation that can be physically performed within $O$. This should not be a surprise: for instance, the Lagrangians that describe local fields with local interactions constitute a very specific (and small) subset of all possible Lagrangian field theories."\footnote{In response to a version of the `impossible measurements' problem, Earman and Valente \cite[p.14]{earman2014relativistic} also question the assumption that all unitary operators in a local algebra $\mathfrak{A}(O)$ correspond to unitary evolutions in $O$.} 

One might worry that this result does not fully address Sorkin-type impossible measurement scenarios. Bostelmann et al. \cite{PhysRevD.103.025017} prove that when the FV measurement framework is applied, all system observables in $O_3$ are independent of which non-selective measurement is performed in $K_1$, but this is not reassuring if there are physical system observables that are not measurable by any probe representable in the FV measurement framework. Fewster, Jubb and Ruep \cite{Fewster2023} addresses this issue by proving that, for the case of a real scalar field theory, every local system observable can be asymptotically measured by some collection of probes in the FV framework. That is, for every local system observable $A \in \mathcal{S}(N)$ for $N \subseteq M$, there is a set of system observables and FV measurement schemes such that the induced system observable $\varepsilon^{\mathcal{C}_\alpha}_{\sigma_\alpha}(B_{\alpha})$ converges to $A$\footnote{The proof demonstrates convergence in the strong operator topology of the GNS representation associated with any quasi-free state with distributional two-point function.} (implying that $A$ can be measured to arbitrary precision). They expect that with ``merely mild technical effort" their existence proof for asymptotic measurement schemes could be extended to multiple real scalar fields, Wick powers of real scalar fields, and other types of fields \cite[pp.24, 25]{Fewster2023}. van der Lugt \cite{van_der_lugt_relativistic_2021} approaches the problem of showing that physical observables are measurable within the FV framework from the perspective of standard non-relativistic quantum information theory. He introduces a `hybrid model' that implements the FV framework using the simpler Hilbert space models of NRQM (i.e., Type I von Neumann algebras with a natural tensor product structure) and then uses results from quantum information theory to show that in this hybrid model all operations that do not permit superluminal signalling can be measured in the FV framework.

\subsection{Physical interpretation of the state update rules in the FV measurement framework}\label{stateupdatesection} 

How should we physically interpret the FV state update rules? More specifically, can the state update rules be interpreted as representing physical processes or are they merely epistemic in the sense that they are only calculational devices that allow us to derive probabilistic predictions for measurement outcomes? In this subsection, we will go beyond the context that is strictly relevant to the Sorkin-type impossible measurement scenarios and consider selective as well as non-selective measurements. To foreshadow, we will argue that the FV framework can be interpreted as representing the physical process of measurement. However, there are two fundamental respects in which the interpretation of the FV state update rules differs from the interpretation of their counterparts in QMT:  the states $\omega$ and $\omega ^\prime$ are counterfactual states (not actual states of the system) and for selective measurements there is no region of spacetime in which a transition from $\omega$ to $\omega ^\prime$ must occur. Our interpretation of the FV framework in this section is compatible with many of the brief interpretative remarks offered by Fewster and Verch, but we do disagree with a few of their interpretative comments.

The state update rules are expressed in terms of scattering morphism $\Theta$. The sequence of morphisms that compose $\Theta^*$ (see the list on p.\pageref{scatteringlist}) suggests the following intuitive, chronological interpretation of this scattering theory for the states, ordered from the past to the future:
\begin{align}\label{scatteringpicture} 
    \omega \otimes \sigma (\mathcal{U}(M^-)) \rightarrow \varpi (\mathcal{C}(J^{+}(K) \bigcap J^{-}(K))) \rightarrow \nu (\mathcal{U}(M^+))
\end{align}

\noindent Of course, $\Theta$ is actually an isomorphism, but roughly speaking the morphisms vicariously map via the algebras the prepared state $\omega \otimes \sigma$ to the state $\varpi$ of $\mathcal{C}$ in the causal hull of the system-probe interaction region to the final system-probe state $\nu$. The fact that $\omega \otimes \sigma$ is a product state and $\nu$ is not is a reflection of a time-asymmetry that is built into the measurement scheme:  we assume that the prepared system and probe states are uncorrelated and that the measurement interaction correlates the system-probe states \cite[p.10]{fewster2020quantum}. How should we interpret the arrows in (\ref{scatteringpicture})? Clearly, they cannot represent the time evolution of the actual system because, as we have been emphasizing, the dynamical isomorphisms represent local embeddings and not time evolution (of either states or operators). Furthermore, these three states are not even defined on the same algebra:  $\omega \otimes \sigma$, $\nu$ are defined on $\mathcal{U}$ and $\varpi$ is defined on $\mathcal{C}$.

The interpretation of these states depends on the interpretation of the associated algebras. The actual world is represented by the coupled algebra $\mathcal{C}(M)$. The uncoupled algebra $\mathcal{U}(M)$ is best interpreted, as Fewster and Verch \cite[p.8]{fewster2020quantum} suggest, as representing ``the counterfactual world in which the interaction does not occur." Fewster and Verch \cite[p.8]{fewster2020quantum} also describe $\mathcal{U}(M)$ as representing a ``control situation." Accordingly, the actual state of the system-probe is given by $\varpi$ on $\mathcal{C}(M)$. $\varpi$ on $\mathcal{C}(M)$ is a global state in the sense that this state represents the entire actual history of the combined system-probe system. Likewise, $\omega \otimes \sigma$ and $\nu$ over $\mathcal{U}(M)$ are each counterfactual global states. 

The interpretations of $\varpi$, $\omega \otimes \sigma$, and $\nu$ differ from the interpretations of their counterparts in both QMT and conventional scattering theory for QFT. In QMT, the initial state $\rho \otimes \sigma$ is taken to be the \textit{actual} system-probe state, not a counterfactual state like $\omega \otimes \sigma$. $U(t)$ represents the actual time evolution of the system-probe (setting aside the question of what happens upon measurement). In conventional scattering theory in particle physics, the measurement probe is not included in the representation, but there is a similar contrast between the status of the the initial and final states: in conventional scattering theory, the initial and final states are also typically taken to represent the actual (free) states of the system at asymptotically early and late times, not counterfactual states.\footnote{More careful representations of scattering in particle physics such as LSZ or Haag-Ruelle scattering theory recognize that at asymptotic times the systems are not actually free so cannot be represented using a (free) Fock representation (see \cite{EarmanandFraser2006} for discussion). Our main point is that, in contrast, the FV framework introduces entire counterfactual histories in which no interaction between the system and probe occurs.}

Consider the non-selective state update rules. Like the states $\omega \otimes \sigma$ and $\nu$ from which they are derived by tracing out the probe, $\omega$ and $\omega_{ns} ^\prime$ are counterfactual states. As a result, the state update $\omega \rightarrow \omega_{ns} ^\prime$ cannot be interpreted as representing a physical change of the actual state of the system from $\omega$ to $\omega_{ns} ^\prime$. $\omega$ and $\omega_{ns} ^\prime$ are the appropriate states to assign in $M^-$ and $M^+$ respectively insofar as they predict the correct pre- and post-measurement conditional probabilities.\footnote{Regions $M^-$ and $M^+$ overlap in $K^\perp$. For non-selective measurements, $\omega(A) = \omega_{ns} ^{\prime} (A)$ for all A localizable in $K^\perp$, as one would expect. However, for selective measurements, this is not the case, as we discuss below.} Fewster and Verch \cite[p.7]{fewster_measurement_2023} describe state updates as ``an exercise in bookkeeping" that ``provides an effective description of a physical process." Is the FV measurement framework then merely a calculational device for deriving predictions, or does it also describe the physical process of measurement? While the state update rules are a calculational device, the FV framework is also equipped with the resources to describe actual physical processes. The state update rules are derived using $\varpi(\mathcal{C})$, which represents the the actual state and (presumably) actual physical changes in the values of the physical quantities that are associated with local regions.

Consideration of selective measurement (i.e., measurement given that probe effect $B$ is observed) yields two additional arguments against interpreting the state update rules as representing physical changes of state. These arguments concern how state updates are associated with spacetime regions, not the counterfactual versus actual status of the states. First, for state update for selective measurements, the expectation values given by $\omega_{s}^\prime$ do not in general agree with the expectation values given by $\omega$ in the necessary spacetime regions. As Fewster and Verch establish, if $\omega$ has the Reeh-Schlieder property, then $\omega_{s} ^\prime (A) \neq \omega (A)$ for any non-trivial observable $A$ that either (a) is localizable in the causal complement of $K$ or (b) is localizable in any region $O$ of the causal past of $K$ that has no null geodesics connecting it to $K$.\footnote{And assuming that the system obeys Huygens' principle} This is not surprising because the Reeh-Schlieder property is an indication that there are Bell correlations between regions, and an intermediate step in these results is that $\omega (A) = \omega_{s} ^{\prime} (A)$ for $A$ satisfying (a) or (b) only when $A$ and induced observable $\varepsilon _{\sigma} (B)$ are uncorrelated in state $\omega$. Fewster and Verch \cite[p.16]{fewster2020quantum} argue for the reasonable conclusion that, since $\omega_{s} ^\prime (A) \neq \omega (A)$ for $A$ in either the causal complement or the causal past of $K$, ``there seems to be no purpose in envisaging a transition from $\omega_{s}$ to $\omega_{s} ^{\prime}$ occurring along or near some surface in spacetime (whether a constant time surface as in non-relativistically inspired accounts of measurement, or e.g., along the backward light cone of the interaction region as in the proposal of Hellwig and Kraus \cite{PhysRevD.1.566}, or an earlier proposal of Schlieder)."

Second, and more compellingly, Fewster and Verch's conclusion that there is no reason to assume that an evolution from $\omega$ to $\omega_{s} ^\prime$ occurs in any region of spacetime is supported by a theorem about successive selective measurements \cite[Corollary 6]{fewster2020quantum}. Consider two probes in interaction regions $K_1$ and $K_2$ that are causally orderable (i.e., $K_2 \bigcap J^{-}(K_1)=\emptyset$). (See Fig. \ref{FigBostelmann} for an example; however, unlike the Borsten et al. \cite{Borsten:2019cpc} measurement scenario, these measurements will be selective.) Assume that the causal factorization property holds: $\hat{\Theta}=\hat{\Theta}_{1} \circ \hat{\Theta}_{2}$ where $\hat{\Theta}_{1}$ and $\hat{\Theta}_{2}$ are defined as above. This is a natural assumption, but it can also be verified for concrete models of system-probe interactions. Fewster and Verch show that application of the selective rule for state update produces the same result regardless of whether the interactions are modeled as successive selective measurements in regions $K_1$ and $K_2$ or as a single selective measurement in region $K_{1} \bigcup K_{2}$ in which probe effects $B_1$ and $B_2$ are both observed. More precisely, for probe preparation states $\sigma_1$ and $\sigma_2$ and probe effects $B_1$ and $B_2$, let $\omega ^{\prime} _1$ be the state conditioned on $B_1$ being observed in initial state $\omega$, $\omega ^{\prime \prime} _{12}$ be the state conditioned on $B_2$ being observed in state $\omega ^\prime _1$, and $\omega ^{\prime} _{12}$ be the state conditioned on $B{_1} \otimes B_2$ being observed in initial state $\omega$. Then $\omega ^{\prime \prime} _{12}=\omega ^{\prime} _{12}$.\footnote{Assuming that $B_1$ has nonzero probability of being observed in $\omega$ and $B_2$ has nonzero probability of being observed in $\omega ^{\prime}_1$.} Fewster and Verch \cite[p.17]{fewster2020quantum} ``emphasize that we have have not needed to invoke any reduction of the state across geometric regions." They consider this corollary (and the accompanying theorem for the corresponding pre-instruments) to be the ``core result" of their article \cite[p.27]{fewster2020quantum}. The upshot is that the selective state update rule in the FV framework ``renders moot the discussion of where and when a state change of the system takes place as a consequence of measurement" \cite[p.27]{fewster2020quantum}. In other words, in QMT the selective state update rule can be interpreted literally as representing a measurement-induced collapse that occurs somewhere in the world. Even if one believes that this is not a compelling interpretation of NRQM, it is an admissible interpretation of the formalism. In contrast, the FV selective state update rule does not admit a literal interpretation as representing a measurement-induced collapse that occurs somewhere in the world. 

Corollary 6 is a direct result of the requirement that the FV measurement theory reflect relativistic spacetime structure. As Fewster and Verch \cite[p.17]{fewster2020quantum} note, Theorem 5---which is the counterpart for instruments of Corollary 6 for state update---does not hold for non-relativistic theories such as Euclidean QFT. Essentially, what Theorem 5 and Corollary 6 establish is that the measurement theory enshrines both the symmetry between the ordering of measurement operations when $K_1$ and $K_2$ are causally disjoint (i.e., $K_{2} \subset K_{1} ^{\perp}$) \textit{and} the broken symmetry between the ordering of the measurement operations when $K_1$ and $K_2$ are strictly causally ordered (i.e., $K_{2} \subset J^{+}(K_{1})$). In contrast, Sorkin's relativistic causal ordering was a requirement that had to be put in by hand as P3(c) when applying non-relativistic quantum measurement theory to a relativistic case. Again, the contrast with QMT is also illuminating. There is no analogue of Corollary 6 for QMT because successive measurements are strictly temporally ordered:  there is no alternative to representing the measurements as $\omega ^{\prime \prime} _{12}$ (i.e., with successive state updates) because the time evolution operator $U$ appears in the state update rule.

This raises the question of whether the FV measurement framework is inherently epistemic in the sense that it only admits an interpretation in terms of the processing of information by observers, and does not admit an interpretation in terms of physical processes in the world. We have already suggested that the representation of the actual state of affairs by $\varpi(\mathcal{C})$ affords the FV framework the opportunity to represent actual physical processes, and not only the counterfactual descriptions underwritten by the state update rules. There are also two respects in which the state update rules themselves are not merely epistemic:  they are not observer-dependent in any significant respect and they do not need to be given a strong operationalist interpretation. An epistemic interpretation of the FV measurement framework is possible, but not required.

Some remarks by Fewster and Verch suggest that state update has an observer-dependent interpretation:  for example, in the context of selective measurements, ``there seems no reason to invoke a physical process of state reduction occurring at points or surfaces in spacetime, rather, the updated state reflects the observer’s filtering of the system by conditioning on measurement outcomes" \cite[p.16]{fewster2020quantum}. However, there is nothing especially observer-dependent about the state update rules. An updated state $\omega ^\prime$ (minimally) serves the purpose of generating conditional probabilities (i.e., conditional on a specified probe effect). Of course, conditional probabilities can be calculated for conditions that are not known by an observer to obtain---or even conditions that are known by an observer not to obtain. Moreover, the updated state $\omega ^\prime$ is spacetime-independent. This is trivially true because the algebras of observables, and not the states, are associated with regions of spacetime. More substantively, when states are associated with spacetime regions via the algebras, $\omega ^\prime$ can be defined over $\mathcal{U}(M)$; the state is associated with all of spacetime, so clearly is not dependent on the location of the observer. Of course, $\omega ^\prime$ is the correct state to assign to the system only in $M^+$ insofar as it only gives the correct conditional probabilities for measurements performed in this region. But this is not in any way observer-dependent---$M^+$ is picked out by the system-probe interaction region $K$. $M^+$ is \textit{not} defined with respect to the observer's location in spacetime, including with respect to where the observer chooses to perform a measurement on the probe to obtain information about the value of an induced system observable.   

As just noted, a selective state update can yield an updated state $\omega_{s} ^\prime(A)$ that differs from the prepared state $\omega(A)$ in the causal past of $K$. Fewster and Verch \cite[p.16]{fewster2020quantum} endorse Hellwig and Kraus' \cite{PhysRevD.1.566} strongly operationalist interpretation of this fact:  ``whether the state actually remains unchanged or not in the past of the coupling region is a `pure convention' with no operational significance as the region is no longer accessible to further experiment." While it is true that the observer in the causal future of $K$ who assigns state $\omega_{s} ^{\prime} (A)$ cannot experimentally test the predictions of this state assignment in $M^-$, it is not necessary to adopt this strong version of operationalism in order to defend the state assignment. The counterfactual interpretation of states $\omega$ and $\omega ^{\prime}$ defended above also permits observers in $M^-$ and $M^+$ to adopt different global states. Both $\omega$ and $\omega_{s} ^{\prime}$ are counterfactual states that represent states of affairs in possible worlds. It is not meaningful to ask which of these states represents the true state of the actual world in \textit{any} region of spacetime. (A state $\varpi$ on $\mathcal{C}(M)$ is the true state of the actual world.) The state update rules perform their function of predicting expectation values for measurements by supplying counterfactual states that are only appropriate to use for calculational purposes in suitable spacetime regions (e.g., before another measurement of the system is made). 

Finally, the FV measurement framework is an ongoing research program, so there are outstanding formal and interpretative questions. How the formalism gets developed will have consequences for how the interpretation of the formalism discussed in this section can be elaborated. For example, one formal development is that the recent extension of the FV measurement framework from compact probe-system coupling regions to non-compact ones for the case of a quantized real linear scalar field in \cite{Fewster2023}. Some abstract *-algebraic models have been constructed for particular types of field systems, probes, and interactions between them (e.g., for the simple case in which field and probe are modeled using free (massive or massless) scalar fields and are linearly coupled in \cite{fewster2020quantum}. Another formal project is to implement the FV measurement framework using C*-algebras or concrete von Neumann algebras. \cite{Fewster2023} applies the FV framework to C*-algebras. von Neumann algebras have not yet been explicitly constructed, but \cite[p.28, fn. 18]{Fewster2023} contains an argument that, for quasi-free Hadamard states, the asymptotic measurement scheme that is defined is implementable on the von Neumann algebra. (See Ruep \cite{ruep_thesis_2022} for more detail.)\footnote{Thanks to Maximilian Ruep for helpful correspondence about this.} 

On the interpretative side, Fewster and Verch \cite[p.13]{fewster_measurement_2023} note that there are outstanding questions about how to interpret superpositions and transition probabilities in the algebraic framework because these concepts are native to algebras of bounded operators on separable Hilbert spaces, which are Type I von Neumann algebras. Local algebras in AQFT are typically Type III von Neumann algebras. As \citet{ruetsche_earman_2011} explain, this creates problems for justifying the core interpretive assumption that quantum states represent probabilities \textit{of something} (e.g., measurement outcomes or hidden variables). In quantum mechanics with a finite number of degrees of freedom, the standard justification for assuming that states represent probabilities of measurement outcomes relies on the existence of atoms (i.e., minimal projectors). The Type III von Neumann algebras that are characteristic of local QFT do not contain atoms. Ruestche and Earman \cite{ruetsche_earman_2011} consider several strategies for extending the standard argument that states represent probabilities of either events (e.g., measurement outcomes) or value states (e.g., hidden variables) to Type III von Neumann algebras, including a version of the `funnel' proposal suggested in Fewster and Verch \cite[p.13]{fewster_measurement_2023},\footnote{Fewster and Verch \cite{fewster_measurement_2023} cite a more recent and further developed proposal in Buchholz and St\o{}rmer \cite{buchholz_superposition_2015}.} but find all of the arguments deficient. They conclude that justifying the interpretation of states as representing probabilities of either events or value states is an open problem for Type III von Neumann algebras.

In summary, the FV measurement framework begins by positing a set of axioms for AQFT. These physical principles inform the measurement theory that FV introduce. This measurement theory is also constrained by the fact it is set up in an analogous way to QMT from NRQM, and by the use of a scattering theoretic framework to model the measurement process. When Sorkin-type `impossible measurement' scenarios are modeled in the FV framework they do not lead to superluminal signalling. The FV measurement theory for AQFT has state update rules that take a different form and have a different interpretation from QMT. In particular, the `in' $\omega$ and `out' $\omega^\prime$ algebraic states that feature in the state update rules are best interpreted as counterfactual states. The actual state of the system is represented by $\varpi$ on $\mathcal{C}(M)$. Moreover, the state update rules cannot be literally interpreted as representing a physical change of state upon measurement that occurs in some region of spacetime.

\section{Pragmatic approach:  Detector models }\label{detectormodelssec}

\subsection{The use of Unruh-DeWitt-type detector models in RQI}\label{nrdetectorsec}

Sorkin-type examples of impossible measurements indicate that it is problematic to na{\" i}vely extend ideal measurement theory from NRQM within a minimal framework for relativistic quantum theory. From a practical point of view, this is distressing because measurement is obviously central to the use of the theory. This is an especially pressing concern in relativistic quantum information (RQI), which is concerned with the theoretical and experimental treatment of finite-time processes and realistic detectors. This makes it appealing to hang on to ideal measurement theory and to give up one of the other assumptions in the reductio argument instead. Detector models implement this strategy by introducing non-relativistic quantum mechanical detector systems and coupling them to relativistic quantum fields represented using QFT. The  relativistic quantum fields are not `directly' measurable, but can only be measured indirectly through their dynamical coupling to a \textit{controllable} detector system, or probe\footnote{Often the terms `detector' and `probe' are used interchangeably, especially if it is not clear from the context whether we are modeling a macroscopic or microscopic detector coupled to the field. A microscopic quantum mechanical system (like a spin or an atom) is commonly called a `probe' of the field, while this term is not used for explicitly macroscopic detector systems (like a superconducting qubit, see below).}. This von Neumann-like approach involves modeling the measurement as a dynamical process, where the quantum field and the probe are suitably coupled for a finite (or possibly infinite) time, the measurement duration. After the coupling has been switched off (or becomes negligible), the probe can be directly measured, and quantum measurement theory is applied to the detector system. The outcomes can be translated into statements about the quantum field, at least in principle. In this sense, the quantum field is measured by the detector system. If the detector-field coupling is local then the detector can be thought of as a local probe that is locally measuring the quantum field. If we want to retain ideal measurement theory from NRQM, this dynamical understanding of the measurement process may be the only feasible option in Minkowski spacetime, given Sorkin's no-go result. 

Detectors are typically defined as \textit{controllable} localised systems that are locally probing the quantum field. The requirement that the detector system is controllable is more naturally fulfilled if the detector is chosen to be a non-relativistic system. This means that it is well-described by non-relativistic quantum mechanics and the measurement theory that comes with it. Crucially, we can consider projective measurements over the detector system with the usual L{\"u}ders state update rule and probability assignments that correspond to each possible outcome. This is an advantage because the notion of a measurement outcome that is associated to a finite-rank projector is typically not available in QFT, as an implication of the Reeh-Schlieder theorem and arguments that local algebras are generically Type III von Neumann algebras, which by definition do not contain finite-rank projectors \cite{fewster2019algebraic,fewster2020quantum, RevModPhys.90.045003}. It is not clear that generalizing from projectors to POVMs addresses this problem because the spectral theorem no longer holds, and therefore cannot be appealed to as support for the interpretation of the probabilities as probabilities of measurement outcomes \cite{ruetsche_earman_2011}. Even though it can be a relief that the usual notion of a measurement outcome can be maintained through the introduction of a detector system, the association of detector outcomes with induced field observables is far from straightforward.  Partial answers to the question `what do detectors detect?' have been given \cite{PhysRevD.98.105011, deRamonRiveraJose2021,SmithAlexander2017, PhysRevD.103.125021}, but a systematic account is still missing. 

Originally particle detector models were introduced to extract particle phenomenology in QFT (in curved spacetimes) related to the Unruh and Hawking effects \cite{PhysRevD.14.870,PhysRevD.29.1047}. Since quantum field theories do not permit a particle ontology \cite{wald1994quantum,fraser2021particles}, this motivated the operational approach that `a particle is what a particle detector detects' advocated by Davies and others \cite{PhysRevD.14.870,1984qtg..book...66D,fraser2021particles}. The Unruh-DeWitt detector model has become a paradigm example in the field of Relativistic Quantum Information (RQI). RQI was born out of the need to merge quantum information theory with with relativity theory, and a core commitment of the approach is that relativistic QFT is a necessary ingredient (see e.g. \cite{Hu_2012,ANASTOPOULOS2023169239}). See Peres and Terno \cite{RevModPhys.76.93} for a pioneering defense of this approach. RQI describes quantum communication through quantum fields (e.g. \cite{PhysRevA.81.012330,PhysRevLett.114.110505}) and the entanglement structure of QFT by locally coupling multiple detectors to the quantum field (e.g.\cite{Reznik2003,PhysRevD.94.064074}). 

In the realm of quantum information, the notion of operations performed in local regions that is informally used in the application of quantum mechanics becomes central. Detector models are introduced to extend the use of QFT from high energy physics to relatively low energy systems probed in quantum information or quantum optics \cite{davies1976quantum,PhysRev.130.2529}. Many variations upon the Unruh-DeWitt model\footnote{Recent reviews of the variants of Unruh-DeWitt-type detector models in QFT (in flat and curved spacetimes) can be found in \cite{PhysRevD.108.045015,PhysRevD.106.025018}.} have been developed and applied to probe many different types of relativistic quantum systems described by QFT.


\subsection{Constructing non-relativistic detector models}\label{detectorfieldinteraction}

Like any other model, detector models are an addition to the underlying theory and, as a result, they are not a priori guaranteed to comply with its premises. Detector models raise a major concern when the underlying theory is relativistic QFT: are the predictions of the non-relativistic model respectful of relativistic causality? This is a justified concern, especially because the detector is chosen to be a non-relativistic quantum-mechanical system and, as such, alien to Minkowski spacetime. From this perspective, the non-relativistic quantum-mechanical nature of the detector seems like a serious drawback. On the other hand, thanks to its non-relativistic nature, the detector system is localizable in the usual quantum-mechanical terms. First-quantised non-relativistic systems admit a position representation, which implies that their states will be representable, and localizable, by means of their spatial wavefunction. Such a representation is known not to be available for relativistic systems \cite{malament1996defense}. Relativistic quantum fields are localized in a different sense: they are operator-valued objects that are locally defined over space and time (e.g., in AQFT by associating algebras of observables to bounded spacetime regions \cite{haag2012local}). Since the field (relativistic) and the detector system (non-relativistic) enjoy very different notions of localization, it is first important to clarify the sense in which they can be locally coupled.


The simplest version of the Unruh-DeWitt (UDW) model involves a scalar quantum field coupled to a non-relativistic quantum system (e.g. an atom, a harmonic oscilator, or a two-level system). There have been attempts to extend the model beyond the scalar field, e.g., to spinor fields \cite{PhysRevD.93.024019}, but this complication is not relevant for our purposes. Also, for simplicity, we will only refer to the case of linear coupling between the detector and the field, even though more complicated couplings, e.g., quadratic, have been investigated in the literature \cite{PhysRevD.93.024019, Takagi, PhysRevD.104.105021}. A careful treatment of the modeling of light-matter interaction with UDW-type detectors beyond the scalar approximation can be found in \cite{PhysRevA.103.013703}. 

 Perhaps the most well-known detector model that has been considered in the literature is the pointilke UDW detector model, in which it is assumed that the detector is coupled to the field over a timelike trajectory. This is prescribed by the interaction Hamiltonian that generates translations with respect to the proper time $\tau$ associated to the detector's trajectory. In the interaction picture, this Hamiltonian is given by
 \begin{equation}\label{UDWpoint}
    \hat H_{\text{int}}=\lambda\chi(\tau)\hat \mu(\tau) \otimes\hat\phi(\mathsf{x}(\tau)).
\end{equation}
 Here $\lambda$ is the coupling strength, $\chi (\tau)$ is the switching function, which is usually assumed to be  integrable, and $\mathsf{x}(\tau)$ is the spacetime trajectory of the detector parametrized by its proper time $\tau$. The Hamiltonian couples the field along the wordline of the detector to an internal degree of freedom of the detector $\hat{\mu}$ \footnote{In the case of a two-level system with energy gap $\omega$, $\hat{\mu}$ is the monopole operator $\hat{\mu}(t)= e^{i\omega t}\hat{\sigma}^++ e^{-i\omega t} \hat{\sigma}^-$ where $\hat{\sigma}^{\pm}$ the operators that map between the ground and excited state.}. The point-like model can exhibit ultraviolet divergences related to the coincidence limit of the time-ordered $n$-point functions. One strategy for avoiding the divergences of the point-like model is to introduce a finite extension of the detector-field interaction through a smearing, but as we will see this introduces issues with the covariance and the causality of the model that are due to the extension of the interaction and are absent in the point-like model \cite{2020broken,deramon2021relativistic}.

The simplest generalisation of the point-like interaction Hamiltonian \eqref{UDWpoint} involves a linear coupling between the detector observable $\hat{\mu}(t)$ and the scalar field operator $\hat{\phi}(t,\bm{x})$ 
\begin{equation}
    \hat{H}_{\text{int}}= \lambda \chi(t) \hat{\mu}(t)\otimes \int \d \bm{x} F(\bm{x}) \hat{\phi}(t,\bm{x}) \label{hint}
\end{equation}
where the switching function $\chi(t)$ models the duration of the interaction between field and detector and the smearing function $F(\bm{x})$ specifies the spatial extension of the interaction (in the proper frame of the detector system) \cite{PhysRevD.97.105026,PhysRevD.101.045017}. The support of these functions specifies the spacetime region $O$ over which the detector is coupled to the field, i.e., $O=\text{supp}\chi(t)F(\bm{x})$. If both the smearing and the switching functions are compactly supported, then the interaction region $O$ is bounded. Note that the interaction region need not coincide with the (initial) localization region of the quantum-mechanical detector system. Commonly both functions (switching and smearing) are introduced as a phenomenological input of the model, especially when the detector system is macroscopic. The smearing is modeling the `size' of the interaction, which in general will not coincide with the apparent size of the detector, and the switching is modeling the mechanism for switching the interaction on and off (whenever such mechanism is available\footnote{For elementary interactions the switching function is harder to motivate, since, for example, the coupling of an electron coupled to the electromagnetic field cannot be `switched off'.}).



It is perhaps curious that even in the case of an explicitly macroscopic detector system (e.g. in superconducting circuits \cite{PhysRevA.96.052325}) the physical intuition that `the interaction happens where the detector is' is not fulfilled. In \cite{PhysRevA.96.052325} the authors investigate the model-dependence of the predictions for different smearing functions and different cut-off functions that determine `how many' field mode functions are relevant for the detector-field interaction. The result suggests that, in this case, the real shape and size of the macroscopic detector does not affect the prediction as much as the choice of a UV cutoff. This means that one can directly model the feature of finite extension, based on mathematical convenience, without worrying about how the microscopic details of the detector affect the smearing function. In other studies, where the detector is explicitly a microscopic probe (e.g. the electron of an atom coupled to the quantum electromagnetic field), the smearing function has been associated with the microscopic nature of the probe (e.g., the orbitals of a hydrogen atom interacting with the electromagnetic field in a light-matter interaction \cite{PhysRevD.94.064074}). It is common to attribute the smearings to the detector (e.g. `smeared' detector as opposed to point-like detector) even though smearing the field operator is more accurate mathematically. Conceptually, it is preferable to attribute the smearing (the `shape' of the interaction) to neither the detector nor the field but to their joint interaction. As McKay, Lupascu, and Mart{\'i}n-Mart{\'i}nez \cite{PhysRevA.96.052325} put it,  ``the shape of the qubit cannot be determined just with an individual description of the qubit itself. Rather, this shape belongs neither to the qubit nor to the line but to the both of them in interaction with each other, constituting a property that becomes evident and relevant in and through interactions between the relevant quantum systems.'' 

Overall, the choice of switching and smearing functions is a crucial input of the model that can critically affect its predictions. This choice can be motivated by the underlying (i.e., microscopic) physics, first-principles, mathematical convenience, or even aesthetics. In the spirit of the pragmatic approach, it is common to investigate all possible (calculable) choices without particular emphasis on motivating each possible choice. This model-dependence poses an extra challenge when detectors are used to study universal effects like the Unruh effect \cite{EARMAN201181}. On mathematical grounds, the smearing functions were first introduced as a cure to the UV divergences of the point-like model \cite{DeWitts} where the detector interacts with the field in a point-like manner. The UV divergences of the point-like model come from the distributional character of the `field at a point'. Concretely, the response function of a detector at leading order in perturbation theory is a function of the field's Wightman function and can be regulated in different ways through the introduction of suitable switchings and smearings \cite{Schlicht, Louko2006HowOD}. In this literature, it is typical that the smearing depends on a regulator $\epsilon$ (e.g., Gaussian/Lorentzian function) for the purpose of regularising the response of a point-like detector (e.g., excitation probability) in the limit $\epsilon \rightarrow 0$ \cite{Schlicht}. Without taking the limit, an infinitely extended smearing function is unphysical since it implies a `non-local' coupling between the field and the detector in all space. 


Finally, let us consider the `covariant' generalisation of the Unruh-DeWitt interaction Hamiltonian \cite{PhysRevD.101.045017,2020broken}, where the switching and the smearing come together to form a \textit{spacetime} smearing function $\Lambda(\mathsf{x})$ e.g. $H_{int}(t)=\lambda\int \d V  \Lambda(\mathsf x) \hat{\mu}(\tau(\mathsf{x}))\otimes \hat{\phi}(\mathsf x)$. In this interaction Hamiltonian both the field and the detector operator are `smeared' by $\Lambda$ in the sense that the detector inherits spatial dependence through its proper time $\tau=\tau(\mathsf x)$ in a general reference frame with coordinates $\mathsf x$. For example, if we are considering Lorentz boosts in Minkowski spacetime, $\hat{\mu}(\tau)=\hat{\mu}(\gamma (t-\bm{v}\bm{x}))$ in a boosted frame with coordinates $(t,\bm{x})$. The detector observable $\hat{\mu}$ is only time- (and not space-) dependent in its proper frame, where the spacetime smearing function factorizes like $\Lambda(\mathsf{x})=\chi(\tau)F(\bm{x})$ in terms of the switching and the smearing functions. The time duration and the spatial extension of the interaction can only be defined separately in the detector's proper frame (e.g. Fermi normal coordinates in curved spacetime \cite{2020broken}), while they mix in a general reference frame \cite{PhysRevD.97.105026,PhysRevD.101.045017,PhysRevD.106.025018}. This `covariant' form of the interaction Hamiltonian was proposed in \cite{PhysRevD.101.045017} for a consistent description of detector physics in curved spacetimes, even though the model fails to be fully covariant due to the non-relativistic nature of the detector \cite{2020broken}. 

To give a definition of what we mean by a non-relativistic detector model, it will be useful to write the interaction Hamiltonian (density) in the following general form that was introduced in \cite{deramon2021relativistic} (see also \cite{PhysRevA.86.012111})
\begin{equation}
  \hat h(\mathsf x)=\lambda \Lambda(\mathsf x) \hat {\mathcal{J}}(\mathsf x)\otimes\hat \phi(\mathsf x) \label{cov}
  \end{equation}
where $\hat {\mathcal{J}}(\mathsf x)$ is a current operator that is associated with the detector. This form of the interaction Hamiltonian covers the zoo of detector models that one finds in the contemporary literature (see \cite{PhysRevD.108.045015}). In principle, the detector current (through which the particle detector couples to the field) could be derived using an effective field theory approach (e.g.\cite{PhysRevA.86.012111,PhysRevD.102.093003},\cite{PhysRevLett.125.213603}). We say that the detector system is non-relativistic if the detector current is not microcausal over the extension of the interaction region: 
\begin{equation}
    \left[\hat {\mathcal{J}}(\mathsf x), \hat {\mathcal{J}}(\mathsf x')\right] \neq 0 \quad\text{for spacelike separated} \quad \mathsf x, \mathsf x' \in O.
\end{equation}
In general, the microcausality condition will not be satisfied by the current operators when one considers spacelike separated points within the extension of the interaction region $O=\text{supp}\Lambda$ due to the non-relativistic dynamics of the detector system \cite{deramon2021relativistic}. This observation will become important when analysing the frictions with relativistic causality in the following subsections. \footnote{Note that, based on this criterion, the pointlike model is fully causal since all points in the support of the interaction (a timelike trajectory) are causally connected.}


Due to the coupling of a non-relativistic detector model to a relativistic system, it is not a priori guaranteed that the resulting model for measurement will comply with relativistic causality. In general, one deals with three distinct layers: the underlying theory (i.e., QFT), the chosen model (i.e., the non-relativistic detector model), and the relevant mathematical approximations (for more discussion see \cite{RUIZDEOLANO202282}). If an empirical prediction is inconsistent with the underlying theory (e.g., signalling at spacelike separation), it can be the fault of the model and/or the approximation. Approximations that are known to be at odds with relativistic causality in the particle detector literature are the rotating wave approximation, the non-relativistic approximation and the relevant zero-mode approximation \cite{PhysRevD.100.065021, Papageorgiou_2019,PhysRevD.99.065005}. This is due to the non-locality associated with the field observables that these approximations introduce. 

Approximations are not to blame for violations of relativity theory in Sorkin-type scenarios (even though they do play a role \cite{Benincasa_2014}), but they do illustrate the pragmatic approach taken in detector modeling:  approximations that violate relativity theory can be tolerated in regimes in which the violations are negligible. To address Sorkin-type scenarios, the goal is not to rule out superluminal signalling in principle, but to argue on a model-by-model basis that violations are negligible in the intended domain of application. Furthermore, superluminal signalling is not restricted to Sorkin-type impossible measurement scenarios within the detector models program. The next subsection discusses superluminal signalling in bipartite measurement scenarios, and then we will turn to superluminal signalling in tripartite, Sorkin-type measurement scenarios in Sec. \ref{rulingoutIMsec}.

\subsection{Frictions with relativistic causality: Superluminal signalling with two detectors}

As a preliminary to analyzing Sorkin-type impossible measurement scenarios, consider two detectors in spacelike separated regions. For example, consider two two-level systems A,B coupled to the field through the interaction Hamiltonian 
\begin{equation}
    \hat{H}_{int} =\sum_{\nu=\textsc{a},\textsc{b}} \lambda_{\nu} \chi_{\nu}(t) \hat{\mu}_{\nu}(t)\otimes \int \d \bm{x} F_{\nu}(\bm{x}) \hat{\phi}(\bm{x},t) \label{mu}
\end{equation}
 Since the two detectors are not directly coupled to each other, the question is: how much signalling can be `transmitted' through their coupling to the quantum field? Since the field is relativistic, is there any causality condition for the field that blocks superluminal signalling if the two detectors (i.e., the two interaction regions) are placed in spacelike separation?


In  Mart{\'i}n-Mart{\'i}nez \cite{PhysRevD.92.104019} it was shown that after A and B have interacted with the field (assuming that A interacts with the field before B in some reference frame) the state of detector B at leading order in perturbation theory is
\begin{equation}
    \hat\rho^{(2)}_{\textsc{b}}= \lambda_{\textsc{a}} \lambda_{\textsc{b}} \hat{\rho}^{(2)}_{\textsc{b},\text{signal}}+ \sum_{\nu=\textsc{a},\textsc{b}} \lambda^2_{\nu} \hat{\rho}^{(2)}_{\nu,\text{noise}} \label{rho2}
\end{equation}
where the noise term is local on detector B, and all the influence of the presence of detector A on detector B's density matrix is captured by the `non-local' term that is proportional to $\lambda_{\textsc{a}} \lambda_{\textsc{b}}$. This signalling part of the density matrix can be written as
\begin{equation}
    \hat{\rho}^{(2)}_{\textsc{b},\text{signal}} =2 \int \d t \d t'  \chi_{\textsc{a}}(t) \chi_{\textsc{b}}(t') \mathcal{C}(t,t') \hat{d}(t,t')
\end{equation}
where
\begin{equation}
    \mathcal{C}(t,t'):= \int \d \bm{x} \d \bm{x}' F_{\textsc{a}}(\bm{x}) F_{\textsc{b}}(\bm{x}')\langle[\hat{\phi}(t,\bm{x}),\hat{\phi}(t',\bm{x}')] \rangle \label{smearedcommutator}
\end{equation}
and $\hat{d}$ is an operator that depends on the states and the internal frequencies $\omega_{\textsc{a}},\omega_{\textsc{b}}$ of the detectors. If both smearings are compactly supported, the integration is performed only over the two disjoint and spacelike separated spacetime regions, supp$\chi_{\textsc{a}} F_{\textsc{a}}$ and supp$\chi_{\textsc{b}} F_{\textsc{b}}$. Microcausality guarantees that the field commutator vanishes in spacelike separation, and there is no  superluminal signalling between the two detectors at second order in perturbation theory.\footnote{In the case of point-like interactions, this argument can be extended to higher orders in perturbation theory \cite{PhysRevA.81.012330}. A non-perturbative argument can be found in \cite{deramon2021relativistic}.} This behaviour has also been studied in the general case, using the Hamiltonian density \footnote{$\mathcal{E}(\tau)$ is a one-parameter family of spacelike surfaces, where $\tau$ is a global function whose level curves represent the planes of simultaneity of the detector's center of mass and (under some assumptions \cite{PhysRevD.101.045017}) $\tau$ is the detector's proper time. $\d\mathcal{E}$ denotes the family of induced measures on the surfaces $\mathcal{E}(\tau)$. Note that we have assumed that the two detectors share the same proper time.} \eqref{cov}
\begin{equation}
    \hat H_{int}(\tau)= \sum_{\nu=\textsc{a},\textsc{b}} \lambda_{\nu} \int_{\mathcal{E}(\tau)}\d\mathcal{E}  \hat{J}_{\nu}(\mathsf x) \otimes \hat{\phi} (\mathsf x). \label{hintAB}
\end{equation}
Note that, for convenience, we have absorbed the spacetime smearing function in the definition of the detector current operator, i.e., $\hat{J}_{\nu}(\mathsf x):=\Lambda_{\nu}(\mathsf x)\hat{\mathcal{J}}_{\nu}(\mathsf x)$ (comparing with \eqref{cov}).  
 If we assume that the state is initially uncorrelated, i.e. $\hat\rho_{\text{initial}}=\hat\rho_\textsc{a} \otimes \hat\rho_\textsc{b}\otimes \hat\rho_{\phi}$, the  general expression for signalling is \cite{PhysRevD.108.045015}
\begin{align}
    \hat\rho^{(2)}_{\textsc{b},\text{sign}}=-\ii[\hat\Sigma,\hat\rho_\textsc{b}] \label{sigma1} \,\,\,
     \text{where}\,\,\, \hat\Sigma=\int\int\d V\d V' \braket{\hat J_\textsc{a}( \mathsf x')}G_\textsc{r}(  \mathsf x,\mathsf x')\hat J_\textsc{b}( \mathsf x)
\end{align}

\noindent and where $G_\textsc{r}(\mathsf x,\mathsf x')$ is the retarded Green's function\footnote{$ G_\textsc{r}(  \mathsf x,\mathsf x')=-\ii \theta(\tau(\mathsf{x})-\tau(\mathsf{x'}))\braket{[\hat\phi(\mathsf{x}),\hat\phi(\mathsf{x}')]}$.}. We see that, for general switching functions (dropping the assumption that the switching functions are compactly supported and non-overlapping), the role of the field commutator in \eqref{smearedcommutator} is played by the field's retarded Green's function in \eqref{sigma1}.

The operator $\hat{\Sigma}$ can be understood as the current associated with detector B smeared by the propagated expectation value of the current associated with detector A. In the case of the massless Klein-Gordon field in a 3+1 dimensional flat spacetime, for instance, the propagator takes the familiar form of the Lienard-Wiechert potentials
\begin{align}
     G_{\textsc{r}}[\braket{\hat J_\textsc{a}}](t,   \bm{x})=\int \d^3\bm{x}' \frac{\braket{\hat J_\textsc{a}}(t_{\textsc{r}},\bm{x}' )}{2|\bm{x}-\bm{x}'|}
\end{align}
where $t_{\textsc{r}}=t-|\bm{x}'|$ is the retarded time. We see that the operator $\hat{\Sigma}$ carries all the information about the signalling from detector A to B. In \cite{PhysRevD.108.045015} it was shown that the variance of $\hat{\Sigma}$ bounds the Fisher information of B, i.e., the information that detector B can `learn' about the coupling of A to the same quantum field. Again, we notice that if the `source' $\braket{\hat J_\textsc{a}( \mathsf x')}$ is spacelike separated from the `receiver' $\hat{J}_\textsc{b}$, $\hat{\Sigma}$ is the zero operator and there is no superluminal signalling. This is because $G_\textsc{r}(  \mathsf x,\mathsf x')\braket{\hat J_\textsc{a}( \mathsf x')}$ is supported in the future lightcone of A's interaction region. Nevertheless, it is quite common in the detector literature to use smearing functions that are not compactly supported. For example,  Gaussian smearings are chosen for the sake of computational convenience and analytical results. 


Moreover, there are cases in which the use of non-compactly supported functions is not optional. When using detector models to represent the light matter interaction, the smearings are associated with the atomic wavefunctions that are generally not compactly supported, unless confined in an infinite square well. In their seminal paper on the Unruh effect \cite{PhysRevD.29.1047}, Unruh and Wald introduce the coupling of the position operator $\hat{ \bm{x}}_t$ (e.g. of an electron) to the field as 
\begin{equation}
   \hat{H}_{int}= \lambda \chi(t)\int \d \bm{x} \, \hat{\phi}(t,\bm{x}) \otimes  \delta (\bm{x}-\hat {\bm{x}}_t).
\end{equation}
The field operator is defined over the spectrum of the position operator of the non-relativistic particle. This type of interaction Hamiltonian can resemble the dipole coupling in the light-matter interaction \cite{PhysRevD.94.064074,PhysRevD.87.064038}. In this case the expectation value of the current is
\begin{equation}
    \braket{ \hat{J}_{\textsc{a}}}(\mathsf{x})=\chi_{\textsc{a}}(t)\braket{\delta (\bm{x}-\hat{\bm {x}}^\textsc{a}_t)}=\chi_\textsc{a}(t)|\psi_\textsc{a}(t, \bm{x})|^2
\end{equation}
If we plug this current into \eqref{sigma1} we see that there is non-zero signalling even if the detectors are `centered' in spacelike separation. Intuitively,  the detectors are `overlapping' even when in spacelike separation due to the quantum-mechanical `tails'. These `tails' are obscuring relativistic causality when the detectors are put in contact with the underlying relativistic QFT. This contact between the non-relativistic detector system and the relativistic quantum field is a unique feature of non-relativistic detector models. Roughly, the scales that one can define in the interface between the relativistic and non-relativistic theory (e.g., the Compton scale or the light crossing time of an atom) can be thought of as the regimes of validity of the non-relativistic models; however, the additional, system-specific scales that are introduced by the model will play a role too. This means that regime of validity of each model has to be evaluated on a case-by-case basis.

The apparent causality violations introduced by two detectors that are \emph{mostly} spacelike separated  (when the overlap of their `tails' is `small') has been analysed and quantified in \cite{PhysRevD.108.045015} from the perspective of quantum metrology. Perhaps counterintuitively, the causal `overlap' depends not only on how fast the tails decay and on the characteristics of the spacetime, but also on the internal characteristics of the detector systems (e.g., the internal frequencies $\omega_{\textsc{a},\textsc{b}}$). Nevertheless,  one can derive frequency-independent bounds to the information that B can gain for detector's A interaction with the field \cite{PhysRevD.108.045015}. Quantifying this cross-talk between distant detector systems is important in the analysis for entanglement harvesting, for which one needs to distinguish between genuine harvesting and the correlations that are established through communication \cite{PhysRevD.106.045014,  PhysRevD.104.125005}.

Overall, in the weak coupling regime and to leading order in perturbation theory, the apparent causality violations introduced by non-compact detector-field interactions seem manageable and can be argued to be outside the regime of validity the model, based on the relevant scales of each problem. Beyond perturbation theory, for compactly supported detector-field interactions, there is a non-pertutbative argument for blocking superluminal signalling based on the causal factorisation of the scattering operator of the model. We denote by $S_{\textsc{a}+\textsc{{b}}}$ the scattering operator of the total system:  i.e., the time-ordered exponential of the interaction Hamiltonian \eqref{hintAB} and $S_{\textsc{a}}$, the scattering matrix representing the interaction between detector A and the field (similarly for B). In \cite{deramon2021relativistic} it was shown that 
\begin{equation}
    S_{\textsc{a}+\textsc{{b}}} = S_{\textsc{b}}S_{\textsc{a}} \label{fact}
\end{equation}
if the two interaction regions $O_{\textsc{a},\textsc{{b}}}$ are compactly supported and causally orderable (that is, if $O_{\textsc{b}}$ does not intersect the causal past of $O_{\textsc{a}}$).  Causal factorisation also guarantees that the final state of the field after both interactions does not depend on their order (and so it does not depend on the reference frame) as long as they are spacelike separated, since in this case $ S_{\textsc{b}}S_{\textsc{a}}= S_{\textsc{a}}S_{\textsc{b}}$. Causal factorisation is sufficient for blocking superluminal signalling in bipartite scenarios but, as we will see in the next section, it will not suffice for blocking superluminal signalling in the set-up of the Sorkin-type problem.

\subsection{Impossible measurements induced by detector-field interactions}\label{rulingoutIMsec} 

In the case of three (or more) detectors coupled to the field, Sorkin-type impossible measurement problems can arise \cite{deramon2021relativistic,Benincasa_2014}. In de Ram{\'o}n, Papageorgiou, and Mart{\'i}n-Mart{\'i}nez \cite{deramon2021relativistic}, it was shown that this type of acausal behaviour persists for the most general kind of detector models that use the general Hamiltonian density in \eqref{cov} (i.e., for both compactly and non-compactly supported detector models, with the exception of the point-like model).  The extension of the detector-field interaction and the non-relativistic dynamics of the detector system are responsible for this acausal behaviour.  Nevertheless, the advantage of the detector approach is that one can quantify this causality violation in terms of the scales that are introduced by the detectors and the detector-field interactions. 

Following the demonstration in \cite{deramon2021relativistic}, consider the impossible measurement scenario for local regions $O_1$, $O_2$, and $O_3$ depicted in Fig. \ref{fig2}. A unitary `kick' is implemented over region $O_1$, possibly through the coupling of a detector to the field, which then can be disregarded. In particular, the initial state of the detectors plus field has the form $\hat\rho_{\text{initial}}=\hat U \hat\rho_{0}\hat U^{\dagger}$ where $\hat\rho_{0}$ is an arbitrary state of the joint system, and $\hat U =\mathbb{1}_\textsc{a}\otimes\mathbb{1}_\textsc{b}\otimes\hat U_{\phi}$ is an arbitrary unitary acting on the field's Hilbert space. Two detectors A,B interact with the field over the regions $O_{2,3}$, respectively. If detector A were not coupled to the field, the expectation values of observables of detector B, denoted as $\hat{D}_{\textsc{b}}$, would not depend on $U_{\phi}$ since B only interacts with the field in the causal complement of $O_1$. In the presence of detector A, the condition that B's observables are not sensitive to the local `kick' $\hat{U}$ is (for derivation see \cite{deramon2021relativistic}) 
\begin{equation}
\hat{V}^\dagger\hat D_\textsc{b}\hat{V}=\hat D_\textsc{b}, \label{condition}
\end{equation}
where
\begin{equation}
    \hat V={\hat S}_{\textsc{a}+\textsc{b}}\hat U{\hat S}^{\dagger}_{\textsc{a}+\textsc{b}}.
\end{equation}
Condition \eqref{condition} is equivalent to $[\hat D_\textsc{b},\hat{V}]=0$. Using the causal factorisation  condition \eqref{fact} (since $O_{2,3}$ are causally orderable), \eqref{condition} becomes\footnote{where we have omitted the tensor product with the identities $\mathbb{1}_{\textsc{a}}$ (in the first input) and $\mathbb{1}_{\textsc{b}}$ (in the second) to simplify the notation.}
\begin{equation}
    [{\hat S}^{\dagger}_{\textsc{b}}\hat D_\textsc{b}{\hat S}_{\textsc{b}},{\hat S}_{\textsc{a}}\hat{U}{\hat S}^{\dagger}_{\textsc{a}}]=0. \label{cnd}
\end{equation}
To make sense of \eqref{cnd} we can think of ${\hat S}^{\dagger}_{\textsc{b}}\hat D_\textsc{b}{\hat S}_{\textsc{b}}$ as an induced observable that resides on region $O_3$ and ${\hat S}_{\textsc{a}}\hat{U}{\hat S}^{\dagger}_{\textsc{a}}$ as the local 'kick' propagated through the coupling to A. 

Next we have to examine the localisation of ${\hat S}_{\textsc{a}}\hat{U}{\hat S}^{\dagger}_{\textsc{a}}$. That is, how does the coupling to detector A `propagate' the local `kick' over region $O_1$ to region $O_3$? Crucially, it turns out that the localisation of ${\hat S}_{\textsc{a}}\hat{U}{\hat S}^{\dagger}_{\textsc{a}}$ includes the forward lightcone of region $O_2$ (see Fig. \ref{fig2}) and, as a result, the expectation values for detector B in $O_3$ will depend on the local `kick'. By expanding condition \eqref{cnd} one finds that this is because $[\hat{J}_{\textsc{a}}(\mathsf x), \hat{J}_{\textsc{a}}(\mathsf x')]\neq 0$ for spacelike separated points within the extension of region $O_2$ (i.e., $\text{supp}\Lambda$). As de Ram{\'o}n et al. \cite{deramon2021relativistic} put it, ``[this result] links superluminal signalling with superluminal propagation within the device that is implementing the measurement...The physical intuition is that, when a detector is spatially extended, the information propagating inside the detector is not constrained to travel subluminally since the detector is a non-relativistic system." In Sec. \ref{detectorfieldinteraction} we argued that $\text{supp}\Lambda$ cannot be straightforwardly interpreted as the region occupied by the detector, but the main point is that if the detector current $\hat{J}_{\textsc{a}}$ were another relativistic field, and as such obeyed Microcausality, then its coupling to the field over region $O_2$ would not change the localisation of the local `kick' over $O_1$ and no observable in $O_3$ would be sensitive to the `kick'. Note that condition \eqref{cnd} is a special case of the causality condition \eqref{Borstencondition} proposed in Borsten et al. \cite{Borsten:2019cpc},\footnote{Because \eqref{cnd} is equivalent to $[{\hat S}^{\dagger}_{\textsc{a}}{\hat S}^{\dagger}_{\textsc{b}}\hat D_\textsc{b}{\hat S}_{\textsc{b}}{\hat S}_{\textsc{a}},\hat{U}]=0$, which takes the same form as the Borsten et al. condition \eqref{Borstencondition}.}, which is not satisfied in general in the detector models approach. We will return to this point when we review the detector-based measurement theory in the next subsection. 

This structural issue of using non-relativistic detector models, namely that they are defined using currents that do not obey a microcausality condition, can be tolerated by conducting a rigorous analysis of the regimes of validity of the models. That is, the severity of the causality violations in physically reasonable situations can be quantified. This is not only necessary for justifying the use of the models, but also for making sense of this abstract type of causality violations in concrete scenarios that can represent `realistic' detection experiments. de Ram{\'o}n et al. \cite{deramon2021relativistic} note that ``since for point-like detectors there is not superluminal propagation, one can disregard this kind of faster-than-light signalling for `small enough' detectors. Whether a detector is small or not will depend, of course, on the parameters of the problem."  One can also argue, in terms of the coupling strength, that in the weak coupling limit the Sorkin-type problem is of at least $\mathcal{O}(\lambda^n)$ when $n$ detectors are involved. As explained above, the signalling between any two detectors A and B is of second order, i.e., of order $\lambda_{\textsc{a}}\lambda_{\textsc{b}}$ (see Eq. \eqref{rho2}). This is because the $\lambda^2_{\textsc{a}}$ and $\lambda^2_\textsc{b}$ terms are `local' to each detector and do not allow for the detectors to `see each other'. Similarly, in the tripartite case of detectors A, B and C in the Sorkin-type configuration, the coupling constants have to be combined for detector C to `see' A through B, and so the superluminal signalling is of at least third order $\lambda_{\textsc{a}} \lambda_{\textsc{b}} \lambda_{\textsc{c}}$. In fact, in \cite{deramon2021relativistic} it was shown explicitly that, for UDW-type detectors in the tripartite scenario, the superluminal signalling is of \textit{fourth} order in perturbation theory \cite{deramon2021relativistic} while most relevant calculations are of second order in the coupling constant.

To summarize, the use of non-relativistic detector models to represent measurements of relativistic systems introduces the possibility of superluminal signalling. This phenomenon is not confined to Sorkin-type impossible measurement scenarios; for example, superluminal signalling is also possible in models for bipartite measurement scenarios. In terms of the `impossible measurement' reductio argument in Sec. \ref{reductio}, the detector models approach rejects assumption P3 that ideal measurement theory is applied directly to the field system. Instead, projective measurements (modeled by rank-1 projection operators) are only performed on detectors, and L{\"u}ders' rule is only applied as a state update rule following measurements on detectors. This strategy is combined with a pragmatic approach: a detector model may only be used when non-relativistic effects such as superluminal signalling can be shown to be negligible. That is, superluminal signalling is ruled out FAPP in the regime of applicability of a given detector model. The analysis of the magnitude of non-relativistic effects is carried out on a case-by-case basis for concrete detector models used under specified conditions. For Sorkin-type impossible measurement scenarios, this analysis pinpoints the source of superluminal signalling as being that the current associated with the detector violates Microcausality. As an example of the case-by-case analysis that rules out superluminal signalling FAPP, when the impossible measurement scenario is modeled using UDW-type detectors, the effects of superluminal signalling appear at fourth order in the perturbation series in the coupling constant, while the results that are taken to be physically significant are at second order.

\subsection{The Polo-G{\'o}mez, Garay, and Mart{\'i}n-Mart{\'i}nez detector-based measurement theory}\label{PG measurement theory sec} 


In the previous section we sketched the \textit{dynamical} understanding of how the `impossible measurements' arise in extended detector-field interactions in concrete models, without making explicit use of any measurement theory. In this section we will analyse the consequences of this for the detector-based measurement theory, by checking to what extent the detector-induced state updates satisfy the causality condition \eqref{Borstencondition} by Borsten et al. A detector-based measurement theory for QFT that specifies the state update rules for the field that are induced by projective measurements on the detectors has been developed by Polo-G{\'o}mez, Garay, and Mart{\'i}n-Mart{\'i}nez in \cite{Polo_Gomez_2022}. We will summarize this detector-based measurement theory here to set up a comparison with the measurement theory of the FV framework in Sec. \ref{comparisonsection}. 

As before, the general set up is that measurements on the field are carried out by first allowing the detector and field to interact in some region, and then measuring the detector in the causal future of this region when the detector and field are no longer coupled. Assume that the initial state of the detector-field system is a separable state represented by the density operator $\rho= \rho_d  \otimes \rho_{\phi}$. Given the interaction Hamiltonian $\hat{H}_{int}$ between the field and the detector, the evolved state is $\hat{S}_1 \rho \hat{S}^{\dagger}_1$, where $\hat{S}_1= \mathcal{T}\text{exp}\left[-i \int_{-\infty}^{t_1} \d t \hat{H}_{int}(t)\right] $ and $t_1$ is a time after which the detector-field interaction is turned off. At a later time $t_2 \geq t_1$ a projective measurement $\hat{P}(t_2)$ (denoted as $\hat{P}_2$) is applied to the detector system and the total state is updated as follows 
\begin{equation}
    \rho'= \frac{(\hat{P}_2\otimes \mathds{1} ) \hat{S}_1\rho \hat{S}_1^{\dagger}(\hat{P}_2\otimes \mathbb{1}) }{\text{tr}\left( (\hat{P}_2\otimes \mathbb{1}) \hat{S}_1\rho \hat{S}_1^{\dagger}\right)} \label{update}
\end{equation}
Note that the unitary scattering operator $\hat{S}_1$ is supported over the interaction region, while the projection operator depends only on time since the detector operators have no explicit spatial dependence. Also, the states $\rho,\rho'$ are in general not spacetime-dependent. This becomes important for the interpretation of the induced state update for the field, as we explain below.  Assume that the initial state of the detector is $\rho_d=\ket{\psi}\bra{\psi}$ and that after the interaction with the field the detector is projected by means of the rank-one projector $\ket{i}\bra{i}$ (e.g. onto the $i-$th energy eigenstate of the detector), and then trace out the detector system in \eqref{update} to get
\begin{equation}
    \rho'_{\phi}= \frac{\hat{M}_{i,\psi}\rho_{\phi}\hat{M}^{\dagger}_{i,\psi}}{\text{tr}_{\phi}\left(\rho_{\phi}\hat{M}_{i,\psi}\hat{M}^{\dagger}_{i,\psi}\right)}
\end{equation}
where 
\begin{equation}
   \hat{M}_{i,\psi}:= \langle i | \hat{S}_1| \psi \rangle.
\end{equation}


In the regime of weak coupling between the field and the detector one can use the Dyson expansion for the scattering operator $\hat{S}_1$, using $\hat{H}_{int}$ from \eqref{mu}:
\begin{equation}
   \hat{M}_{i,\psi}= \langle i|\psi \rangle \mathsf{1}+ \lambda   \hat{M}^{(1)}_{i,\psi} + \lambda^2   \hat{M}^{(2)}_{i,\psi}+ \mathcal{O}(\lambda^3) \label{Dyson}
\end{equation}
where 
\begin{align}
    &\hat{M}^{(1)}_{i,\psi}= -i \int \d t \d\bm{x} \chi(t) F(\bm{x}) \langle i| \hat{\mu}(t) | \psi \rangle \hat{\phi}(t,\bm{x}) \\
    &\hat{M}^{(2)}_{i,\psi}= - \int \d t\d t' \theta (t-t') \chi(t) \chi(t') \int \d \bm{x} \d\bm{x}' F(\bm{x}) F(\bm{x}') \langle i| \hat{\mu}(t) \hat{\mu}(t')| \psi \rangle \hat{\phi}(t,\bm{x})  \hat{\phi}(t',\bm{x}') . 
\end{align}
In the case of non-selective measurements, Polo-G{\'o}mez et al. \cite{Polo_Gomez_2022} show that
\begin{align}
    \rho^{(ns)}_{\phi} &= \sum_i \hat{M}_{i,\psi}\rho_{\phi}\hat{M}^{\dagger}_{i,\psi} \nonumber\\
    &= \text{tr}_d \left( \hat{S}_1 (\ket{\psi} \bra{\psi} \otimes \rho_{\phi}) \hat{S}^{\dagger}_1 \right) \label{ns2}.
\end{align}

We see that, by summing over all possible outcomes $i$, the updated state only depends on the dynamical coupling between the field and the detector. Polo-G{\'o}mez et al. \cite[p.4]{Polo_Gomez_2022} explain that ``this is because the projective measurement acts only on the detector once the interaction has been switched off, and it does not provide additional information since being non-selective the outcome is not known". This point is important for showing that the expectation values of observables $\hat{A}$ that are defined in spacelike separation from the detector-field interaction region do not change due to the non-selective measurement. That is,
\begin{equation}
    \text{tr}_{\phi} (\rho_{\phi} \hat{A}) = \text{tr}_{\phi} (\rho^{(ns)}_{\phi} \hat{A})
\end{equation}
since $[\hat{S}_1, \hat{A}]=0$ thanks to the fields obeying the Microcausality condition. 

Nevertheless, strictly speaking, a non-selective state update of the type \eqref{ns2} can enable `impossible measurements'. Consider the impossible measurement scenario in Fig. \ref{fig2}, with $\hat{A}\in \mathcal{A}(O_3)$ and the non-selective measurement happening over region $O_2$ (that is, $\hat{M}_{i,\psi} \in \mathcal{A}(O_2)$). Call $\rho^{(ns)}_{\phi}:=\mathcal{E}_2[\rho_{\phi}]$. Then the map $\mathcal{E}_2$ does not satisfy the condition \eqref{Borstencondition} above that was introduced by Borsten et al. \cite{Borsten:2019cpc} to exclude impossible measurements. Note that, $\mathcal{E}_2$ (as defined in \eqref{ns2}) does not correspond to an idealized projective measurement (see the generalized L\"uders' rule  \eqref{luders}). The criterion \eqref{Borstencondition} by Borsten et al still applies, since as they mention in \cite{Borsten:2019cpc} the criterion holds for any valid state update map $\mathcal{E}_2$ (any map that leads to well-defined expectation values $\langle \hat{A}\rangle$ over region $O_3$). This includes state update maps that follow from particular probe prescriptions, as in the detector-based measurement theory that we consider here, or the FV framework \cite{PhysRevD.103.025017}. In our case, since the effect of the non-selective measurement only depends on the unitary scattering map ($\hat{S}_1$ in \eqref{ns2}, ${\hat S}_{\textsc{a}}$ in what follows) the dynamical analysis of the previous section applies almost straightforwardly.

In particular, we will show that the condition \eqref{cnd} on $\hat{S}_{\textsc{a}}$ that is shown to block the Sorkin-type problem in de Ram{\'o}n et al. \cite{deramon2021relativistic} is a special case of Borsten et al.'s  condition \eqref{Borstencondition} on $\mathcal{E}_2$. 
To see this explicitly, using the notation of the previous section, we consider the following equivalent of equation \eqref{cnd}:
\begin{equation}
    [{\hat S}^{\dagger}_{\textsc{a}}{\hat S}^{\dagger}_{\textsc{b}}\hat D_\textsc{b}{\hat S}_{\textsc{b}}{\hat S}_{\textsc{a}},\hat{U}]=0 \label{recnd}
\end{equation}
where $\hat{U}$ represents the unitary `kick' in $O_1$. Taking the trace of the action of this operator on any state $\rho_{\textsc{a}} \rho_{\textsc{b}} \rho_{\phi}$ of the total system yields
\begin{align}
     &\text{tr}_{\textsc{a,b},\phi}\left([{\hat S}^{\dagger}_{\textsc{a}}{\hat S}^{\dagger}_{\textsc{b}}\hat D_\textsc{b}{\hat S}_{\textsc{b}}{\hat S}_{\textsc{a}},\hat{U}] \,\rho_{\textsc{a}}  \rho_{\textsc{b}} \rho_{\phi}\right)=0 \\
     & \text{tr}_{\textsc{a},\phi}\left( [{\hat S}^{\dagger}_{\textsc{a}}(\text{tr}_{\textsc{b}}{\hat S}^{\dagger}_{\textsc{b}}\hat D_\textsc{b}{\hat S}_{\textsc{b}}\rho_{\textsc{b}}){\hat S}_{\textsc{a}},\hat{U}] \,\rho_{\textsc{a}} \rho_{\phi}\right)=0.
\end{align}
Defining 
\begin{equation}
   \hat{\Phi}_{\textsc{b}}:=\text{tr}_{\textsc{b}}\left({\hat S}^{\dagger}_{\textsc{b}}\hat D_\textsc{b}{\hat S}_{\textsc{b}}\rho_{\textsc{b}}\right), \label{inducedobs}
\end{equation}
the induced field observable that corresponds to the measurement of the expectation value of the detector observable $ \hat{D}_{\textsc{b}} $ (see \cite{SmithAlexander2017}), we have that 
\begin{equation}
\text{tr}_{\textsc{a},\phi} \left([{\hat S}^{\dagger}_{\textsc{a}}\hat{\Phi}_{\textsc{b}}{\hat S}_{\textsc{a}},\hat{U}] \,\rho_{\textsc{a}} \rho_{\phi}\right)=0 .
\end{equation}
Performing the trace over detector A, this equation can be written as 
\begin{equation}
    \text{tr}_{\phi} \left([\mathcal{E}^d_2(\hat{\Phi}_{\textsc{b}}),\hat{U}]\, \rho_{\phi}\right)=0 \label{dual}
\end{equation}
where 
\begin{equation}
  \mathcal{E}^d_2(\hat{\Phi}_{\textsc{b}}):= \text{tr}_{\textsc{a}}\left( {\hat S}^{\dagger}_{\textsc{a}}\hat{\Phi}_{\textsc{b}}{\hat S}_{\textsc{a}} \rho_{\textsc{a}} \right)  \label{edual}
\end{equation}
is the dual non-selective map. If we demand that Eq. \eqref{dual} holds for all states of the field $\rho_{\phi}$ we get the condition  
\begin{equation}
    [\mathcal{E}^d_2(\hat{\Phi}_{\textsc{b}}),\hat{U}]=0, \label{induced}
\end{equation}
which is precisely in the form of the criterion \eqref{Borstencondition} for the induced maps and observables \eqref{edual} and \eqref{inducedobs} that we defined above. What we showed is that in this case, condition \eqref{induced} is equivalent to the dynamical condition \eqref{recnd} that we derived in the previous section, and fails to be satisfied due to the non-local dynamics of the non-relativistic detector system (i.e., the current $\hat{J}_{\textsc{a}}(\mathsf x)$ associated with the detector does not obey Microcausality). This diagnosis is directly relevant for the non-selective state update derived in the detector-based measurement theory. The violation of \eqref{induced} in this context shows that, due to the non-local dynamics, the (dual) state update map $\mathcal{E}^d_2$ does not define an observable in the causal complement of $O_1$ (where the unitary `kick' $\hat{U}$ is supported).


Now we turn to the case of the state update rule for selective measurements \eqref{update}. In contrast to non-selective measurements, expectation values of observables that are spacelike-separated from the detector-field interaction region are affected. That is,
\begin{equation}
    \text{tr}_{\phi} (\rho_{\phi} \hat{A}) \neq \text{tr}_{\phi} (\rho'_{\phi} \hat{A})
\end{equation}
For this reason, Polo-G{\'o}mez et al. argue that for selective measurements $\rho'_{\phi}$ cannot be used for calculating expectation values of observables that are causally disconnected from the causal future of the interaction region.\footnote{Polo-G{\'o}mez et al. use the term \textit{contextual} to refer to the dependence of the state update rule on whether or not the spacetime points of the fields in the $n$-point functions are in the causal future of the detector measurement region. We refrain from using this terminology because contextuality has a different meaning in quantum foundations.} The physical explanation they offer is that after the dynamical interaction between the field and the detector is switched off, the detector gets entangled with the field. In general the state of the field exhibits spacelike correlations (see the discussion in Sec. \ref{stateupdatesection}), so projecting the detector selectively destroys some of these spacelike correlations. As they put it, ``The entanglement between the detector and the field generated by their interaction thus hinders the possibility of applying the selective update outside the causal future of the detector in a way consistent with the relativistic framework of QFT" \cite[p.15]{Polo_Gomez_2022}.

As a result, Polo-G{\'o}mez et al. conclude that the selective state update rule can only be applied in regions that are causally connected to the causal future of the detector measurement region. At the same time, they argue that it is not satisfactory to represent this selective state update by the restriction of the density operator $\rho'_{\phi}$ to the forward lightcone because ``a density operator does not naturally depend on points of the spacetime manifold" \cite[p.5]{Polo_Gomez_2022}. They also point out that this restriction on $\rho'_{\phi}$ would be ambiguous for the purpose of updating the field's $n-$point functions $w(\mathsf x_1,...,\mathsf x_n):= \langle \hat{\phi}(\mathsf x_1)...\hat{\phi}(\mathsf x_n) \rangle$ in the case in which some of the spacetime points $\mathsf x_i$ are inside of and some outside of the forward lightcone of the interaction region. Hence, they shift their attention to updating all possible $n-$point functions of the field rather than the density operator representing the state of the field. This is because the $n-$point functions, in contrast to the density operator, naturally depend on the spacetime points. Also, for practical purposes, the state of the field is equivalent to the set of $n-$point functions. In the case of the detector-based measurement theory, there is an extra pragmatic motivation for this shift, which is that, in the weak coupling regime of a linearly coupled interaction Hamiltonian, the $n-$th order response of the detector depends on the field $n$-point functions (making use of the Dyson expansion of $\hat{S}$ as in \eqref{Dyson}). For example, the leading order (second order in the coupling constant) excitation probability (of a two-level system) is one-to-one with the field two-point function \cite{PhysRevD.98.105011}. For these reasons, Polo-G{\'o}mez et al. argue that the $n$-point functions, and not the density operator, should be regarded as the primary means of representing the state of the field.

As a result of these considerations, in the detector-based measurement theory the state update rule following a selective measurement is spacetime-dependent in two respects. First, whether the selective or non-selective state update rule applies to an $n$-point function for the field $\langle \hat{\phi}(\mathsf x_1)...\hat{\phi}(\mathsf x_n) \rangle$ depends on the locations of the spacetime points $\mathsf x_i$. Second, the updated $n$-point functions for the field system depend on the detector measurement region in which a selective measurement on the detector is performed. In contrast, in the FV framework the selective state update rule applies to `early' state $\omega$ in a spacetime-independent way in both respects. In particular, the scattering morphism depends on the detector-field interaction region, but is entirely independent of the detector measurement region.

The detector-based measurement theory raises the same question as the FV framework:  is state update merely epistemic in some sense? Polo-G{\'o}mez et al. contend that the detector model update rules can only be interpreted as representing a change in an observer’s state of information about the field, not as representing a physical change in the state of the field (see Sec. V and Appendix A of \cite{Polo_Gomez_2022}). Whether the detector-based measurement theory requires an epistemic interpretation in this strong sense that state update is merely an update of an observer's state of information is an interpretative issue that is beyond the scope of this paper. However, the fact that the state update rules for the field depend on the spacetime region in which the measurement on the detector is performed does make the state update rules observer-dependent and constrains their interpretation. By assumption, the detector and the field system only interact in the detector-field interaction region. The observer can choose to perform a projective measurement on the detector at any time after the detector and field cease interacting. As a result, it does not make sense to literally interpret the state update rules as representing a physical change of state that is brought about by measurement. This is an unnatural interpretation of this measurement theory because, by assumption, the field and detector are no longer interacting in the detector measurement region; therefore, according to this measurement theory, the projective measurement on the detector does not cause (bring about) a change in the physical state of the field in this region. Of course, physical changes in the state of the field in the detector measurement region could be attributed to entanglement between the field and detector or the measurement theory could be modified to include a non-trivial interaction between the detector and field in the detector measurement region; however, these moves would be counter to the main goal of modeling \textit{local} measurements using field theory. Locality in this sense is ensured by the assumption that the measurement interaction is confined to the detector-field interaction region. Therefore, the state update rules in the detector-based measurement theory cannot literally be interpreted as representing physical changes to the field system that occur in the detector measurement region.

To briefly summarize this section, the detector models approach is pragmatic. The detector is modeled using NRQM, which allows projective measurements on the detector to be represented. Constructing a concrete model of a detector coupled to a field system involves choices such as a smearing and a switching function, which may be made on pragmatic grounds. The introduction of a non-relativistic detector introduces non-relativistic effects. Superluminal signalling in Sorkin-type impossible measurement scenarios can be attributed to the fact that the current associated with the detector, that goes into the detector-field interaction Hamiltonian, does not satisfy Microcausality. However, from a pragmatic perspective, this is not problematic as long as the effect is negligible in the domain of applicability of the model. Reassurance that this is the case can be obtained on a case-by-case basis, by carrying out the calculations for a concrete model. Polo-G{\'o}mez, Garay, and Mart{\'i}n-Mart{\'i}nez have proposed a detector-based measurement theory with state update rules for the induced field observables. Borsten et al.'s condition on physical observables \eqref{Borstencondition} holds only approximately due to the violation of Microcausality by the detector current. The selective state update rule only applies to regions causally connected to the causal future of the detector measurement region in which the projective measurement on the detector is performed. This is an observer-dependent feature of the state update rules that Polo-G{\'o}mez et al. interpret this state update rule as representing an update to the observer's state of information.

\section{Discussion: Comparing the FV framework and the detector models approach}\label{comparisonsection}

The reductio argument generated by Sorkin-type impossible measurement scenarios helps to clarify the important differences between the detector models approach and the FV framework. We again emphasize that neither approach was designed to address Sorkin's problem; however, Sorkin's problem is a useful analytical tool for comparing the approaches. The main differences arise from different strategies for addressing the problem of impossible measurements, although the diagnoses of the problems are broadly similar. The detector models approach identifies as one important source of the problem the premise that ideal measurement theory (and L{\"u}ders' rule in particular) can be applied directly to relativistic quantum field systems. Furthermore, according to the detector models approach, the `impossible measurements' reductio argument leaves out many model-specific details about the detectors that are actually used to make measurements of quantum fields. Properly modeling the detectors and their interactions with the field system can provide reassurance that superluminal signalling is negligible (i.e., FAPP does not occur) in the domain of applicability of the detector models. 

The FV framework also blames the impossible measurement scenarios on the application of ideal measurement theory (including L{\"u}ders' rule) to relativistic quantum systems. Their response also involves modeling the probe as well as the field system. However, the FV framework proposes a general, abstract framework for representing both the probe and the field using AQFT that is not model-specific. As a result, additional principles for AQFT need to be posited. In particular, the Time-Slice Property axiom needs to be added to the premises of the `impossible measurements' reductio argument in order to block impossible measurement scenarios. Furthermore, a new measurement theory for AQFT is needed in this approach. In contrast to the detector models approach, the application of L{\"u}ders' rule to represent a projective measurement of the detector is not taken as a starting point for determining the measurement theory for the field system. Fewster and Verch instead start with a representation of measurement based on the axioms of AQFT and the scattering isomorphism and follow the strategy (but not the physical interpretation) of QMT to derive a new measurement theory for AQFT. Both the detector models approach and the FV framework end up with new state update rules for field systems, but their methods for deriving them are different.   

It would be easy to focus only on the differences between the detector models and FV framework, but there are some general similarities between the approaches that are worth drawing attention to because they shed light on the form taken by measurement theory in QFT and fundamental features of QFT itself. These two approaches are very different in spirit, so it seems plausible that points of agreement between them could reveal genuine features that an ultimate, complete theory of relativistic quantum fields that includes a measurement theory will have. We take it that neither the detector models approach nor the FV framework is the final account of local measurement for QFT because each satisfies some desiderata and not others, as we explain below. From this perspective, the detector models and FV framework are not inherently incompatible approaches. To facilitate a direct comparison, our discussion in this section will focus on the Fewster and Verch framework with the Polo-G{\'o}mez, Garay, and Mart{\'i}n-Mart{\'i}nez detector-based measurement theory discussed in Sec. \ref{PG measurement theory sec}.

\subsection{Similarities}\label{similarities section}

\begin{itemize}

\item \textbf{The role of dynamics}:  Regarding the response to the `impossible measurement' problem, the FV framework and detector models approaches agree in general terms on the two main problems with the reductio argument in Sec. \ref{reductio}:  the premise that L{\"u}ders' rule applies to relativistic systems (P3(b)) must be rejected and additional assumptions about the dynamics of relativistic systems undergoing measurement must be introduced. In the FV framework, L{\"u}ders' rule is abandoned entirely and new state update rules for relativistic QFT are derived. The detector models approach also refrains from applying L{\"u}ders' rule directly to relativistic quantum fields, though it is applied to non-relativistic detectors. On the dynamical side, it is the Local Time-Slice Property that is crucial for blocking Sorkin-type impossible measurement scenarios in the FV framework. The Local Time-Slice Property is a general principle that ensures that the quantum fields propagate subluminally and deterministically. In the detector models approach, careful attention to how the dynamics of the relativistic field and its coupling to the detectors is modeled is crucial for addressing the `impossible measurements' reductio. In this case, the non-relativistic coupling of the detector to the field via currents that do not satisfy microcausality is the source of superluminal signalling in impossible measurement scenarios. This problem is addressed on a case-by-case basis by performing calculations that involve the interaction Hamiltonian to assess the magnitude of the non-relativistic effects. From this perspective, one thing that goes wrong in Sorkin's argument is that these model-specific dynamical details are not properly taken into account.     ;'

    \item \textbf{Localization regions for observables}:  Abandonment of the \textit{prima facie} operational interpretation of a local algebra of observables $\mathcal{A}(O)$ as representing operations that it is possible to carry out \textit{in region $O$}. In both the detector models approach and the FV framework, this change in interpretation is made possible by the introduction of detectors or probes. In practical approaches to applying measurement in relativistic quantum theory such as Sorkin's, the traditional interpretation of a smeared field operator is that the smearing reflects the spacetime region over which the operation represented by the field operator is performed. In the detector models approach, the explicit representation of the detector that is coupled to the field system complicates the interpretation of the smearing function. As discussed in Sec. \ref{detectorfieldinteraction}, the role of the smearing function is ultimately pragmatic; therefore, as long as the predictions are not affected, there is no reason to choose a smearing that is supported only in the interaction region between the detector and field system (e.g., may choose a Gaussian). Furthermore, as was observed in \cite{McKay_2017} and is further supported by considering the covariant Hamiltonian density representing the interaction between the detector and field system, the most natural interpretation of the spacetime smearing function is that it is a holistic property of the detector-field interaction rather than a property of either one by itself. In the FV framework, the system observables can similarly be associated with different localization regions. As a consequence of (Local Time-Slice Property), an algebra of observables $\mathcal{A}(O)$ can be localized in any region in the domain of dependence of $O$. This is a change from the traditional operational interpretation of an algebra of observables $\mathcal{A}(O)$ in AQFT as representing a set of operations that it is possible to perform in region $O$. (See \cite[p.35]{Fewster2023}). 
    
    \item \textbf{States are primarily represented using expectation values of fields at different times}: In AQFT, an algebraic state $\omega$ is a positive, normalized, linear functional from $\mathcal{A}(O)$ (or $\mathcal{A}(M)$) to $\mathbb{C}$. For self-adjoint $A \in \mathcal{A}(O)$, $\omega(A)$ represents the expectation value of $A$ in state $\omega$. Algebraic states are the primary representatives of the physical states and they represent expectation values. As Polo-G{\'o}mez et al. \cite[p.7]{Polo_Gomez_2022} point out, their measurement theory for the detector approach similarly treats expectation values as the primary means of representing the state of the field and density operators as secondary. More precisely, the central quantities are the $n$-point functions. Their main rationale for shifting to the $n$-point functions is that the state update rules are spacetime-dependent and the density operators are not. Therefore, the $n$-point functions must be the primary vehicle for representing the state.

    In both the FV framework and the detector-based measurement theory, the representation of local measurements involves expectation values of fields at different times. In the FV framework, the state $\varpi$ on $\mathcal{C}(M)$ for the coupled probe and field system is a global state in the sense that it encodes the expectation values of the field over all local regions. The scattering isomorphism facilitates the representation of `in' and `out' algebraic states (and hence expectation values) in regions $M^-$ and $M^+$. In the detector-based measurement framework, the $n$-point functions directly involve fields at different times. As we noted in Sec. \ref{introduction}, this shift away from the instantaneous states that play a central role in NRQM has longstanding historical roots in QFT in lines of theoretical development that led to the formulation of scattering theory (again, see \cite{blum_state_2017}). Here we see the same theme emerging in the treatment of local measurements. As we will discuss in Sec. \ref{consistenthistories}, the problems raised by instantaneous states in QFT are also an important motivation for histories-based approaches to relativistic quantum theory, including the Quantum Temporal Probabilities program \cite{Anastopoulos_2021}.    
    
    \item \textbf{State update rules for relativistic field systems cannot be literally interpreted as representing a physical change of state that occurs in some spacetime region}: Our discussion of the FV framework emphasized this point that (as Fewster puts it) ``no geometric boundaries across which the state reduction occurs are needed" \cite[p.10]{Fewster2019generallycovariantscheme}. In the FV framework, this is a consequence of two features:  the counterfactual interpretation of the `in’ and `out’ algebraic states over the uncoupled algebra and Corollary 6, which establishes that when successive selective measurements are performed they can either be evaluated jointly or sequentially. As a result, the state update rule in the FV framework cannot be literally interpreted as representing a physical change of state that happens in any spacetime region. The detector-based measurement theory similarly posits state update rules that do not admit a literal interpretation in terms of a physical change of state that occurs in the region in which the state update rules are applied. However, the arguments for this conclusion differ from those for the FV framework. Polo-G{\'o}mez et al. argue that the detector-based state update rules need to be interpreted as an update of the observer's state of information about the field system, and not as a change in the observer-independent state of the field \cite[p.5]{Polo_Gomez_2022}. Even if one does not go so far as endorsing an epistemic interpretation in this strong sense, the fact that the state update \textit{for the field} depends on the spacetime location in which the selective measurement \textit{on the detector} is performed is not compatible with interpreting the state update rule as representing a physical change of state of the field that is brought about by measurement. 
    
    \item \textbf{Nobody solves the Measurement Problem}: Proponents of the FV framework and the Polo-G{\'o}mez detector-based measurement theory explicitly agree that their respective measurement theories for QFT do not solve the Measurement Problem that arises in NRQM. The Measurement Problem is an issue that arises after a theory for (non-relativistic or relativistic) quantum systems and an accompanying measurement theory is formulated. Proponents of both the detector models approach and the FV framework agree that they are engaged in the preliminary task of formulating a physical theory plus a compatible measurement theory for relativistic systems. 

\end{itemize}

\subsection{Differences}

\begin{itemize}
    \item \textbf{Pragmatic vs principled approaches have different goals}: As we have stressed, the detector models approach is pragmatic in spirit, while the FV framework adopts a more principled approach. The adoption of different approaches means that the detector models approach and the FV framework prioritize different goals. A central goal of the detector models framework is to construct models that adequately describe realistic detectors, including detectors that can actually be built in a lab. In contrast, the FV framework has the primary goal of supplying a framework for measurement in QFT that is generally applicable. Many pragmatic choices are made in the course of constructing a model for a particular detector, including the use of NRQM to model the detector, the smearing functions and field-detector couplings, and the acceptance of FAPP arguments ruling out impossible measurements. In contrast, the FV framework focuses on formulating a fully relativistic measurement theory based on the general physical principles of AQFT in which impossible measurement scenarios cannot arise at all.
    
    \item \textbf{Different scopes of applicability}: The detector models approach and the FV framework have different scopes of applicability. While the FV framework aspires to provide an entirely general account of measurement in QFT, the framework introduces simplifying assumptions that exclude some physically realistic detectors. For example, the FV framework assumes that the region $K$ in which the probe and system interact is compact.\footnote{In this recent paper \cite{Fewster2023} the FV measurement framework is extended to non-compact probe-system coupling regions for the case of a quantized real linear scalar field.} This assumption does not apply to all of the models of detectors described in Sec. \ref{detectormodelssec}. The detector models approach is not a special case of the FV framework. Its physical assumptions are more flexible than the FV framework in virtue of its case-by-case approach. The FV framework also faces the limitation that, in order to apply it to a situation, field and detector models that satisfy the axioms of AQFT need to be constructed. However, it should not be assumed that the detector model approach is always the one that is more useful for practical applications, for example, when probes are not modeled as two-level systems. The FV framework may be more appropriate for representing field-field couplings and how information is transferred from one field to another (e.g. see \cite{PhysRevLett.125.213603}). 
    
    There is interesting work to be done investigating the relationship between the two approaches, particularly in areas of potential overlap. One of the initial motivations of the FV framework was to make contact with the traditional perturbative analysis of the Unruh-DeWitt detector (see Section 5.3. in \cite{fewster2020quantum}). There has also been a recent effort to introduce relativistic corrections to the Unruh-DeWitt model by introducing second-quantized versions of the model \cite{PhysRevD.105.125001,PhysRevD.107.056023,  perche2023particle}. Interestingly, in the seminal paper by Unruh \cite{PhysRevD.14.870} one already finds a version of the model that involves a field-field coupling.
   
    \item \textbf{Derived vs. posited status of state update rules}: In the FV framework, the state update rules are derived from the axioms of AQFT and the properties of the scattering morphism. The derivation of the state update rules is further constrained by executing the steps in parallel with the standard derivation in Quantum Measurement Theory. A measurement scheme is prescribed and CP-instruments are used to derive both the selective and non-selective state update rules. In contrast, the state update rules in the Polo-G{\'o}mez et al. detector-based measurement theory are based on plausible assumptions about how to model the detector, the field, and their interactions and plausible arguments based on relativity theory. As Polo-G{\'o}mez et al. \cite[p.6]{Polo_Gomez_2022} put it, ``[t]his update rule respects causality by fiat." 
    
    \item \textbf{Application of state update rules is spacetime-dependent vs. spacetime-independent}: In the Polo-G{\'o}mez et al. detector-based measurement theory, the state update rules are spacetime-dependent in two respects. First, whether the $n$-point function is updated according to the selective state update rule or the non-selective state update rule is dependent on the locations of the spacetime points in the $n$-point functions. Second, the state update rules are also dependent on the region in which the projective measurement on the detector is performed. In contrast, the FV framework is not spacetime-dependent in either of these respects. In the FV framework, the selective state update rule yields an updated algebraic state that applies to all of spacetime. As discussed above, this state should be interpreted as a counterfactual state, but it is still the case that the selective state update yields a single, global state. Furthermore, the state update rules in the FV framework are independent of the spacetime region in which the probe is measured. The instrument is defined using the scattering morphism, which is spacetime-dependent on the field-probe coupling region only.
    
    \end{itemize}

\subsection{Further Comparison}\label{FurtherCompSec} 

While both the physical theory of NRQM and its accompanying measurement theory are well-established parts of physics, both the formulation of the physical theory of QFT and the formulation of its accompanying measurement theory are works in progress. The flurry of recent papers on how to represent measurements on relativistic quantum systems is one indication of this. Both the FV framework and the detector models approach are part of this larger ongoing research program. In our view, the FV framework and the detector models approach should be viewed as complementary projects rather than rivals. As we have emphasized, they adopt different strategies and have different goals. In their present incarnations, they also have different scopes of applicability. Applying the two approaches to model the same measurement scenario can also lead to apparently different predictions. For example, entanglement harvesting from the vacuum by spacelike-separated local detectors is investigated in \cite{ruep_weakly_2021,grimmer_measurements_2021}. Ruep \cite{ruep_weakly_2021} applies the FV measurement framework and concludes that weakly coupled detectors cannot harvest entanglement due to noise associated with the inescapable mixedness of the states of the local probe. In response, Grimmer et al. \cite{grimmer_measurements_2021} argue that when a concrete detector model is applied that includes an assumption about the scale of the detector, this effect is negligible and there is no obstacle to weakly coupled detectors harvesting entanglement. As in other cases in science, different models can be used to give complementary descriptions of the same situation. Discrepancies in predictions can be useful for investigating the conditions under which effects occur and the relationships among our models. 

While we view the FV framework and detector models approach as complementary, it should be noted that there are differences of opinion on this issue in the literature. One point of disagreement concerns whether it is possible \textit{in principle} to model detectors using QFT, as the FV framework sets out to do. Here is an example of an assumption that is made to set up the FV framework:
    
\begin{quote}
    
We also do not claim to solve the Measurement Problem of quantum theory. Rather, we take it for granted that the experimenter has some means of preparing, controlling and measuring the probe and sufficiently separating it from the QFT of interest – which we will call the ‘system’ – the question is what measurements of the probe tell us about the system. That is, our interest is in describing a link in the measurement chain, in a covariant spacetime context. \cite[p.3]{fewster2020quantum}

\end{quote}
    
\noindent This assumption that measurement theory describes ``a link in the measurement chain" is a standard one within Quantum Measurement Theory applied to NRQM (see, for example, \cite[p.225]{busch_book_2016}), but its use in QFT is criticized by Grimmer, Torres, and Mart{\'i}n-Mart{\'i}nez \cite[p.5]{grimmer_measurements_2021} and Grimmer \cite{Grimmer2023}, who argue that the assumption that ``someone, somewhere, knows how to measure something" is not warranted when the probe is taken to be a relativistic quantum field. For practical reasons, the detector models approach is the only one that we can use right now to model some realistic detectors. Measurement apparatuses are low energy systems that would need to be described using interacting QFTs with bound states \cite[p.1]{Polo_Gomez_2022}. However, one might expect (as Fewster and Verch do) that eventually a treatment of systems and some types of detectors will be possible within QFT. Grimmer et al. \cite{grimmer_measurements_2021} make the strong claim that detector models formulated using NRQM are needed in principle to model measurements of QFT systems \footnote{A related point of disagreement between proponents of the detector models approach and the FV framework is whether finite-rank projectors are needed to represent measurements. If so (as proponents of the detector models approach believe), then such objects are not available in the Type III von Neumann algebras that are typical in QFT, which would be problematic for the FV framework (see Sec. \ref{stateupdatesection} and \ref{nrdetectorsec}). }. Grimmer \cite{Grimmer2023} presents a stronger argument that, in order to acquire empirical significance, QFT observables must be  appropriately related to a non-QFT model, with the Unruh-DeWitt detector model being a paradigm example. Furthermore, Grimmer believes that these models of measurement must be constructed on a case-by-case basis and is skeptical about the viability of a general measurement theory for QFT such as the FV measurement framework or even the Pol{\'o}-Gomez et al. detector-based measurement theory.

\section{Another approach: Histories-inspired responses to Sorkin-type impossible measurement scenarios}\label{consistenthistories}


It is important to point out that Sorkin's motivation for formulating the `impossible measurements' issue is to advocate for the sum-over-histories approach to quantum theory. As he puts it in the abstract of \cite{sorkin1993impossible}, ``It is argued that this problem leaves the Hilbert space formulation of quantum field theory with no definite measurement theory, removing whatever advantages it may have seemed to possess vis a vis the sum-over-histories approach, and reinforcing the view that a sum-over-histories framework is the most promising one for quantum gravity." {Histories-inspired approaches do not set out state update rules; they instead assign probabilities directly to histories. In this section, we will comment on the form that the `impossible measurements' problem takes in histories-based formalisms. To the best of our knowledge, one cannot find a complete response in the histories literature, even though the problem is clearly articulated in older \cite{isham1994quantum} and more recent \cite{https://doi.org/10.48550/arxiv.2109.03187,albertini2023ideal} literature. Different tools that are being used in different variants of histories-based approaches, can have consequences for causality. In this section, we focus on the role of the so-called consistency condition in the consistent histories approach.\footnote{In the sum-over-histories approach, the `Persistence of Zero' condition was recently introduced in \cite{dowker2023intrinsic} as a causality principle.}


As there are many variants of the histories-based approaches, first we will sketch the main tools and ideas that are common in the histories literature. Roughly, the possible `histories' of a given system are all the possible time-extended propositions that one can assign to the system; that is, all possible `paths' of the system in the (underlying classical) sampling space. The goal of the formalism is to assign probabilities to all possible paths. These probabilities are `quantum' in the sense that in general they `overlap' and they are not guaranteed to satisfy the usual additivity conditions. Histories-based formalisms can be viewed as generalisations of the path-integral \cite{Hartle1993SpacetimeQM,Omnes:1988ek}. As histories-based approaches have mostly been applied to quantum cosmology \cite{https://doi.org/10.48550/arxiv.1803.04605}, the formalism typically refers to histories of a closed system (e.g. the universe), which is why the notion of agency, or `external' measurement, is not part of the formalism. This is one of the conceptual advantages that is highlighted by Sorkin for using histories-based approaches to resolve the impossible measurements issue, or for re-evaluating the measurement problem in a spacetime context. In Sorkin's words, ``With the formal notion of measurement compromised as it seems to be already in quantum field theory, the greatest advantage of the sum-over-histories may be that it does not employ measurement as a basic concept. Instead it operates with the idea of a partition (or `coarse-graining') of the set of all histories, and assigns probabilities directly to the members of a given partition, using what I would call the quantum replacement for the classical probability calculus" \cite[p.11]{sorkin1993impossible}.

 
 In non-relativistic quantum theory, a history of a quantum system is simply a chain of time-ordered single-time propositions that are represented by projection operators (in the Heisenberg picture) and are typically associated with possible values of the observables of the system. Such a chain $\alpha=(\alpha_1, \alpha_2, ..., \alpha_n)$ is represented by a class operator
\begin{equation}
    \hat{C}_{\alpha}= \hat{P}_{\alpha_1}(t_1)\hat{P}_{\alpha_2}(t_2)...\hat{P}_{\alpha_n}(t_n) \label{class}
\end{equation}
where the $\alpha_i$'s are indexing the eigenvalues of some observable (or more generally the closed subspaces of the Hilbert space). The set of time-points $t_1,t_2,...,t_n$ is called the history's temporal support \cite{isham1994quantum}.

It is worth emphasizing that the role of the projectors in \eqref{class} is propositional, i.e., encoding the possible propositions that can be attributed to a closed system. It would be natural to think of a history as a sequence of actual events, e.g. measurement outcomes, but this interpretation is not straightforward as the definition \eqref{class} has nothing to do with measurements a priori. The use of projectors here is similar in spirit to how they are used in quantum logic, where one can define `or' and `and' operations for combining the propositions that can be assigned to a physical system \cite{isham1994quantum}. The joint probability of $\alpha$ (the sequence of propositions $\alpha_1$ at $t_1$, ..., $\alpha_n$ at $t_n$) is given by
\begin{equation}
    p(\alpha)= \text{tr}\left( \hat{C}^{\dagger}_{\alpha} \rho_0 \hat{C}_{\alpha}\right) \label{prob}
\end{equation}
where $\rho_0$ the state of the system (in the Heisenberg picture). In general, these probabilities do not satisfy the Kolmogorov additivity condition. That is, if $\alpha, \beta$ are exclusive histories and $\alpha \vee \beta$ denotes their conjunction, then it does not hold in general that 
\begin{equation}
    p(\alpha \vee \beta)=p(\alpha)+p(\beta). \label{add}
\end{equation}
If we define the decoherence functional to be the following complex-valued functional of two histories
\begin{equation}
     d(\alpha, \beta) := \text{tr}\left( \hat{C}^{\dagger}_{\alpha} \rho_0 \hat{C}_{\beta}\right),
\end{equation}
then $  p(\alpha)=d(\alpha,\alpha)$. 

In the consistent histories approach, one enforces the additivity condition \eqref{add} by demanding that 
\begin{equation}
 \text{Re} \,d(\alpha, \beta)=0,  
\end{equation}
which is called the \textit{consistency condition} \cite{dowker1996consistent}. In the context of standard path-integral approaches the decoherence functional can be calculated as a double path-integral \cite{dowker2010hilbert}. 
In a sense, the decoherence functional quantifies the `overlap' of two histories and the most obvious way of satisfying the consistency condition is to demand that the `overlap' vanishes: i.e.,
\begin{equation}
   \text{tr}\left( \hat{C}^{\dagger}_{\alpha} \rho_0 \hat{C}_{\beta}\right)=0 \label{consistent}
\end{equation}
for two exclusive histories $\alpha, \beta$. The definition of exclusiveness is that for at least one time-step $t_i$ the two corresponding projectors are orthogonal, i.e., $\hat{P}_{\alpha_i}(t_i)\hat{P}_{\beta_i}(t_i)=0$.  In histories-based approaches the decoherence functional (rather than the state) can be considered the primary object from which the probabilistic predictions of the theory are extracted \cite{dowker2010hilbert}. This approach gives more general probability rules (than the Born rule) and the issue of state update does not come up since the formalism operates directly at the level of whole histories \cite{Omnes:1988ek}. In a relativistic setting, one is not looking for state-updates that can be consistently applied `step-by-step' without leading to causality violations, but rather the question is: which histories of the system can be assigned a probability that does not entail superluminal influences between the history's `nodes'? In what follows, we will elaborate on this question.

Adjusting the history-based formalism to quantum field theory is not trivial, mostly because in a relativistic set-up there is no fixed time-ordering of events. The definition of the history's temporal support as a set of time-points $\{t_1, t_2,...,t_n\}$ is too restrictive in this case. Even in globally hyperbolic spacetimes there are many foliations that one can choose, not to mention the challenges in general non-hyperbolic spacetimes \cite{isham1994quantum}. Also, in a field theory it is unnatural to associate events with spacetime points, and one would like to consider localised propositions associated with spacetime regions. Isham \cite{isham1994quantum} proposed that the `basic' regions that one can consider are connected (causally convex) regions with compact closure. A temporal support of a history is then a collection of basic spacetime regions $\{ O_1, O_2, ..., O_n\}$ such that their closures are disjoint and for every pair $O_i,O_j$ it holds that $O_i \prec O_j$, or $O_j \prec O_i$, or $O_i$ and $O_j$ are spacelike separated (where $\prec$ is the same ordering relation that was assumed by Sorkin and was defined above as premise P3(c) of the reductio argument in Sec. \ref{reductio}).\footnote{Recall that $O_j \prec O_k$ iff some point of $O_j$ causally precedes some point of $O_k$ in the spacetime. Take the transitive closure of $\prec$. Regions must be chosen such that this extended $\prec$ is a partial order (i.e., cannot have both $O_j \prec O_k$ and $O_k \prec O_j$).}

One would like to associate quantum field propositions with these local regions. This is challenging for various reasons. Questions that appeal to the Fock space structure of the Hilbert space, like `is the field in the ground state?' cannot be asked locally (but can, for example, on a hypersurface, as in the `impossible measurement' example due to Sorkin described in Sec. \ref{sorkinsec}). Due to the Type III nature of the local algebras there are no finite rank projectors, and the `yes-no'-type questions familiar from quantum logic cannot be asked locally \cite{haag2012local,Redhead1995}. Isham \cite{isham1994quantum} suggests that the elementary propositions of the field theory over a spacetime region $O$ should be of the form $P(f)$ that corresponds to the possible values of the smeared field operator $\phi(f)$ where $\text{supp}(f) \subseteq O$, and he argues that one would have to specify the way in which a local lattice of propositions $\mathcal{L}_{O}$ associated to region $O$ is generated from these basic propositions. In the recent work \cite{https://doi.org/10.48550/arxiv.2109.03187}, a map $O\rightarrow E_O$ from regions to projective effect valued measures is assumed, but it is not clear in general which statement about the quantum field the effect valued measures correspond to. These difficulties with extending the notion of local propositions to a QFT set up are also related to the difficulties in applying the modal interpretation to QFT \cite{CLIFTON2000167, Earman2005-JOHRIA}.

When considering how to structure basic regions and propositions in QFT so as to solve the `impossible measurements' problem, Sorkin presents the following dilemma: one can either further restrict the allowed-measurement regions and the corresponding ordering relation, or else select the allowed observables on ``some more ad hoc basis." In his words, the problem is ``foreshadowed by our need to take a transitive closure in defining $\prec$" (see discussion in section \ref{sorkinsec}) and as a result one could ``further restrict the allowed measurement regions $O_j$ in such a manner that the transitive closure we took in defining $\prec$ would be redundant. For example, we could require that for each pair of
regions $O_j,O_k$ all pairs of points $x \in O_j$ and $y \in O_k$ be related in the same way" \cite[p.9]{sorkin1993impossible}. Of course, this would block the Sorkin problem by excluding the configuration of regions in figures \ref{fig1} and \ref{fig2}. This further restriction of $\prec$ would imply that one can only consider temporal supports that consist of spacetime regions that are pairwise related like two `thickened' spacetime points (fully spacelike, fully timelike, or fully lightlike), blocking the possibility that a region partially invades the forward lightcone of another region.    This is a global restriction, perhaps an `all-at-once' constraint as in \cite{adlam2022laws}, that is hard to reconcile with the local perspective. As Sorkin puts it ``it is difficult to see how the ability to perform a measurement in a given region—or the effect of that measurement on future probabilities—could be sensitive to whether some other measurement was located totally to its past, or only partly to its past and partly spacelike to it" \cite{sorkin1993impossible}. It is also noteworthy that the initial definition of $\prec$ already excludes cases that might be of physical interest, such as overlapping regions, which were considered by Bohr and Rosenfeld in \cite{bohr1933frage}, and regions that intersect the causal past of each other, which are relevant for the study of possible spacetime embeddings of general process matrices \cite{https://doi.org/10.48550/arxiv.2203.11245}. In general, by restricting $\prec$ one might exclude physically interesting cases. Finally, regarding the second possibility of restricting the allowed observables, Sorkin points out that the inability of two coupled subsystems to signal through the measurement of an observable that is additive suggests that one could still allow integrals of spatially smeared observables over a spatial subset of a hypersurface (see the discussion in Sec. \ref{Secreductiostrategies}, also \cite{albertini2023ideal}). However, \textit{spacetime} smeared fields do not possess this additive character due to the time-extension (for a treatment of time-extended propositions in history-based approaches see, e.g., \cite{anastopoulos2007quantum}).


Overall, it is not clear how the histories machinery can be used to eliminate the impossible measurements, but it offers some tools that can be used for salvaging the framework. One technical tool that is usually not emphasized in the `standard' single-time formalism, is the consistency condition \eqref{consistent} that gives rise to well-defined multi-time probabilities. In \cite{https://doi.org/10.48550/arxiv.2109.03187}, Fuksa argues that the consistency condition can be used to characterise the causal behaviour of the probabilistic predictions of the formalism. To briefly demonstrate the point, consider two propositions $P_{\alpha_1}$ associated to a region $O_1$ and $\hat{P}_{\alpha_2}$ associated to a region $O_2$, and say that $P_{\alpha_1}$ corresponds to some observable $\hat{A}_1$ and $\hat{P}_{\alpha_2}$ corresponds to some observable $\hat{A}_2$. The observers associated to the two regions cannot signal if the non-selective measurement of $\hat{A}_1$ does not affect the statistics in $O_2$, that is, if 
\begin{equation}
 p(\alpha_2| A_1)= p(\alpha_2)   \label{outcomes}
\end{equation}
This holds if $[\hat{P}_{\alpha_1}, \hat{P}_{\alpha_2}]=0$.\footnote{The more common no-signalling condition based on the Microcausality condition states that the expectation value of $\hat{A}_2$ (and not the probability distribution over the outcomes per se) does not depend on the non-selective measurement of $\hat{A}_1$ if $[\hat{A}_1,\hat{A}_2]=0$ (see e.g. \cite{earman2014relativistic}). Equation \eqref{outcomes} is a stronger condition since $[\hat{A}_1,\hat{A}_2]=0$ does not always imply $[\hat{P}_{\alpha_1}, \hat{P}_{\alpha_2}]=0$.}. The observation is that, even if the two projectors do not commute, \eqref{outcomes} holds thanks to the consistency condition (with respect to $\alpha_1$). To see this (following \cite{https://doi.org/10.48550/arxiv.2109.03187}), from \eqref{prob} we have that 
\begin{equation}
   p(\alpha_1,\alpha_2)= \langle  \hat{P}_{\alpha_1} \hat{P}_{\alpha_2} \hat{P}_{\alpha_2} \hat{P}_{\alpha_1}  \rangle 
\end{equation}
Assuming the consistency condition \eqref{consistent} with respect to $\alpha_1$, that is
\begin{equation}
   \langle  \hat{P}_{\alpha'_1} \hat{P}_{\alpha_2} \hat{P}_{\alpha_2} \hat{P}_{\alpha_1} \rangle=0 \,\,\, \text{for} \,\,\,\alpha'_1 \neq \alpha_1,
\end{equation}
 we get that 
\begin{align}
    p(\alpha_2|A_1)&= \sum_{\alpha_1} \langle \hat{P}_{\alpha_1} \hat{P}_{\alpha_2}\hat{P}_{\alpha_2} \hat{P}_{\alpha_1} \rangle \nonumber\\
    &= \sum_{\alpha_1, \alpha'_1} \langle \hat{P}_{\alpha'_1} \hat{P}_{\alpha_2}\hat{P}_{\alpha_2} \hat{P}_{\alpha_1}  \rangle \nonumber\\
    &= \sum_{\alpha_1} \langle  \hat{P}_{\alpha_2} \hat{P}_{\alpha_2} \hat{P}_{\alpha_1}  \rangle \nonumber\\
    & = \sum_{\alpha_1} \langle  \hat{P}_{\alpha_2}^2  \rangle \nonumber \\
    &= p(\alpha_2).
\end{align}
This means that, thanks to the consistency condition with respect to $\alpha_1$, a non-selective measurement of $\hat{A}_1$ in $O_1$ does not affect the statistics of $\hat{A}_2$ in $O_2$ even if $O_1,O_2$ are not fully spacelike separated.  Similarly, in \cite{https://doi.org/10.48550/arxiv.2109.03187} Fuksa considers the three-step history that corresponds to the Sorkin-type problem, and is represented by the class operator $\hat{C}_{(\alpha_1,\alpha_2,\alpha_3)}= \hat{P}_{\alpha_3} \hat{P}_{\alpha_2} \hat{P}_{\alpha_1}$ where $\hat{P}_{\alpha_1}$ commutes with 
$\hat{P}_{\alpha_3}$, but $\hat{P}_{\alpha_2}$ does not commute with either $\hat{P}_{\alpha_1}$ or $\hat{P}_{\alpha_2}$ ($\hat{P}_{\alpha_3}$ is associated with region $O_3$ that is spacelike separated from $O_1$, and $O_2$ is not spacelike separated from either $O_1$ or $O_3$, see figure \ref{fig2}). Now, since region $O_1$ is not spacelike separated from the union of $O_2$ and $O_3$, the consistency condition with respect to $\alpha_1$ gives that  
\begin{equation}
   p(\alpha_2,\alpha_3| A_1)= p(\alpha_2, \alpha_3) \label{threestepprob}
\end{equation}
which means that a non-selective measurement of $\hat{A}_1$ at $O_1$ does not affect the joint statistics of $\alpha_2,\alpha_3$ for \textit{any} state $\psi$ \footnote{Note that demanding \eqref{threestepprob} for every state to end up with the algebraic relation \eqref{cond} for the class operator is a non-trivial move since, in general, whether the consistency condition is satisfied depends on the state.}if the class operator satisfies 
\begin{equation}
   \hat{C}^{\dagger}_{(\alpha'_1,\alpha_2,\alpha_3)}  \hat{C}_{(\alpha_1,\alpha_2,\alpha_3)} =0   \,\,\,\text{for} \,\,\, \alpha_1 \neq \alpha'_1. \label{cond}
\end{equation}
 Fuksa \cite{https://doi.org/10.48550/arxiv.2109.03187} points out that if we were to `squeeze' another intermediate region $O_2'$ between $O_1$ and $O_3$ (partially invading their future and past lightcone respectively) the condition that blocks `impossible measurements' becomes
\begin{equation}
   \hat{C}^{\dagger}_{(\alpha'_1,\alpha_2,\alpha'_2,\alpha_3)}  \hat{C}_{(\alpha_1,\alpha_2,\alpha'_2,\alpha_3)} =0 \,\,\,\text{for} \,\,\, \alpha_1 \neq \alpha'_1, 
\end{equation}
which is a stricter condition than \eqref{cond}. By introducing more intermediate regions, more conditions are added to the list of conditions that must be satisfied for blocking the Sorkin-type problem, and there is no obvious way in which they are redundant (reduce to one another). So it is not obvious how to block `impossible measurements' in general, but this analysis offers a recipe of `who can signal to whom' given a particular set of regions and propositions. In \cite{https://doi.org/10.48550/arxiv.2109.03187} they point out that given a chain of regions $O_i$, a region $O_j$ cannot signal to a `subsequent' region $O_k$ as long as the class operator `decoheres' (satisfies the consistency condition) with respect to the $j$-th variable, and that propositions before $j$ and after $k$ do not contribute to the signalling between $j$ and $k$.

Conditions of the type \eqref{cond} are ad hoc and global in nature, i.e., depend on the whole set of regions and propositions that one is considering so, as we said above, they are hard to motivate from the local perspective of observers embedded in spacetime. One way to motivate the consistency condition is through the introduction of decohering pointer variables that are locally coupled to the field \cite{Hu1993}.  Roughly, the idea is that sufficiently decohered propositions (or measurements) of the pointer variables are correlated with quantum field histories that satisfy the consistency condition and would not lead to (significant) causality violation. Nevertheless, there is a strong dependence of the pointer variable multi-time probabilities on the chosen measurement scheme, i.e., the chosen measurement resolution \cite{Anastopoulos2006}.

A histories-based formalism that explicitly treats QFT measurement through the introduction of coarse-grained pointer variables is the Quantum Temporal Probabilities (QTP) formalism \cite{PhysRevA.86.012111, ANASTOPOULOS2023169239}. Joint probabilities of the pointer variables are defined by means of unequal-time correlation functions, and the consistency condition is satisfied for a certain degree of coarse-graining (see also \cite{Anastopoulos2006}). A connection of this formalism to the closed-time-path (CTP) integral was recently established in \cite{ANASTOPOULOS2023169239}. In general, as is also pointed out in \cite{https://doi.org/10.48550/arxiv.2109.03187}, it is not obvious in general how to establish standard causality conditions in the path-integral formalism (`in-out' formalism) beyond scattering theory (see discussion in \cite{blum_state_2017}). The CTP formalism (Schwinger-Keldysh or ‘in- in’ formalism \cite{schwinger1961brownian,Keldysh:1964ud}) is better suited for analysing the causal behaviour of \textit{local} QFT measurement, thanks to its emphasis to real-time causal evolution \cite{Anastopoulos_2021}. Also, the QTP program demonstrates how the CTP formalism provides the `right' correlation functions that go into the definition of joint probability distributions over outcomes of coarse-grained pointer variables that are locally coupled to the field \cite{ANASTOPOULOS2023169239}. Time is also treated as a random variable (in analogy to stochastic processes) and time-of-arrival problems can be described accordingly. It is work in progress to evaluate the causal behaviour in bipartite scenarios and in multi-partite Sorkin-type set-ups using this framework, and to fully analyse how the possibilities of signalling are encoded in the CTP correlation functions.

\section{Conclusion}\label{conclusion}


Sorkin's `impossible measurements' problem serves as both a motivation for formulating an account of measurement that is compatible with QFT and a guide to possible strategies for carrying out this program. Sorkin \cite{sorkin1993impossible} and Borsten, Jubb, and Kells \cite{Borsten:2019cpc} present examples that show that the na{\"i}ve application of L{\"u}ders' rule to model non-selective measurements in relativistic quantum theory can lead to superluminal signalling. We explicitly presented these examples as a reductio argument in Sec. \ref{reductio}. Identifying the problematic set of premises is useful for distinguishing the `impossible measurements' problem from other foundational issues raised by QFT. `Impossible measurements' are not caused by L{\"u}ders' rule failing to be manifestly Lorentz covariant due to the relativistic temporal ordering assumption P3(c). Making L{\"u}ders' rule manifestly Lorentz covariant does not solve the `impossible measurements' problem; however, the approaches to responding to the `impossible measurements' problem surveyed in Sec. 3-6 all seek to replace the relativistic temporal ordering assumption P3(c) in the course of developing manifestly Lorentz covariant alternatives to applying L{\"u}ders' rule directly to the field system. The `impossible measurements' problem is also unrelated to state-dependent features of QFT. In particular, the initial state of the field system is not required to be the vacuum state or any other state that satisfies the assumptions of the Reeh-Schlieder theorem. Of course, these features of QFT do need to be taken into account to have a complete understanding of measurement in QFT, as the discussions of selective measurement of vacuum states and the role of Type III von Neumann algebras throughout this paper indicate. However, these issues are not a cause of the `impossible measurements' problem.  

Sorkin-type `impossible measurement' scenarios clearly illustrate that Microcausality is not by itself sufficient to rule out superluminal signalling in relativistic quantum theories when L{\"u}ders' rule is used to model non-selective measurements. Assuming that the practical ability to signal superluminally is an unacceptable consequence of a relativistic quantum theory, responding to the `impossible measurements' reductio argument in Sec. \ref{reductio} requires rejecting or revising at least one of the premises and/or adding at least one premise that is sufficient to block the conclusion. Strategies for formulating an account of measurement for QFT can be distinguished according to how they respond to the reductio argument. The FV measurement framework for AQFT introduces additional physical principles for QFT and formulates a new measurement theory for QFT that is informed by QMT for non-relativistic quantum mechanics but differs both formally and interpretationally. It can be shown that superluminal signalling does not occur when Sorkin-type measurement scenarios are modeled using the FV framework \cite{PhysRevD.103.025017}. The detector models approach rejects the reductio argument's premise that L{\"u}ders' rule is directly applied to the field system; instead, measurements are modeled by coupling the field system to a detector that is represented using NRQM and then performing projective measurements on this detector that may be evaluated using L{\"u}ders' rule. Detector models are constructed on a case-by-case basis. For Unruh-DeWitt-type models, superluminal signalling can be ruled out FAPP for Sorkin-type scenarios. The histories-inspired approach that is preferred by Sorkin takes the more radical approach of turning away from the representation of measurement processes `step by step' and instead assigning probabilities directly to entire histories. As far as we are aware, histories-based approaches have not yet achieved a complete resolution of the `impossible measurement' problem, but progress has been made. Of course, these three approaches are not the only options for either responding to the `impossible measurements' reductio argument or accounting for measurement in QFT.

There are important differences between these three approaches to accounting for measurement in QFT. The FV measurement framework proceeds axiomatically in a general fashion; in contrast, the detector models approach focuses on constructing concrete models of detectors and the systems to which they are coupled on a case-by-case basis. The differences between the principled and pragmatic attitudes that are adopted as well as in the goals of these two research programs lead to differences in the resulting models for measurement. Most obviously, the state update rules for the field system differ in the FV framework and the detector-based measurement theory. The state update rules are derived from more basic principles within the FV framework and posited based on plausibility arguments in the detector-based measurement theory. On the other hand, the detector-based approach has a wider scope of applicability at present than the FV framework (though there may also be measurement scenarios treatable within the FV framework but not using detector-based models). A further difference of opinion concerns how best to deal with Type III von Neumann algebras when representing measurements on quantum fields. The histories-informed approaches differ from both of the other approaches insofar as they do not model the measurement process using state update rules at all.

While the differences are important, the similarities between these approaches are also revealing. The development of a fully satisfactory account of measurement in QFT is still a work in progress, so similarities may offer a glimpse of general features of solutions to the `impossible measurements' problem or even of a measurement theory for QFT. An important moral is that the dynamics is crucial for diagnosing and addressing the `impossible measurements' problem. The FV framework's exclusion of superluminal signalling relies on the Local Time-Slice Property, which is a dynamical principle of AQFT. In detector models, the fact that the currents associated with the detector modeled using NRQM do not satisfy microcausality is the source of superluminal signalling in impossible measurement scenarios. This problem can be addressed on a case-by-case basis by performing calculations that involve the interaction Hamiltonian to determine the regime in which non-relativistic effects are negligible. In the histories-inspired approach, the decoherence functional includes information about the local dynamics of the system as well as the kinematics. Solving the `impossible measurements' problem is still an open problem in this approach, but a possible solution would rely on the decoherence functional (see e.g. \cite{albertini2023ideal}). Overall, histories-based approaches can lead to notions of causality that go beyond scattering. 

Another moral is that both the FV framework and the detector models approach dispense with the traditional operational interpretation of a local algebra of observables $\mathcal{A}(O)$ as representing operations that it is possible to carry out in region $O$. In the FV framework, observables can generally be localized in many different regions. The fact that an operation can be performed in a local lab region is instead represented by explicitly identifying a region $K$ in which the field and probe interact and noting that the coupled and uncoupled algebras may only be related by isomorphisms outside of the causal hull of $K$. The detector models approach does involve introducing smearings, but the support for the smearing function need not be interpreted as representing the region in which the operation represented by a smeared field operator is performed. The choice of a smearing function is a pragmatic one that is not limited to functions with support in the detector-field interaction region; furthermore, the most natural interpretation of the spacetime smearing function is as a holistic property of the detector-field interaction. The same detector system coupled to the field through different physical interactions can lead to different interaction regions. Finally, further work is required for interpreting the local propositions over local regions in histories-based approaches.

The historical context for contemporary programs for developing a measurement theory for local measurements in QFT was set by the choice to formulate QED in terms of asymptotic scattering theory in the 1940's \cite{blum_state_2017}. This was a departure from NRQM, in which instantaneous states at a (finite) time or stationary states are primary. It is interesting to note that none of the three approaches examined in Sec. 4-6 re-introduce instantaneous states-at-a-time (see discussion in \cite{Fraser2023}). Instead of regarding a state in a Hilbert space as the paradigm representative of the physical state, expectation values (or correlation functions) are the primary representatives of the physical state. In the FV framework, the algebraic state $\omega(A)$ represents the expectation value of $A$ in state $\omega$. Similarly, in the detector-based measurement theory, the $n$-point functions are the primary objects that figure in the state update rules for the field system. In the histories-inspired approaches, the decoherence functional is the primary object that generates the probabilistic predictions of the theory. Furthermore, in all three approaches expectation values at different times are needed. In the FV framework, the scattering isomorphism implements a finite-time scattering theory that facilitates the representation of `in' and `out' expectation values. Furthermore, the state $\varpi$ on $\mathcal{C}(M)$ for the coupled probe-field system encodes the expectation values of the field over all local regions. In the detector-based measurement theory, the $n$-point functions directly involve fields at different times. The state update rules also introduce a scattering operator. And, of course, in histories-inspired approaches, multi-time histories are the central quantities of physical significance. 
The final moral is that the state update rules for the field system cannot be literally interpreted as representing a physical change of state that occurs in some spacetime region in either the FV framework or the detector-based measurement theory. In the former case, the `in' and `out' states over the uncoupled algebra are best understood as having a counterfactual interpretation, as we argued. Moreover, state updates for successive selective measurements may be evaluated either sequentially or jointly. This is a reflection of the fact that the FV state update rules are manifestly Lorentz covariant. The detector-based state update rules are interpreted by Polo-G{\'o}mez, Garay, and Mart{\'i}n-Mart{\'i}nez as representing an update of the observer's state of information about the field system, not as representing an observer-independent change in the physical state of the field system. Even if this interpretation of the state update rules is contested, the fact remains that, for a selective measurement, the state update for the field depends on the spacetime location in which the measurement on the detector is performed, which is incompatible with interpreting the state update as representing a physical change in the field that is brought about by measurement.

This final moral is, of course, highly suggestive. Sorkin-type `impossible measurements' are not a symptom of the Measurement Problem. However, conversely, solutions to the `impossible measurement' problem could affect how the Measurement Problem is framed in QFT. More concretely, one version of the Measurement Problem in NRQM is that a literal interpretation of the dynamical evolution of the state is in general inconsistent with a literal interpretation of the state update rule following a measurement. The interpretative consequences of proposed accounts of measurement in QFT is an important direction for future research. (See \cite{FraserPhilosophy} for a starting point.) Given that all of the proposals reviewed in this paper are works in progress with open questions and challenges, this is only one of many directions for future research. We hope that this paper encourages continued development of a diverse range of approaches, as well as further investigation of the relationships among them.

\bibliography{refs}
\end{document}